\documentclass[oneside,11pt]{article}

\usepackage{mathrsfs,amsfonts,amssymb,amsmath,amsthm,latexsym,verbatim,epsfig}
\usepackage{color,url,pgf,tikz}
\usepackage{graphicx,amsbsy,amscd,bm}
\usepackage{floatrow}
\usepackage{subfigure}

\newcommand{\R}{{\mathbb R}}

\DeclareMathOperator*{\var}{Var}

\DeclareMathOperator*{\E}{\mathbf{E}}
\DeclareMathOperator*{\argmin}{argmin}

\DeclareMathOperator*{\conv}{conv}

\DeclareMathOperator*{\p}{\mathbf{P}}
\providecommand{\wt}[1]{\widetilde{#1}}
\providecommand{\wh}[1]{\widehat{#1}}

\providecommand{\norm}[1]{\left \lVert#1 \right \rVert}
\providecommand{\nnorm}[1]{ \lVert#1 \rVert}
\newcommand{\scp}[2]{\left\langle#1, #2\right\rangle}

\providecommand{\mc}[1]{\mathcal#1}
\providecommand{\T}{\top}
\newcommand{\sign}{\operatorname{sign}}

\newcommand{\bou}[2]{\Big\lvert _{{#1}}^{#2}}

\newcommand{\blanco}[1]{  }

\newcommand{\deriv}[3]{%
\ifthenelse{#1 = 1}{\frac{d\,#2}{d\,#3}}{\frac{d^{{#1}} #2}{d{#3}^{{#1}}}}
}

\newcommand{\partials}[3]{%
\ifthenelse{#1 = 1}{\frac{\partial\,#2}{\partial\,#3}}{\frac{\partial^{#1}
    #2}{\partial#3^{#1}}}
}
\def\sm{\sum_{i=1}^m}

\def \coloneq{\mathrel{\mathop:}=}

\def \eps{\varepsilon}

\newtheorem{theo}{Theorem}
\newtheorem{propo}{Theorem}

\newtheorem{lemmachen}{Theorem}

\newtheorem{definitioA}{Theorem}[section]

\newtheorem{lemmaA}{Theorem}[section]

  \newtheorem{prop}[propo]{Proposition}
  
  \newtheorem{lemma}[lemmachen]{Lemma}

\newtheorem{defnApp}[definitioA]{Definition}
\newtheorem{lemmaApp}[lemmaA]{Lemma}

\newenvironment{bew}{\begin{proof}[Proof]}{\end{proof}}

\newfloatcommand{capbtabbox}{table}[][\FBwidth]

\begin{document}

\title{\Large {Linear signal recovery from $b$-bit-quantized linear
measurements: precise analysis of the trade-off between bit depth and number
of measurements}}
\author{
\textbf{Martin Slawski}\\
         Department of Statistics and Biostatistics\\
         Department of Computer Science\\
         Rutgers University\\
          Piscataway, NJ 08854, USA\\
       \texttt{martin.slawski@rutgers.edu}
       \and
       \textbf{Ping Li}\\
         Department of Statistics and Biostatistics\\
         Department of Computer Science\\
         Rutgers University\\
          Piscataway, NJ 08854, USA\\
       \texttt{pingli@stat.rutgers.edu}
}
\date{}
\maketitle

\begin{abstract}
\noindent We  consider the problem of recovering a high-dimensional structured signal from independent Gaussian linear measurements
each of which is quantized to $b$ bits. Our interest is in linear approaches to signal recovery, where ``linear'' means that non-linearity
resulting from quantization is ignored and the observations are treated as if they arose from a linear measurement model. Specifically,
the focus is on a  generalization of a method for one-bit observations due to Plan and Vershynin {\small{[\emph{IEEE~Trans.~Inform.~Theory, \textbf{59} (2013), 482--494}]}}. At the heart of the present paper is a precise characterization of the optimal trade-off between the number of measurements $m$ and the bit depth per measurement $b$ given a total budget of $B = m \cdot b$ bits when the goal is to minimize the $\ell_2$-error in estimating the signal. It turns out that the choice $b = 1$ is optimal for estimating the unit vector (direction) corresponding to the signal for any level of additive Gaussian noise before quantization as well as for a specific model of adversarial noise, while the choice $b = 2$ is optimal for estimating the direction and the norm (scale) of the signal. Moreover, Lloyd-Max quantization is shown to be an optimal quantization scheme w.r.t.~$\ell_2$-estimation error. Our analysis is corroborated by numerical experiments showing nearly perfect agreement with our theoretical predictions. The paper is complemented by an empirical comparison to alternative methods of signal recovery taking the non-linearity resulting from quantization into account. The results of that comparison point to a regime change depending on the noise level: in a low-noise setting, linear signal recovery falls short of more sophisticated competitors while being competitive in moderate- and high-noise settings.
\end{abstract}

%\vspace{-1.5ex}
\section{Introduction}\label{sec:introduction}
% \vspace{-0.9ex}

One of the celebrated results in compressed sensing (CS) states that it is possible to recover a high-dimensional signal $x^* \in \R^n$ from
a small number $m$ of Gaussian linear measurements if 1) $x^*$ exhibits ``low-dimensional structure'' and 2) signal recovery is tailored to the
underlying low-dimensional structure. Moreover, 2) can typically be accomplished in a computationally tractable manner, e.g.,~by solving a linear program. There
is an enormous amount of literature on the subject spanning different areas, in particular mathematics, computer science, and engineering; we
refer to \cite{riceCS} for an overview.

The concept of signal recovery from incomplete data in the sense of having available less measurements than would ordinarily be required has
subsequently been developed further by considering settings in which the linear measurement process is subject to quantization, with the
extreme case of single-bit quantization (e.g.,~\cite{BoufounosBaraniuk2008, Gopi2013, Gupta2010, Jacques2013a, Report:Li_1bitCS15, PlanVershynin2013b, PlanVershynin2013a, Zhang2014}.
In general, one can think of $b$-bit quantization, $b \in \{1,2,\ldots\}$. Assuming that one is free in choosing $b$ and a corresponding scalar quantizer given
a fixed budget of bits $B = m \cdot b$ yields a trade-off between the number of measurements $m$ and the bit depth  $b$ per measurement. An optimal balance of these
two quantities minimizes a criterion of interest like the $\ell_2$-error in recovering the signal. Such trade-off arises naturally in the presence of
communication constraints. For example, signal acquisition and signal recovery may have to be carried out at different locations, and transmitting the acquired
data is subject to a limited rate. The optimal trade-off depends on the signal, the noise mechanism and the noise level, the suitability of the
parameters of the scalar quantizer relative to the signal, and the specific approach used for signal recovery. The dependency on the latter may be sidestepped
by considering information-theoretical lower and upper bounds. The corresponding analysis is valuable as it would yield fundamental limits (see the survey \cite{Boufounos2015}
for such limits in specific settings), but it does not necessarily have immediate practical consequences unless there exists computationally tractable recovery
algorithms achieving those limits. In this paper, we follow a different route by focusing on ``linear'' signal recovery as proposed in \cite{PlanVershynin2013a} with
follow-up work in \cite{PlanVershynin2015, PlanVershyninYudovina2014}. Here, ``linear'' means that non-linearity resulting from quantization is ignored, and that the observations
are treated as if they arose from a linear measurement model. Linear signal recovery may appear overly simple. In fact, it is known to be suboptimal in a noiseless setting (i.e.,~the
only source of distortion is quantization), with an $\ell_2$-error decaying with $m^{-1/2}$ compared to $m^{-1}$ \cite{Boufounos2015}. In spite of that, there is quite some justification
for having a closer look at the linear approach. It turns out that ignoring non-linearity does not have a dramatic effect in a regime where the norm of the signal and the level of
additive noise are comparable, in which case the $\ell_2$-error of linear signal recovery can only be improved in terms of a multiplicative constant. Empirically, as is demonstrated herein, the improvements achieved by more sophisticated methods tend to be rather small. Moreover, linear signal recovery typically comes with minimum requirements in terms of computation and storage. Apart from that, linear signal recovery constitutes a natural baseline. It is thus helpful to understand the aforementioned trade-off between the number of measurements and bit depth in this simple case.

\vspace{-2ex}
\paragraph{Outline and summary of contributions.} In $\S$\ref{sec:linearrecovery}, we provide an overview on linear signal recovery as pioneered in \cite{PlanVershynin2013a, PlanVershynin2015, PlanVershyninYudovina2014} and adopt the specific formulation in \cite{PlanVershynin2013a} for estimating the ``direction'' $x_u^* \coloneq x^*/\nnorm{x^*}_2$. The corresponding analysis of the trade-off between $m$ and $b$ when estimating $x_u^*$ is laid out in $\S$\ref{sec:analysis}. The analysis builds on ideas in \cite{PlanVershynin2013a, PlanVershyninYudovina2014} to a considerable extent, and it is complemented by modifications/extensions
to deal with the specific measurement model of $b$-bit quantization. Out of that analysis, we deduce explicit, easy-to-compute expressions for the relative performance of $b$-bit vs.~$b'$-bit measurements ($b \neq b'$) under three different noise models (additive Gaussian noise before quantization, adversarial bin flips, and random bin flips). We are then in position to
decide on the optimal choice of $b$ given a fixed budget of bits. It turns out that the choice $b = 1$ is optimal in the noiseless case, under additive Gaussian noise as well as under adversarial bin flips. In $\S$\ref{sec:scale}, we discuss an issue that has largely been neglected in the literature, namely the estimation of the ``scale'' $\psi^* \coloneq \nnorm{x^*}_2$. We show that as long as $b \geq 2$, this can be done at a fast rate by maximum likelihood estimation, separately from estimating $x_u^*$. Combining the results for estimating the direction and those for estimating the scale, we conclude that for the specific recovery algorithm under consideration, it does not pay off to take $b > 2$. Along the way we prove that classic Lloyd-Max quantization
\cite{Lloyd1982,Max1960} constitutes an optimal $b$-bit quantization scheme in the sense that it leads to a minimization of an
upper bound on the $\ell_2$-estimation error. Our theoretical results are corroborated by numerical experiments in $\S$\ref{sec:experiments}. Regarding the relative performance of one-bit vs.~two-bit measurements, the experimental results sharply agree with our theoretical predictions. One set of experiments sheds some light on the performance of linear signal recovery under study relative to alternative approaches. While the performance of the former is noticeably inferior to more sophisticated methods in a low-noise setting, it becomes competitive as the noise level increases. Altogether, the findings of this paper point to the conclusion that noisy settings are the domain of a two-fold simple approach, consisting a basic recovery algorithm and measurements of a low bit depth (i.e.,~one- or two-bit measurements). More conclusions can be found in $\S$\ref{sec:conclusion}. Proofs and complementary derivations have been moved to the appendix.

The present paper considerably extends a previous conference publication of the authors \cite{Slawski2015}. In particular, the analysis therein is limited to sparse signals, whereas the
present paper covers general low-complexity signals as quantified by the Gaussian width. In addition, we provide an account on the asymptotic sharpness of our analysis, discuss an extension
to anisotropic measurements, draw additional connections to existing literature and present a more comprehensive set of numerical results.
\vspace{-2ex}
\paragraph{Related work.} There is a plethora of papers discussing various aspects of compressed sensing with quantization. We refer to \cite{Boufounos2015} for an excellent overview on the problem
including basic performance limits, different approaches to quantization and signal recovery, and the associated references. On the other hand, comparatively little seems to be known about the trade-off between the number of measurements and bit depth as it is in the focus of the present paper. An important reference in this regard is \cite{LaskaBaraniuk2011} where this very trade-off is studied
for sparse signals. The analysis in \cite{LaskaBaraniuk2011} concerns ``oracle-assisted'' least squares (i.e.,~least squares with knowledge of the set of non-zero entries of the signal) which is of
theoretical interest, but not a practical approach to signal recovery. The authors point out the role of the signal-to-noise ratio (SNR) for the optimal trade-off that leads to the distinction of
two basic regimes: the so-called measurement compression regime with high SNR, small $m$ and large $b$ as opposed to the quantization compression regime with low SNR, large $m$ and small $b$. Some of the results in our paper can be related to this finding.

We here focus on linear signal recovery. An alternative that is more principled as it uses knowledge about the quantizer is ``consistent reconstruction''. This approach has been studied in a series
of recent papers by Jacques and collaborators \cite{Jacques2013b, Moshtaghpour2106, Jacques2016}. At the moment, this line of research only addresses the case without noise. The trade-off between
$m$ and $b$ appears in \cite{Jacques2013b}, where a specific version of Iterative Hard Thresholding \cite{Blumensath2009} tailored to quantized measurements is considered. It is shown via experimental results (which are partially reproduced in $\S$\ref{sec:experiments}) that unlike the main result of the present paper, increasing $b$ beyond one or two yields improvements, which underlines that the trade-off can be rather different depending on the recovery algorithm used.

After submitting the conference paper \cite{Slawski2015}, we became aware of the work \cite{Thrampoulidis2015} in which Lloyd-Max quantization is found to be optimal for linear signal recovery as in the present paper. The derivation in \cite{Thrampoulidis2015} is not fully rigorous though as it is only shown that Lloyd-Max quantization yields a stationary point, whereas herein, global optimality is established.

\vspace{-2.2ex}
\paragraph{Notation.} For the convenience of the reader, we here gather notation used throughout the paper. For a positive integer $d$, we use the shortcut $[d] = \{1,\ldots,d\}$. $I(P)$ denotes the indicator function of expression
$P$ with $I(P) = 1$ if $P$ is true and $0$ otherwise. For a matrix $A = (a_{ij})_{1 \leq i \leq m, 1 \leq j \leq n}$, $a_i$ denotes the $i$-th row and
$A_j$ the $j$-th column, $i \in [m]$, $j \in [n]$. We use $\nnorm{A}$ for the spectral norm of $A$. The identity matrix of dimension
$d$ is denoted by $I_d$. For a vector $x$, we write $x_J$ for the sub-vector corresponding to an index set $J$. For $x, x' \in \R^n$, $x \odot x' = (x_j \cdot x_j')_{j=1}^n$. A class of signals is denoted by $\mc{K} \subset \R^n$ and we let $\mc{C} = \mc{K} \cap B_2^n$, where for $q \geq 1$, $B_q^n$ denotes the unit $\ell_q$-ball in $\R^n$. We further write $B_0(r;n) = \{x \in \R^n:\; \nnorm{x}_0 \leq r \}$, where $\nnorm{x}_0 = \sum_{j = 1}^n I(x_j \neq 0)$. The unit
sphere of $\R^n$ is denoted by $\mathbb{S}^{n-1}$. For a set $K \subseteq \R^n$, we write $|K|$ for its cardinality, $\text{conv}(K)$ for its convex hull, and for $a \in \R$, we let $a K = \{a \cdot x, x \in K\}$. The letter $g$ refers to a Gaussian random variable or a canonical Gaussian random vector, i.e.,~$g \sim N(0,1)$ or $g \sim N(0, I_n)$. The probability density function (pdf) and the cumulative
density function (cdf) of the standard Gaussian distribution are denoted by $\phi$ and $\Phi$, respectively. In addition to the usual Landau notation, we occasionally make use of the stochastic order symbol $O_{\p}$: a sequence of random variables $(\eps_n)$ satisfies
$\eps_n = O_{\p}(g(n))$ if for all $\delta > 0$ there is a finite $c$ such that $\p(|\eps_n/g(n)| > c) \leq \delta$ for all $n$.

\section{Linear signal recovery based on quantized linear measurements}\label{sec:linearrecovery}

In this section, we first fix the problem setup and then introduce the approach that will be studied in depth in subsequent sections.

\begin{figure}[h!]
\includegraphics[width = 0.78\textwidth]{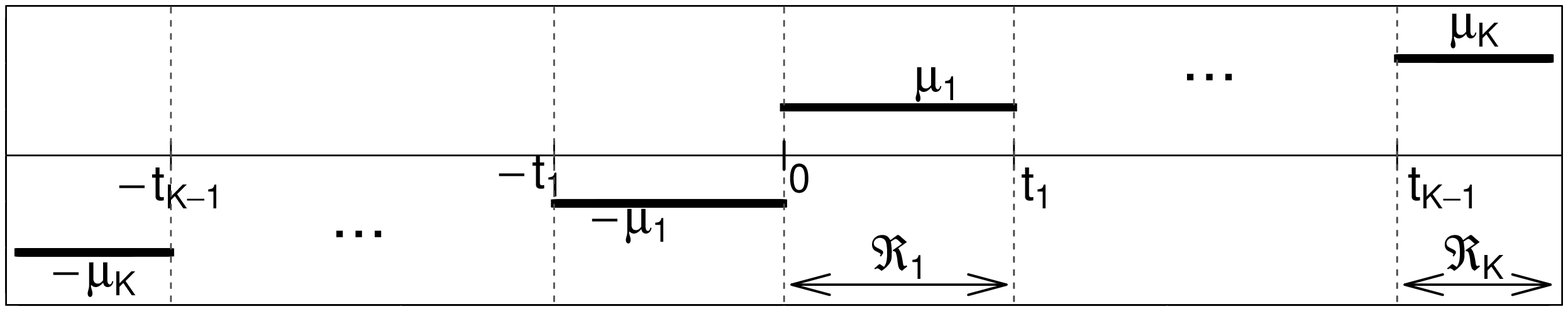}
\caption{Schematic representation of the quantization map and the associated quantities. Note that it suffices to
define $Q$ via a partitioning of $\R_+$, which then extends to a partitioning of $\R$ (as shown here) by symmetry. The map $Q$ is
a step function with levels $-\mu_K,\ldots,-\mu_1,\mu_1,\ldots,\mu_K$ and breakpoints $-t_{K-1},\ldots,-t_1,t_0,t_1,\ldots,t_{K-1}$.}\label{fig:quantization}
\end{figure}

\emph{Measurement model.} Let $x^* \in \mc{K} \subseteq \R^n$ be the signal to be recovered. We think of $\mc{K}$ as a set describing
a class of signals having a certain low-dimensional structure, e.g.,~$\mc{K} = B_0(s;n) \coloneq \{x \in \R^n:\;\norm{x}_0 \leq s\}$, the set of
$s$-sparse signals. More examples are given in $\S$\ref{sec:classes_signals}. The set $\mc{K}$ is assumed to be known. Let $A = (a_{ij})_{1 \leq i \leq m,  1 \leq j \leq n}$ be
a random matrix with i.i.d.~$N(0,1)$ entries whose rows and columns are denoted by $\{ a_i \}_{i = 1}^m$ and $\{ A_j \}_{j = 1}^n$, respectively.
The observations $y = (y_i)_{i = 1}^m$ arise from the model
\begin{equation}\label{eq:nonlinearCSmodel}
y_i = Q(\scp{a_i}{x^*} + \sigma \eps_i), \;\, i=1,\ldots,m,
\end{equation}
where $\eps = (\eps_i)_{i = 1}^m$ has i.i.d.~$N(0,1)$ entries and $\sigma \geq 0$ is referred to as ``noise level''; when $\sigma = 0$, we speak
of a noiseless setting. The map $Q: \R \rightarrow \mc{M}^{\pm}$ is called quantization map or quantizer, which is piecewise constant,
monotonically increasing, and odd. It partitions the real axis into $2^b$ bins where $b \in \{1,2,\ldots \}$ is the bit depth per measurement.
Because of symmetry, it suffices to define a partitioning of $\R_+$ into $K = 2^{b-1}$ bins $\{\mc{R}_{k} \}_{k = 1}^{K}$ resulting from distinct thresholds $\mathbf{t} = (t_1,\ldots,t_{K-1})^{\T}$ (in
increasing order) and $t_0 = 0$, $t_K = +\infty$ such that $\mc{R}_{1}=[t_0, t_1),\ldots,\mc{R}_{K} = [t_{K-1}, t_K)$. Each bin is assigned a distinct representative from the codebook $\mc{M}^{\pm} = -\mc{M} \cup \mc{M}$ with $\mc{M} = \{\mu_1,\ldots,\mu_K  \}$. Accordingly, $Q$ is defined as
\begin{equation}\label{eq:quantizer}
z \mapsto Q(z) = \sign(z) \sum_{k = 1}^K \mu_k I(|z| \in \mc{R}_k).
\end{equation}
For convenience, Figure \ref{fig:quantization} visualizes this definition. Since noise (if any) is added before $Q$ is applied, we speak of ``additive noise before quantization''. Other noise mechanism acting after quantization are possible, too; see $\S$\ref{sec:tradeoff} for specific examples.
%Note that for $b = 1$, one can only hope to recover the 'direction '$x_u^* = x^*/\nnorm{x^*}_2$.

\emph{Linear Signal Recovery.} Linear signal recovery is carried out by means of an estimator $\wh{\theta}(A,y) = \wh{\theta}(\mc{L}_A y)$ for
some linear map $\mc{L}_A:\,\R^m \rightarrow \R^n$ depending on $A$ and $\wh{\theta}$ only depending on $\mc{K}$ but not directly on $Q$. It is perhaps
surprising that even when restricting oneself to this class, it is possible to construct consistent estimators of the ``direction'' $x_u^* = x^* / \nnorm{x^*}_2$. This was first established by Brillinger \cite{Brillinger1982} in the traditional setting of asymptotic statistics with fixed parameter set and $m$ tending to infinity. It turns out that Gaussianity
of $A$ plays a crucial role here. Brillinger's result has been generalized recently to modern high-dimensional settings in various ways  \cite{PlanVershynin2013a, PlanVershynin2015,  PlanVershyninYudovina2014, Thrampoulidis2015}. We follow this line of research in the present paper. Linear signal recovery is limited to estimating $x_u^*$ as estimating the ``scale'' $\psi^* = \nnorm{x^*}_2$ entails using knowledge of $Q$. Indeed, expanding model \eqref{eq:nonlinearCSmodel}, we obtain
\begin{align}\label{eq:scaleissue}
y_i &= \sign(\scp{a_i}{x^*} + \sigma \eps_i) \textstyle\sum_{k=1}^K  \mu_k I(\,|(\scp{a_i}{x^*} +
\sigma \eps_i)| \in \mc{R}_k) \notag \\
&=  \sign(\scp{a_i}{x_u^*} + \tau \eps_i) \textstyle\sum_{k=1}^K \mu_k
I(\,|(\scp{a_i}{x_u^*} + \tau \eps_i)| \in \mc{R}_k/\psi^*), \; i \in [m],
\end{align}
where $\tau = \sigma / \psi^*$. We conclude that model \eqref{eq:nonlinearCSmodel} can always be transformed into one with $\psi^* = 1$ by re-scaling the thresholds $\mathbf{t}$ accordingly.
Hence, in order to be able to estimate $\psi^*$, the thresholds $\mathbf{t}$ have to be taken into account, while linear signal recovery discards this information. Moreover, a second
consequence is that when studying linear signal recovery with regard to the estimation of $x_u^*$ in the sequel, we may assume w.l.o.g.~that $\psi^* = 1$. Linear estimation of the direction
can trivially be combined with separate non-linear estimation of the scale, an approach whose discussion is postponed to $\S$\ref{sec:scale}.

\emph{Marginal Regression.} The choice $\mc{L}_A = m^{-1} A^{\T}$ gives rise to what we will refer to as the ``canonical linear estimator'':
\begin{equation}\label{eq:canlin}
\wh{x} \in \argmin_{x \in \mc{C}} -\scp{\eta}{x}, \quad \text{where} \; \eta \coloneq \frac{A^{\T} y}{m}, \; \; \, \text{and} \;\
\mc{C} \coloneq \mc{K} \cap B_2^n.
\end{equation}
This estimator is proposed in \cite{PlanVershynin2013a} for $\mc{K} = \sqrt{s} B_1^n$ and $b = 1$. In the conference paper preliminary to the present work \cite{Slawski2015}, $\mc{K} = B_0(s;n)$ and general $b$ are considered. The heading ``Marginal Regression'' is rooted in
the fact that $\eta$ is essentially proportional to the vector of univariate (or marginal) regression coefficients
$\wh{\beta} = (\wh{\beta}_j)_{j=1}^n$ corresponding to $n$ separate linear regressions of $y$ onto $A_j$, $j \in [n]$, which are given by
\begin{equation*}
\wh{\beta}_j = \frac{A_j^{\T} y}{{\nnorm{A_j}_2^2}} = \frac{A_j^{\T} y}{m + O_{\p}(\sqrt{m})}
\end{equation*}
which implies that $\eta_j = (1 + O_{\p}(m^{-1/2})) \wh{\beta}_j$, $j \in [n]$.

A formulation rather similar to \eqref{eq:canlin} is studied in \cite{PlanVershyninYudovina2014}.
\begin{equation}\label{eq:projectedmarginal}
\wh{x}_{\lambda} \in \lambda^{-1} \Pi_{\lambda \mc{K}}(\eta) = \lambda^{-1} \argmin_{x \in \lambda \mc{K}} \frac{1}{2}\nnorm{\eta - x}_2^2 = \lambda^{-1}  \argmin_{x \in \lambda \mc{K}}
\frac{1}{2} \nnorm{x}_2^2 - \scp{\eta}{x}
\end{equation}
where $\lambda > 0$ is defined by the relation $\E[A^{\T} y / m] = \lambda x_u^*$ (cf.~\eqref{eq:lambda} below) and $\Pi_{\lambda \mc{K}}$ denotes the
Euclidean projection on $\lambda \mc{K}$. As derived in Appendix \ref{app:equivalence}, if $\mc{K}$ is a cone (i.e.,~$\alpha \mc{K} = \mc{K}$ for all $\alpha >
  0$), then $\wh{x}_{\lambda}$ is identical to $\wh{x}$ in \eqref{eq:canlin} up to a constant of proportionality. This applies to
all except for one of the examples considered for $\mc{K}$ herein. In general, $\wh{x}/\nnorm{\wh{x}}_2$ and $\wh{x}_{\lambda}/\nnorm{\wh{x}_{\lambda}}_2$ are different though.  A notable disadvantage of \eqref{eq:projectedmarginal} compared to \eqref{eq:canlin} is that it
requires knowledge of $\lambda$ which depends on the (typically unknown) noise level $\sigma$. For this reason, we concentrate on
\eqref{eq:canlin} in the following.\\

\emph{Least Squares.} The choice $\mc{L}_A = (A^{\T} A)^{-1} A^{\T}$ would yield the least squares estimator. Since the inverse does not
exist once $m < n$, one needs to take advantage of low-dimensional structure. This yields the constrained least squares
estimator
\begin{equation}\label{eq:generalizedlasso}
\wh{x}_{\text{LS}} \in \lambda^{-1} \argmin_{x \in \lambda \mc{K}}  \frac{1}{2} x^{\T}  \frac{A^{\T} A}{m} x - \scp{\eta}{x}.
\end{equation}
In \cite{PlanVershynin2015} and \cite{Thrampoulidis2015}, this approach is studied under the names ``Generalized Lasso'' respectively ``$\mc{K}$-Lasso'' in allusion to the popular choice for the set $\mc{K}$ as an $\ell_1$-ball \cite{Tib1996}. Observe that \eqref{eq:projectedmarginal} differs from \eqref{eq:generalizedlasso} only in that the matrix $A^{\T} A / m$ is replaced by its expectation, the identity matrix. While this may appear as minor, an important consequence is that \eqref{eq:generalizedlasso} can achieve \emph{exact} recovery of $x^*$ as $b \rightarrow \infty$ and $\sigma = 0$ unlike \eqref{eq:projectedmarginal}. Outside the high signal-to-noise regime, however, both formulations indeed perform similarly as can be concluded from the analysis of \eqref{eq:projectedmarginal} in \cite{PlanVershyninYudovina2014} and the analysis of \eqref{eq:generalizedlasso} in \cite{PlanVershynin2015}. The latter yields a bound that is essentially of the form $\nnorm{\wh{x}_{\text{LS}} - x_u^*}_2 \leq c^{-1} \nnorm{\wh{x}_{\lambda} - x_u^*}_2$ resulting from a lower bound of the form $\frac{1}{m}\nnorm{A x}_2^2 \geq c \nnorm{x}_2^2$, $0 < c < 1$, for all $x$ in the so-called tangent cone of $\lambda \mc{K}$ at $\lambda x^*$. There are two more aspects that are relevant to a comparison.
\begin{itemize}
\item From the viewpoint of computation, the optimization problems \eqref{eq:canlin} and \eqref{eq:projectedmarginal} tend to be easier than the one in \eqref{eq:generalizedlasso} as the objectives in \eqref{eq:canlin},\eqref{eq:projectedmarginal} are linear respectively coordinate-wise separable.
%If the entries $\{a_{ij} \}$ are known to be i.i.d.~$N(0,1)$, there is apparently no reason to prefer \eqref{eq:generalizedlasso} over the simpler formulations \eqref{eq:canlin} and \eqref{eq:projectedmarginal} that already incorporate the fact that the rows of $A$ are i.i.d.~isotropic Gaussian. In particular,
%First, the analysis in \cite{PlanVershynin2015} yields an inferior bound that is essentially of the form $\nnorm{\wh{x}_{\text{LS}} - x_u^*}_2 \leq c^{-1} \nnorm{\wh{x}_{\lambda} - x_u^*}_2$ resulting from a lower bound of the form $\frac{1}{m}\nnorm{A x}_2^2 \geq c \nnorm{x}_2^2$, $0 < c < 1$, for all $x$ in the so-called tangent cone of $\lambda \mc{K}$ at $\lambda x^*$ (see [X] below).
\item More generally, one can consider the case of i.i.d.~anisotropic measurements, i.e.,~$a_i \sim N(0,\Sigma)$, $i \in [m]$. If $\Sigma$ is known, then \eqref{eq:canlin} and \eqref{eq:projectedmarginal} remain applicable with $\mc{L}_A = m^{-1}\Sigma^{-1} A^{\T}$, cf.~$\S$\ref{sec:anisotropic} below. The approach \eqref{eq:generalizedlasso} has its merits in the situation that $\Sigma$ is not known. In general, estimating $\Sigma$ resp.~its inverse is statistically more difficult and computationally more demanding than estimating $x^*$, hence the use of \eqref{eq:canlin} or \eqref{eq:projectedmarginal} combined with plug-in estimation of $\Sigma^{-1}$ is not a suitable option.
\end{itemize}

\section{Analysis}\label{sec:analysis}

The following section is dedicated to the analysis of the $\ell_2$-error of the canonical linear estimator \eqref{eq:canlin} under
the $b$-bit quantization model as defined by \eqref{eq:nonlinearCSmodel} and \eqref{eq:quantizer}. Our main result is an asymptotic
bound for $n \rightarrow \infty$ that allows for a precise (asymptotically sharp) characterization of the dependence on the bit depth $b$, the thresholds $\mathbf{t}$ and the representatives $\bm{\mu} = (\mu_1,\ldots,\mu_K)^{\T}$ parameterizing the quantization map. Given this result, we are in position to address the trade-off between $m$ and $b$. Along the way, we show that Lloyd-Max quantization \cite{Lloyd1982,Max1960} constitutes an optimal quantization scheme in the sense that it leads to a minimization of the error bound w.r.t.~$\mathbf{t}$ and $\bm{\mu}$. Finally, an extension to two other natural noise models is discussed.

At a technical level, the main ingredients of our analysis appear in related literature, in particular in \cite{PlanVershynin2013a}. Certain adjustments are necessary though to deal effectively with the specific measurement model herein, and to end up with a result suitable
for the purpose of studying the trade-off between $m$ and $b$.

\subsection{Preparations}\label{sec:preparations}

Our main result depends on three quantities and one condition which are given below. Throughout this section, we assume w.l.o.g.~that
$\psi^* = 1$ so that $x^* = x_u^*$ as this can always be achieved by re-scaling $\mathbf{t}$ and $\sigma$, cf.~\eqref{eq:scaleissue}.

\textbf{(Q1)} The first quantity has initially been introduced in
\cite{PlanVershynin2013a}.
\begin{equation}\label{eq:lambda}
\lambda = \lambda_b = \lambda_{b,\sigma} = \lambda_{b,\sigma}(\mathbf{t}, \bm{\mu}) \coloneq \E[\theta(g) g], \; \; g \sim N(0,1),
\end{equation}
where the map $\theta$ is defined by the relation
\begin{equation}\label{eq:theta}
\E[y_1 | a_1] = \theta(\scp{a_1}{x_u^*}).
\end{equation}
At a high level, $\lambda$ quantifies the distortion from linearity caused by quantization. It can be shown that (Appendix \ref{app:linearity}) that $\E[A^{\T} y / m] = \lambda x_u^*$, hence
$\lambda$ also equals the constant of proportionality up to which $x_u^*$ can be recovered by linear estimation.
The quantity $\lambda$ is positive, increases with $b$ and approaches one as $b \rightarrow \infty$. A precise expression for
$\lambda_{b,\sigma}(\mathbf{t}, \bm{\mu})$ is the content of Lemma \ref{lem:Omega_b} below.

\textbf{(Q2)} The second quantity is given by
\begin{equation}\label{eq:Psi}
\Psi = \Psi_{b,\sigma} = \Psi_{b, \sigma}(\mathbf{t}, \bm{\mu}) \coloneq \sqrt{\E[y_1^2]},
\end{equation}
which is simply the (marginal) standard deviation of the $\{ y_i \}_{i = 1}^m$. The error bound of Theorem \ref{theo:mainresult}
below is proportional to $\Psi$. We refer to Lemma \ref{lem:Omega_b} for a more specific expression for
$\Psi_{b, \sigma}(\mathbf{t}, \bm{\mu})$.

\textbf{(Q3)} The third quantity is a measure of complexity of the class of signals $\mc{K}$ under consideration, the
so-called Gaussian width of the tangent cone of $\mc{C} = \mc{K} \cap B_2^n$ at $x_u^*$, a notion which can be considered as standard in
the context of high-dimensional linear inverse problems \cite{Cai2016, Chandrasekaran2012, PlanVershynin2013a, PlanVershynin2015, PlanVershyninYudovina2014}. We set
\begin{equation}\label{eq:tangentcone}
\Delta(\mc{C}; x_u^*) \coloneq \{\alpha (x - x_u^*), \; \alpha \geq 0, \; x \in \mc{C}\}, \qquad \overline{\Delta}(\mc{C}; x_u^*) \coloneq \Delta(\mc{C}; x_u^*) \cap \mathbb{S}^{n-1}.
\end{equation}
We suppress dependence on $x_u^*$, i.e.,~we use $\Delta(\mc{C}) = \Delta(\mc{C}; x_u^*)$ and $\overline{\Delta}(\mc{C}) = \overline{\Delta}(\mc{C}; x_u^*)$ for the tangent cone of $\mc{C}$ at $x_u^*$ and its spherical part, respectively. The latter enters Theorem \ref{theo:mainresult} below via its Gaussian width. For $K \subset \R^n$ compact and $g \sim N(0, I_n)$, the Gaussian width of $K$ is defined by
\begin{equation}\label{eq:gaussianwidth}
w(K) \coloneq \E \left[\sup_{v \in K} \left|\scp{g}{v}\right| \right]
\end{equation}

\textbf{(C)} Finally, we require the following condition. At a technical level, rather than being fixed, we think of $x_u^*$ as a sequence $\{  x_u^{*,n} \}_{n \in \mathbb{N}}$ whose elements are contained in spheres of growing dimension, satisfying $\nnorm{x_u^{*,n}}_{\infty} \rightarrow 0$ as $n \rightarrow \infty$. This condition is easily met for typical signal classes of interest under natural sampling models for $x_u^*$ as is elaborated
in the discussion after Theorem \ref{theo:mainresult} below.

\subsection{Main result}

We now state our main result along with a brief general discussion. Further implications are subsequently discussed in separate subsections.
\begin{theo}\label{theo:mainresult} Consider the measurement model given by \eqref{eq:nonlinearCSmodel},\eqref{eq:quantizer} and consider the canonical linear estimator $\wh{x}$ given by \eqref{eq:canlin}. Under condition \emph{\textbf{(C)}}, as $n \rightarrow \infty$, we have
\begin{equation}\label{eq:mainbound}
\nnorm{\wh{x} - x_u^*}_2 \leq 24 \,\frac{\Psi_{b, \sigma}(\mathbf{t}, \bm{\mu})}{\lambda_{b, \sigma}(\mathbf{t}, \bm{\mu})}  \, \frac{w(\overline{\Delta}(\mc{C}))}{\sqrt{m}}.
\end{equation}
with probability at least $1 - 4(\exp(-w(\overline{\Delta}(\mc{C}))/2) + \exp(-cm)) = 1  - o(1) - 4\exp(-cm)$ as $n \rightarrow \infty$, for some constant $c > 0$.
\end{theo}
It turns out that the bound \eqref{eq:mainbound} does not leave much room for further improvement in general. The term $w(\overline{\Delta}(\mc{C}))/\sqrt{m}$ is identified as the rate of estimation.
It follows from existing literature \cite{Cai2016, Chandrasekaran2012} that in the presence of additive noise, one has a corresponding lower bound in terms of the
so-called Sudakov minoration of $w(\overline{\Delta}(\mc{C}))$ which typically yields matching upper and lower bounds modulo constant factors. For the examples of $\mc{K}$ resp.~$\mc{C}$ given in $\S$\ref{sec:classes_signals} below, the upper bounds on $w(\overline{\Delta}(\mc{C}))$ yield a match to known minimax lower bounds, e.g.,~\cite{CandesDavenport2013, CandesPlan2011, Lounici2011}.

We point out, however, that the rate in \eqref{eq:mainbound} can be suboptimal in the noiseless case. For $\mc{K} = B_0(s;n)$, it is shown in \cite{Jacques2013a} that there exists a (computationally intractable) recovery algorithm that achieves an error decay of the order $m^{-1}$ compared to $m^{-1/2}$ in \eqref{eq:mainbound}.

In the sequel, a lot of attention will be paid to the leading constant $\Psi_{b, \sigma}(\mathbf{t},\bm{\mu})/\lambda_{b, \sigma}(\mathbf{t},\bm{\mu})$. The dependency on
this quantity is asymptotically sharp as it is already encountered in the
traditional asymptotic setup in which $\mc{K} = \R^n$ with $n$ being of a smaller order of magnitude than
$m$. In fact, one can show (cf.~Appendix \ref{app:sharpness}) that under a double asymptotic framework in which $m, n \rightarrow \infty$, $n/m \rightarrow 0$ and \textbf{(C)} holds, for any $j \in [n]$
\begin{equation}\label{eq:asymptotics_univariate}
\sqrt{m} \left(\wh{x}_j  - x_{u,j}^* \right) \overset{\mc{D}}{\rightarrow} N \left(0, \frac{\Psi_{b, \sigma}^2(\mathbf{t},\bm{\mu})}{\lambda_{b, \sigma}^2(\mathbf{t},\bm{\mu})} \right),
\end{equation}
where $\overset{\mc{D}}{\rightarrow}$ denotes convergence in distribution. In other words, the estimation error for any single coordinate is proportional to the leading constant of our bound.
%which thus reads as a product of a complexity
%term depending on the structure of the object one aims to recover and the$\Psi_{b, \sigma}(\mathbf{t},\bm{\mu})/\lambda_{b, \sigma}(\mathbf%{t},\bm{\mu})$.

The numerical constant ``$24$'' in the bound (\ref{eq:mainbound}) does not appear to be optimal. As it comes to the key point of the paper, namely the ratio of estimation errors for different choices of the bit depth $b$,
this is not an issue as all terms not depending on $b$ cancel out.

\subsection{Classes of signals}\label{sec:classes_signals}

Theorem \ref{theo:mainresult} can be specialized to popular signal classes by bounding the associated Gaussian widths. We here provide
several examples including a short discussion of computational aspects. We also discuss condition (\textbf{C}) in light of those examples. Derivations have been relegated to Appendix \ref{app:classes_signals}.

1) \emph{Sparsity.} $\mc{K} = B_0(s;n)$\\[1ex]
Following the argument in the proof of Lemma 2.3 in \cite{PlanVershynin2013a}, one can show that
$$w(\overline{\Delta}(\mc{C})) \leq w(B_0(2s;n) \cap \mathbb{S}^{n-1}) \leq \sqrt{2s} + \sqrt{4 s \log \left(\frac{e n}{2s} \right)} +
\sqrt{\pi/2} \leq 3.5 \sqrt{2s \log \left(\frac{e n}{2s} \right)},$$
i.e.,~we recover the usual rate $\left( \frac{s \log(n/s)}{m} \right)^{1/2}$ in Theorem \ref{theo:mainresult}.

2) \emph{Fused Sparsity.} Let $D:\R^n \rightarrow \R^{n-1}$ denote the first-order difference
operator and set $\mc{K} = \text{PC}(s;n) \coloneq \{x \in \R^n:\, Dx \in
B_0(s;n-1) \}$; this is the set of all signals in $\R^n$ that are piecewise
constant with $s$ breakpoints. One can show that
$w(\overline{\Delta}(\mc{C}))$ satisfies the same upper bound as in 1).

3) \emph{Group Sparsity.} Let $\mc{G} = \{ G_{\ell} \}_{\ell = 1}^L$ be a partition of $[n]$  into $L$ groups. Define $\nnorm{\cdot}_{0,\mc{G}}$ by
$|\{\ell \in [L]: x_{G_{\ell}} \neq 0 \}|$ and for $1 \leq r \leq L$, let $B_{0,\mc{G}}(r) = \{x \in \R^n:\, \nnorm{x}_{0,\mc{G}} \leq r \}$. Consider $\mc{K} = B_{0,\mc{G}}(s)$, in which
case we say that $x^*$ is $s$-group sparse (w.r.t.~the partition $\mc{G}$). Then, $w(\overline{\Delta}(\mc{C})) \leq 2 \sqrt{2s \max_{\ell \in [L]} |G_{\ell}|} + \sqrt{4s \log(e L / 2s)}$.

4) \emph{Low-rank matrices.} Our framework can accommodate a set of matrices $\mc{K} \subseteq \R^{n_1 \times n_2}$ by identifying $\R^{n_1 \times n_2}$ with $\R^{n_1 n_2}$. Consider
$\mc{K} = \{X \in \R^{n_1 \times n_2}:\,\text{rank}(X) \leq s \}$ and accordingly $\mc{C} = \mc{K} \cap \{X \in \R^{n_1 \times n_2}: \, \nnorm{X}_F \leq 1\}$, with
$\nnorm{\cdot}_F$ as the Frobenius norm. Then, $w(\overline{\Delta}(\mc{C})) \leq \sqrt{2 s n_1} + \sqrt{2 s n_2}$.

5) \emph{$\ell_1$-ball constraint.} $\mc{K} = \mc{C} = r^* B_1^n \cap B_2^n$, $\;$$r^* =
\nnorm{x_u^*}_1$. In addition, suppose that $x^* \in B_0(s;n)$. Then $w(\overline{\Delta}(\mc{C})) \leq 2 \sqrt{2}
w(B_0(2s;n) \cap \mathbb{S}^{n-1})$, and we may resort to the bound in 1).\\

In the same way as 5) arises as the convex counterpart to 1), one can consider a total
variation constraint in place of 2), an $\ell_1-\ell_2$ ball (``group lasso'' \cite{Yuan2006})
constraint in place of 3), and a Schatten-one norm ball constraint in place 4). Under a sparsity assumption for $x^*$,
the Gaussian widths of the convex formulations equal -- up to numerical constants -- those of the corresponding non-convex formulations; for the sake of brevity and since this is known in the literature, we omit explicit statements/derivations here.\\

\emph{Computation.} The constraint sets in examples 1) to 4) are non-convex. Nevertheless, due to the simplicity of the objective in
Eq.~\eqref{eq:canlin} and the specific structure of the constraint sets, all of the resulting optimization  problems are computationally tractable; 1) and 3)
even have closed form solutions, 4) can be reduced to a singular value decomposition, and 2) can be solved in $O(n^2 s)$ flops by dynamic programming \cite{Bellman1961}. A formal derivation for 1) is contained in Appendix \ref{app:computation} representative for 1) -- 3). For 2), the convex counterpart has a much better computational complexity which scales only linearly in $n$ \cite{Kim2009}, while achieving comparable statistical performance.\\

\emph{Discussion of \emph{\textbf{(C)}}.} For examples 1) -- 4) above, Condition \textbf{(C)} can be shown to be satisfied with probability tending to one as $n \rightarrow \infty$ when sampling $x_u^*$ uniformly at random according to natural generating mechanisms.

1) For a given sparsity level $s$, pick the support of $x^*$ at
random and sample the non-zero entries i.i.d.~from a distribution with finite
fourth moment, and normalize to unit 2-norm. It follows from Markov's
inequality that \textbf{(C)} is satisfied with probability tending to one as
$n,s \rightarrow \infty$.

2) Pick a random partition $\{1,\ldots, i_{k_1}\}$,$\{i_{k_1 + 1} ,\ldots, i_{k_2}\},\ldots,\{i_{k_{s-1}
  +1},\ldots,n\}$ of $\{1,\ldots,n\}$ such that $n_k / n \rightarrow c_k$,
$\sum_{k = 1}^s c_k = 1$ as $n \rightarrow \infty$, where $n_k$ denotes the number of elements in the
$k$-th element of the partition, $k \in [s]$. For each of those, sample the corresponding entries
at random from a distribution with finite first moment, scale them by the square root of the respective block size and then normalize
to unit $2$-norm.

3) Suppose that as $n \rightarrow \infty$, the sizes of
the non-zero blocks are of the same order. Sampling the non-zero entries as in
example 1), condition \textbf{(C)} is fulfilled with probability tending
to one as the block sizes of the non-zero blocks or the number of non-zero
blocks go to infinity.

4) Draw a random $n_1 \times n_2$ matrix with
i.i.d.~$N(0,1)$-entries, $n_1, n_2 > r$, and compute its SVD. Keep the top $r$
left and right singular vectors and replace the corresponding singular values
by an arbitrary element of $\mathbb{S}^{r-1}$.

Clearly, there may exist different or more general sampling schemes for \textbf{(C)} to be
satisfied.

\subsection{Extension to anisotropic measurements}\label{sec:anisotropic}

We extend Theorem \ref{theo:mainresult} to the case of anisotropic Gaussian measurements. More
precisely, we now suppose that the $\{ a_i \}_{i = 1}^m$ are i.i.d.~from a $N(0, \Sigma)$-distribution where $\Sigma$ is
invertible and assumed to be known. In the anisotropic case, the linear estimator \eqref{eq:canlin} is replaced
by
\begin{equation}\label{eq:canlin_anisotropic}
\wh{x}_{\Sigma} \in \argmin_{x \in \mc{C}} -\scp{x}{\Sigma^{-1} \eta},
\end{equation}
with $\eta = A^{\T} y / m$ as in \eqref{eq:canlin}. We then have the following counterpart to Theorem \ref{theo:mainresult}.
\begin{theo}\label{theo:anisotropic}
  Consider the anisotropic measurement model as above, let $\kappa(\Sigma^{1/2})$ denote the condition number of $\Sigma^{1/2}$,
and let $\wh{x}_{\Sigma}$ be as in \eqref{eq:canlin_anisotropic}. Under  condition \emph{\textbf{(C)}}, as $n \rightarrow \infty$, it holds that
\begin{equation}\label{eq:bound_anisotropic}
\nnorm{\wh{x}_{\Sigma} - x_u^*}_2 \leq 24 \kappa(\Sigma^{1/2})\,\frac{\Psi_{b, \sigma}(\mathbf{t}, \bm{\mu})}{\lambda_{b, \sigma}(\mathbf{t}, \bm{\mu})}  \, \frac{w(\overline{\Delta}(\mc{C}))}{\sqrt{m}},
\end{equation}
with probability at least $1 - 4(\exp(-\kappa(\Sigma^{1/2}) w(\overline{\Delta}(\mc{C}))/2) + \exp(-cm)) = 1  - o(1) - 4\exp(-cm)$ as $n \rightarrow \infty$, for some constant $c > 0$.
\end{theo}
The bound for the anisotropic case thus only involves the additional factor $\kappa(\Sigma^{1/2})$. Setting $\Sigma = I_n$, we recover
Theorem \ref{theo:mainresult}. Regarding the trade-off between $m$ and $b$ to be studied in the next sections, the extra factor does
not have any influence as it does not depend on $b$.

\subsection{Implications for the optimal trade-off between $m$ and $b$}\label{sec:tradeoff}

We now study in detail the implications of Theorem \ref{theo:mainresult} for the central question of this paper. Suppose we have a fixed budget of $B$ bits available and are
free to choose the number of measurements $m$ and the number of bits per measurement $b$ subject to $B = m \cdot b$ such that the $\ell_2$-error of $\nnorm{\wh{x} - x_u^*}_2$ of is as small as
possible. What is the optimal choice of $(m,b)$? At this point, we still confine ourselves to the direction $x_u^*$. In $\S$\ref{sec:scale} below, we provide an answer for $x^*$ in place of $x_u^*$ by linking
the findings of the present section to the results on scale estimation.

In virtue of Theorem \ref{theo:mainresult} and the comments that follow, the asymptotic $\ell_2$-error as $n \rightarrow \infty$ depends on $b$ only
via $\Omega_{b, \sigma}(\mathbf{t}, \bm{\mu}) \coloneq \Psi_{b,\sigma}(\mathbf{t}, \bm{\mu}) / \lambda_{b,\sigma}(\mathbf{t}, \bm{\mu})$. In a first step,
we provide more specific expressions for $\Psi_{b,\sigma}(\mathbf{t}, \bm{\mu})$ and $\lambda_{b,\sigma}(\mathbf{t}, \bm{\mu})$ that will prove useful
in subsequent analysis. Below, $\odot$ denotes the entry-wise (Hadamard) multiplication of vectors.

\begin{lemma}\label{lem:Omega_b} Let $\mc{R}_k = \mc{R}_k(\mathbf{t})$, $k \in [K]$, be as in \eqref{eq:quantizer}. We have $\lambda_{b,\sigma}(\mathbf{t}, \bm{\mu}) =
  \scp{\bm{\alpha}(\mathbf{t})}{\bm{E}(\mathbf{t}) \odot \bm{\mu}} / (1 +
  \sigma^2)$ and $\Psi_{b,\sigma}(\mathbf{t}, \bm{\mu}) = \sqrt{\scp{\bm{\alpha}(\mathbf{t})}{\bm{\mu}\odot
    \bm{\mu}}}$, where
\begin{alignat*}{3}
&\bm{\alpha}(\mathbf{t}) = \left( \alpha_1(\mathbf{t} ),
  \ldots, \alpha_K(\mathbf{t}) \right)^{\T}, \;\;
&&\alpha_k(\mathbf{t} ) = \p \left\{ |\wt{g}| \in \mc{R}_k(\mathbf{t} ) \right\}, \; \, &&\wt{g}\sim N(0, 1+\sigma^2), \; \, k \in [K],\\
&\bm{E}(\mathbf{t}) = \left( E_1(\mathbf{t} ), \ldots,
  E_{K}(\mathbf{t}) \right)^{\T}, \; \; && E_k(\mathbf{t}) = \E[\wt{g}| \wt{g}\in \mc{R}_k(\mathbf{t} )],
\; \, &&\wt{g}\sim N(0, 1+\sigma^2), \; \, k \in [K].
\end{alignat*}
\end{lemma}
%Hence it makes sense to compare two different choices $b$ and $b'$
%in terms of the ratio of $\Omega_b = $ and $\Omega_{b'} = \Psi_{b'} /
%\lambda_{b'}$. Since the bound \eqref{eq:marginal_nonlinear_bound} decays with
%$\sqrt{m}$, for $b'$-bit measurements, $b' > b$,  to improve over $b$-bit measurements with
%respect to the total \#bits used, it is then required that $\Omega_b /
%\Omega_{b'} > \sqrt{b' / b}$. The route to be taken is thus as follows: we
%first derive expressions for $\lambda_b$ and $\Psi_b$ and then minimize
%the resulting expression for $\Omega_b$ w.r.t.~the free parameters $\mathbf{t}$ and
%$\bm{\mu}$. We are then in position to evaluate $\Omega_b / \Omega_{b'}$ for $b
%\neq b'$.\\
\noindent\textbf{Optimal choice of $\mathbf{t}$ and $\bm{\mu}$.} In order to eliminate the dependence on $\mathbf{t}$ and $\bm{\mu}$, we minimize
$\Omega_{b,\sigma}$ w.r.t.~these two quantities. In this manner, we also obtain an optimal parameterization of the quantization map $Q$ yielding
minimum $\ell_2$-estimation error. It turns out that the solution coincides with that of the classical Lloyd-Max quantization problem \cite{Lloyd1982, Max1960} stated
below. Let $h$ be a random variable with finite variance and consider the optimization problem
\begin{equation}\label{eq:lloydmaxproblem}
\min_{\mathbf{t}, \bm{\mu}} \E[\{ h - Q(h;\mathbf{t}, \bm{\mu}) \}^2] =
\min_{\mathbf{t}, \bm{\mu}} \E[\{h - \sign(h) \textstyle \sum_{k = 1}^K \mu_k I(|h|
\in \mc{R}_k(\mathbf{t})\, ) \}^2].
\end{equation}
Problem \eqref{eq:lloydmaxproblem} can be solved by an iterative scheme known as the Lloyd-Max algorithm ($\S$3.2.3 in \cite{Gallager2006})
that alternates between optimization of $\mathbf{t}$ for fixed $\bm{\mu}$ and vice versa. For $h$ from a log-concave distribution (e.g.,~Gaussian) that scheme can be shown to deliver
the global optimum \cite{Kieffer1983}.
\begin{theo}\label{theo:lloydmax} Consider the minimization problem
$\min_{\mathbf{t},\bm{\mu}} \Omega_{b, \sigma}(\mathbf{t}, \bm{\mu})$ with $\Omega_{b,\sigma}(\mathbf{t},\bm{\mu}) = \frac{\Psi_{b,\sigma}(\mathbf{t},\bm{\mu})}{
\lambda_{b,\sigma}(\mathbf{t},\bm{\mu})}$. Its minimizer $(\mathbf{t}^*,
\bm{\mu}^*)$ equals that of the Lloyd-Max problem \eqref{eq:lloydmaxproblem}
for $h \sim N(0, 1 + \sigma^2)$. Moreover,
\begin{equation}\label{eq:Omega_b_star}
\Omega_{b,\sigma}(\mathbf{t}^*, \bm{\mu}^*) = \textstyle\sqrt{(\sigma^2 + 1)/\lambda_{b, 0} (\bm{t}_{0}^*, \bm{\mu}_{0}^*)},
\end{equation}
where $(\bm{t}_{0}^*, \bm{\mu}_{0}^*)$ denotes the optimal choice of $(\mathbf{t},\bm{\mu})$ in the sense of \eqref{eq:lloydmaxproblem} when $h \sim N(0, 1)$ (i.e.,~$\sigma = 0$).
\end{theo}
Regarding the choice of $(\mathbf{t}, \bm{\mu})$ the result of Theorem
\ref{theo:lloydmax} may not come as a surprise as the entries of $y$ are
i.i.d.~$N(0, 1+\sigma^2)$. It is less immediate though that this specific choice can also be
motivated as the one leading to the minimization of the error bound \eqref{eq:mainbound}.

The second part of Theorem \ref{theo:lloydmax} implies that the ratio $\Omega_{b,\sigma}(\mathbf{t}^*, \bm{\mu}^*)/ \Omega_{b',\sigma}(\mathbf{t}^*, \bm{\mu}^*)$ does not depend on $\sigma$. Combining Theorem \ref{theo:mainresult}, Lemma \ref{lem:Omega_b} and Theorem \ref{theo:lloydmax}, we are eventually in position
to determine the optimal trade-off between $m$ and $b$. Theorem \ref{theo:mainresult} yields that the $\ell_2$-error decays with $m^{-1/2}$. Therefore, for
$b'$ to improve over $b$ with $b < b'$ at the level of bits, it is required that $\Omega_b / \Omega_{b'} > \sqrt{b' / b}$. In fact, when using $b$ bits per measurement we may multiply
the number of measurements by a factor of $b' / b$ so that the bit budgets are balanced, i.e.,~$m_b \cdot b = m_{b'} \cdot b'$, where $m_b$ and $m_{b'}$ denote
the number of $b$-bit and $b'$-bit measurements, respectively. Using the Lloyd-Max algorithm to determine $(\mathbf{t}^*, \bm{\mu}^*)$ and invoking \eqref{eq:Omega_b_star} as well
as Lemma \ref{lem:Omega_b}, it is straightforward to evaluate the ratios $\Omega_b / \Omega_{b'}$ numerically. In Table \ref{tab:tradeoff}, we provide the results
for selected pairs of $b$ and $b'$.

\begin{table}[h!]
\begin{center}
\begin{tabular}{|l|l|l|l|}
\hline
                          &  $b=1$, $b' = 2$      & $b=2$, $b' = 3$ & $b=3$,
                          $b' = 4$\\
\hline
$\Omega_{b} / \Omega_{b'}$: & \textbf{1.178}                 &  \textbf{1.046}  & \textbf{1.013}\\
required for $b' \gg b$:  & $\sqrt{2}\approx \textbf{1.414}$ & $\sqrt{3/2} \approx \textbf{1.225}$
& $\sqrt{4/3} \approx \textbf{1.155}$ \\
\hline
\end{tabular}
\vspace*{-0.02\textheight}
\caption{Here $b' \gg b$ means that $b'$ is a superior choice relative to $b$ at the bit level.}\label{tab:tradeoff}
\end{center}
\end{table}

From these figures, we see that increasing $b$ reduces the error as the number of measurements are fixed. However,
the reduction is not substantial enough to yield an improvement when thinking in terms of a budget
of bits instead of measurements. Reducing $b$ from two to one increases the error by
a factor of $1.178$ which is below $\sqrt{2}$, the factor required for $b = 1$ to be inferior compared
to $b = 2$. The reduction factor for increasing $b$ becomes even smaller for the transitions from two to three
and three to four bits, and quickly approaches $1$. The overall conclusion is that for estimating $x_u^*$ the optimal
trade-off is achieved by one-bit quantization -- instead of increasing the bit depth $b$, one should rather
increase the number of measurements. This conclusion is valid regardless of the noise level as a consequence
of Theorem \ref{theo:lloydmax}. Even more, the figures in Table \ref{tab:tradeoff} assume optimal quantization for $b \geq 2$, and in
turn knowledge of $\sigma$, which may not be fulfilled in practice.\\

\noindent\textbf{Beyond additive noise.} Additive Gaussian noise is perhaps the
most studied form of perturbation, but one can of course think of numerous
other mechanisms whose effect can be analyzed along the path used for additive noise as long as it is feasible to obtain the corresponding
expressions for $\lambda$ and $\Psi$. We here do so for the following
mechanisms acting \emph{after} quantization \eqref{eq:quantizer}.

\noindent (I) \emph{Random bin flip.} For $i \in [m]$: with probability $1 - p$, $y_i$ remains unchanged. With probability $p$,
$y_i$ is changed to an element from $\mc{M}^{\pm} \setminus \{ y_i \}$ uniformly at random.

\noindent (II) \emph{Adversarial bin flip.} For $i \in [m]$: write $y_i = q \mu_k$ for
$q \in \{-1,1\}$ and $\mu_k \in \mc{M}$. With probability $1 - p$,
$y_i$ remains unchanged. With probability $p$, $y_i$ is changed to $-q
\mu_K$.

\begin{figure}
\begin{center}
\begin{tabular}{ccc}
\includegraphics[width = 0.36\textwidth]{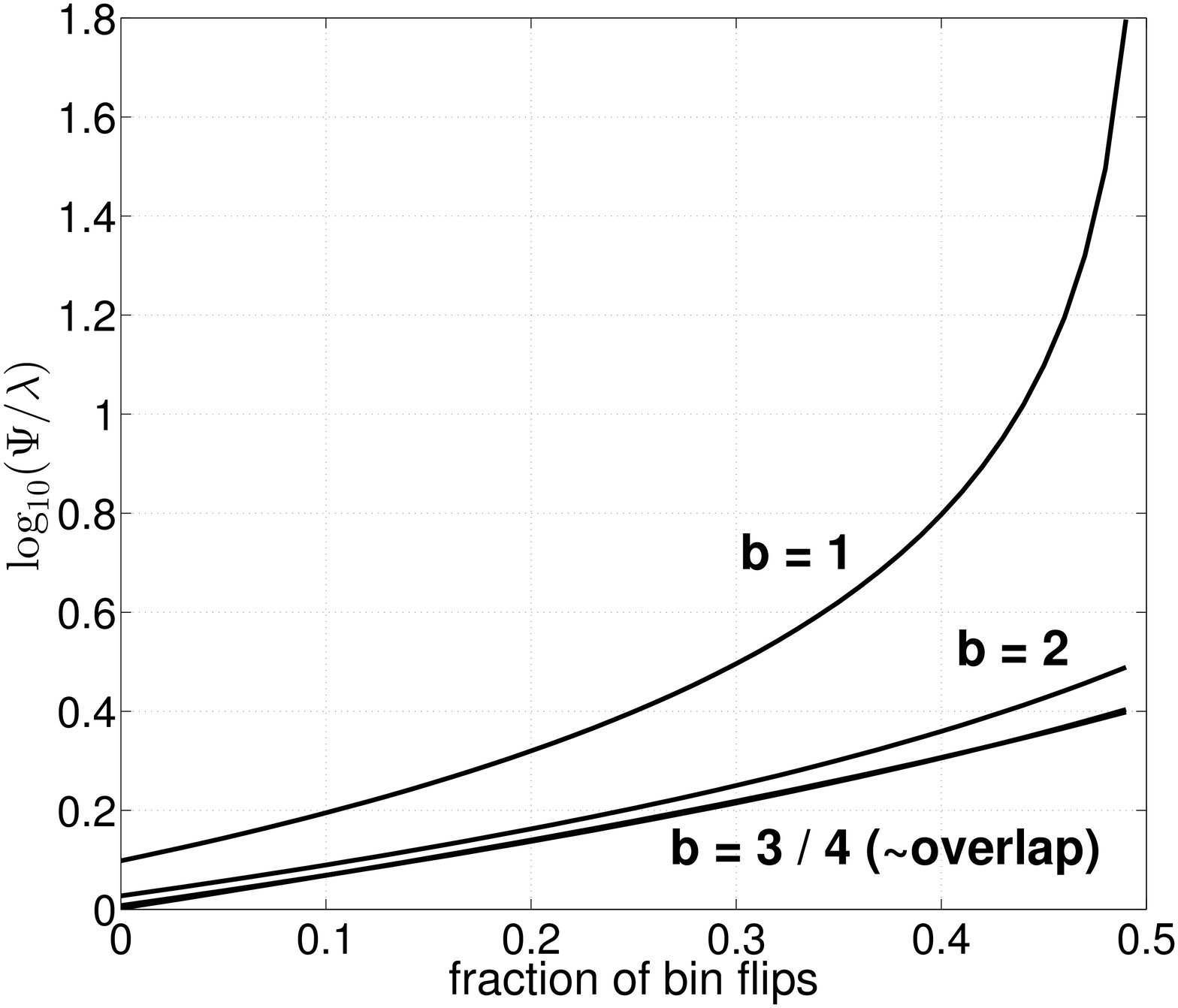} & $\qquad$&
\includegraphics[width = 0.36\textwidth]{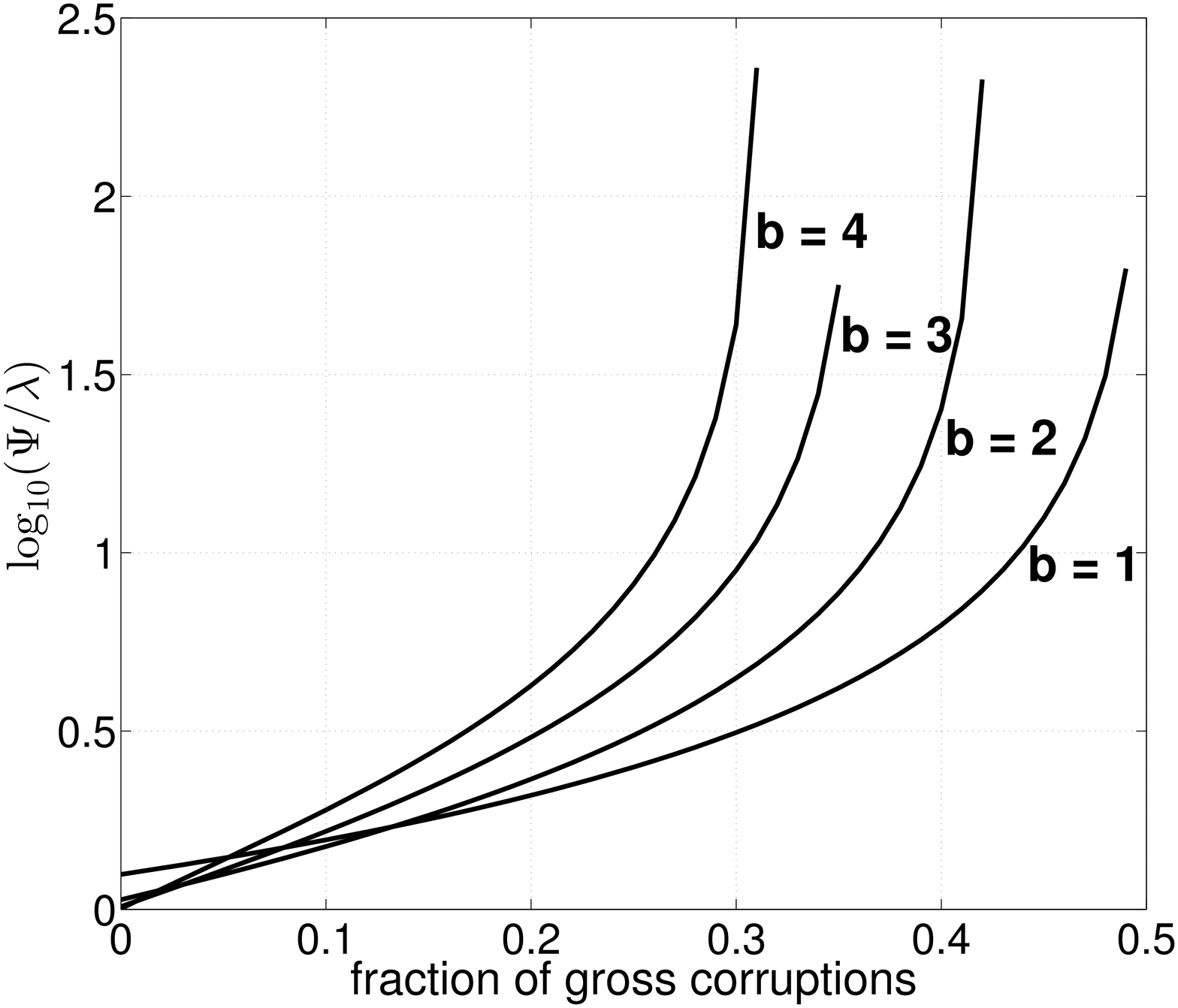}
\end{tabular}
\end{center}
\vspace*{-0.015\textheight}
\caption{$\Psi_{b,p} / \lambda_{b,p}$ (on a $\log_{10}$-scale), $b \in
  \{1,2,3,4\}$, $p \in [0, 0.5]$ for mechanisms (I, left) and (II, right). For (I), the trajectories for $b = 3$ and $b = 4$ are indistinguishable.}\label{fig:flips_theory}
\end{figure}

Note that for $b = 1$, (I) and (II) coincide as both amount to a sign flip with probability $p$. % In our
% analysis we assume for simplicity that $\sigma = 0$ and fix $\mathbf{t} = \mathbf{t}_0^*$ and $\bm{\mu} =
% \bm{\mu}_0^*$ i.e.,~the Lloyd-Max optimal quantization for uncontaminated
% observations as defined in Theorem \ref{heo:lloydmax}.
Depending on
the magnitude of $p$, the corresponding value $\lambda = \lambda_{b,p}$ may
even be negative, which is unlike the case of additive noise. Recall that the error
bound \eqref{eq:mainbound} requires $\lambda > 0$. Borrowing
terminology from robust statistics, we consider $\bar{p}_b = \min \{p: \lambda_{b,p} \leq 0
\}$ as the \emph{breakdown point}, i.e.,~the (expected) proportion of contaminated
observations that can still be tolerated so that
\eqref{eq:mainbound} continues to hold. Mechanism (II) produces
a natural counterpart to gross corruptions when the linear measurements are not subject to quantization. It is not hard to see that among all maps $\mc{M}^{\pm}
\rightarrow \mc{M}^{\pm}$ applied randomly to the observations with a fixed
probability, (II) maximizes the ratio $\Psi_{b,p}/\lambda_{b,p}$, hence the attribute
``adversarial''. In Figure \ref{fig:flips_theory} we display $\Psi_{b,p} /
\lambda_{b,p}$ for both (I) and (II) and $b \in \{1,2,3,4\}$. Table \ref{tab:breakdownpoints}
provides the corresponding breakdown points. For simplicity, $(\mathbf{t}, \bm{\mu})$ are not optimized
but set to the optimal (in the sense of Lloyd-Max) choice $(\mathbf{t}_0^*, \bm{\mu}_0^*)$ in the noiseless case.
The underlying derivations can be found in Appendix \ref{app:beyondadditive}.
\begin{table}[h!]
\begin{center}
 \begin{tabular}{|l|l|l|l|l||l|l|l|l|l|}
 \hline
 (I)                           &  $b=1$ & $b=2$ & $b = 3$ & $b=4$ & (II) &
 $b=1$ & $b = 2$ & $b = 3$ & $b = 4$ \\
\hline
 $\bar{p}_b$           &  $1/2$ & $3/4$ & $7/8$   & $15/16$ & $\bar{p}_b$ &
 $1/2$ & $0.42$ & $0.36$ & $0.31$    \\
 \hline
 \end{tabular}
 \end{center}
\vspace*{-0.02\textheight}
\caption{Breakdown points $\bar{p}_b$ for mechanisms (I, left half) and (II, right half) and $b \in \{1,2,3,4\}$.}\label{tab:breakdownpoints}
\end{table}

Figure \ref{fig:flips_theory} and Table \ref{tab:breakdownpoints} provide one more argument in
favour of one-bit measurements as they offer better robustness vis-\`{a}-vis
adversarial corruptions. In fact, once the fraction of such corruptions
reaches $0.2$, $b = 1$ performs best $-$ on the measurement scale. For the
milder corruption scheme (I), $b = 2$ turns out to the best choice for
significant but moderate $p$.

\section{Scale estimation}\label{sec:scale}

In Section \ref{sec:linearrecovery}, we have decomposed $x^* = x_u^* \psi^*$ into a product of a unit vector $x_u^*$ and a scale parameter
$\psi^* > 0$. We have pointed out that $x_u^*$ can be estimated from a linear map of $y$ separately from $\psi^*$ since the latter can be absorbed into the
definition of the bins $\{ \mc{R}_k \}$. Accordingly, we may estimate
$x^*$ as $\wh{x} \wh{\psi}$ with $\wh{x}$ as in \eqref{eq:canlin} estimating $x_u^*$ and $\wh{\psi}$
estimating $\psi^*$. We here consider the maximum likelihood estimator (MLE) for $\psi^*$.\\

\noindent\textbf{Noiseless case.} To begin with, let us consider
the case $\sigma = 0$, so that the $\{ y_i \}$ are i.i.d.~$N(0,
(\psi^*)^2)$. The likelihood function is then given by
\begin{equation}\label{eq:scalelikelihood}
L(\psi) = \prod_{i = 1}^m \sum_{k=1}^K I(y_i \in \mc{R}_k) \p(|y_i| \in
\mc{R}_k)  = \prod_{k=1}^K \{ 2 ( \Phi(t_k / \psi) - \Phi(t_{k-1} / \psi))
\}^{m_k},
\end{equation}
where $m_k = |\{i:\,|y_i| \in \mc{R}_k \}|, \; k \in [K]$, and $\Phi$ denotes
the standard Gaussian cdf. Note that for $K=1$, $L(\psi)$ is constant
(i.e.,~does not depend on $\psi$) which confirms that for $b = 1$, it
is impossible to recover $\psi^*$. For $K = 2$ (i.e.,~$b = 2$), the MLE has a simple a closed
form expression given by $\wh{\psi} = t_1 / \Phi^{-1}(0.5(1 + m_1/m))$. The
following tail bound establishes fast convergence of $\wh{\psi}$ to $\psi^*$.
\begin{prop}\label{prop:scale} Let $\eps \in (0,1)$ and $c = 2  \{ \phi'(t_1/\psi^*) \}^2$, where
  $\phi$ denotes the standard Gaussian pdf and $\phi'$ its derivative. With
  probability at least $1 - 2 \exp(-c m \eps^2)$, we have $|\wh{\psi}/\psi^*
  - 1| \leq \eps$.
\end{prop}
The exponent $c$ is maximized for $t_1 = \psi^*$ and becomes smaller as $t_1 /
\psi^*$ moves away from $1$. While scale estimation from $2$-bit measurements
is possible, convergence can be  slow if $t_1/\psi^*$ is too small or too large.

One may profit from taking $b \geq 3$ as this introduces multiple thresholds $t_1,\ldots,t_K$, so that
scale estimation is less dependent on the choice of each individual threshold. Moreover, if the thresholds
are well chosen, convergence becomes faster as $b$ increases; see \cite{Li2015} for a detailed analysis regarding
this aspect. On the other hand, for $b \geq 3$, the MLE is no longer available in closed from.\\

\vspace*{-0.01\textheight}
\noindent\textbf{Additive noise.} We now turn to the case $\sigma > 0$. Since the $\{y_i\}$ are now i.i.d.~$\sim N(0, (\psi^*)^2 + \sigma^2)$, the MLE based on \eqref{eq:scalelikelihood} systematically over-estimates $\psi^*$. Therefore, a different approach is needed.
Suppose that $x_u^*$ is known and denote by $[l_i, u_i] \in \{ -\mc{R}_k \}_{k = 1}^K \cup \{ \mc{R}_k \}_{k = 1}^K$
the interval the $i$-th observation is contained in before quantization, $i \in [m]$. Then the joint likelihood for $(\psi^*, \sigma)$  is given by
\begin{equation}\label{eq:scalelikelihood_noisy}
L(\psi, \wt{\sigma}) = \prod_{i = 1}^m \left\{ \Phi\left(\frac{u_i - \psi \scp{a_i}{x_u^*}
    }{\wt{\sigma}} \right) -
  \Phi \left(\frac{l_i - \psi \scp{a_i}{x_u^*}}{\wt{\sigma}} \right) \right \}.
\end{equation}
\emph{Existence and uniqueness of the MLE.} It is easy to see that as $\psi \rightarrow \infty$ or $\wt{\sigma} \rightarrow \infty$
 \begin{equation*}
\Phi \left(\frac{u_i -  \psi \scp{a_i}{x_u^*} }{\wt{\sigma}} \right) \rightarrow 0,  \qquad  \Phi \left(  \frac{l_i - \psi \scp{a_i}{x_u^*}}{\wt{\sigma}} \right) \rightarrow 0, \;\;  i \in [m],
\end{equation*}
and thus $L(\psi, \wt{\sigma}) \rightarrow 0$. Hence, the negative log-likelihood is coercive so that the MLE
always exists. However, the MLE is not necessarily unique. In fact, if there exists
$\psi$ so that $\psi \scp{a_i}{x_u^*} \in [l_i,u_i], \, i \in [m]$ (i.e.,~the bin assignment of the
``linear predictions'' $\{\psi \scp{a_i}{x_u^*}\}_{i=1}^m$ perfectly matches that of the observations), the likelihood \eqref{eq:scalelikelihood_noisy}
attains the maximum possible value of one by setting $\wt{\sigma}$ to zero. There are potentially multiple
values of $\psi$ satisfying the above condition; any of these is a MLE. However, this is a rather unlikely
scenario as long as there is a noticeable noise level.\\

\noindent\textbf{Computation of the MLE.} A straightforward strategy to deal with the resulting two-dimensional optimization problem is coordinate descent, i.e.,~$\psi$ is optimized
for fixed $\wt{\sigma}$ and vice versa. Both these sub-problems are smooth univariate optimization problems which can be solved using methods based on golden section search as implemented in common software packages. Alternating between the two univariate problems until
the objective cannot be further decreased yields a root of the likelihood equation $\nabla L(\psi, \wt{\sigma}) = 0$. It is
known that the negative logarithm of $L$ is convex in $\psi$ \cite{Zymnis2009}, but it is not clear to us whether it is
jointly convex in $(\psi, \wt{\sigma})$. Hence, the above likelihood equation may have multiple roots in general. Empirically, we have not encountered any issue with spurious solutions when using $\psi = 0$ and $\wt{\sigma}$ as the MLE from the noiseless
case as starting point. Moreover, since the optimization problem is only two-dimensional, it is feasible to perform a grid search to locate a smaller region within which the MLE resides, which reduces the chance of getting trapped in an undesired stationary point different from the MLE.\\[-2ex]
%Empirical results, however, suggest the existence of a unique stationary point coinciding
%with the global minimizer, i.e.,~the MLE.
%The MLE has to be computed numerically. It is not clear to us whether the likelihood is log-concave, which would
%ensure that the global optimum can be obtained by convex
%programming. . Better starting values can be obtained from a grid
%search which is feasible as there are only two parameters to be optimized. The
%only issue with \eqref{eq:scalelikelihood_noisy} we are aware of concerns

So far, we have assumed that $x_u^*$ is known. We may follow the plug-in principle and replace
$x_u^*$ by its estimator $\wh{x}$. The resulting estimator for $(\psi^*, \sigma)$ is no longer an MLE,
but it can be a reasonable proxy depending on the distance of $\wh{x}$ and $x_u^*$. The empirical performance of this plug-in approach is discussed in the next section.\\

\noindent \textbf{Error bound when combining estimates for direction and scale.}
Suppose that $\nnorm{\wh{x} - x_u^*}_2 \leq \delta$ and  $|\wh{\psi} - \psi^*| \leq \epsilon \psi^*$. Combining these bounds, we obtain that
\begin{align*}
\nnorm{x^* - \wh{x} \wh{\psi}}_2 = \nnorm{x_u^* \psi^* - \wh{x} (\psi^*  + (\wh{\psi} - \psi^*))}_2
\leq \psi^* \nnorm{x_u^* - \wh{x}}_2 + |\wh{\psi} - \psi^*| \nnorm{\wh{x}}_2  \leq \psi^* (\delta + \epsilon).
\end{align*}
In the noiseless case, $\epsilon$  scales as $O(1/\sqrt{m})$ in light of Proposition \ref{prop:scale}. With $\delta$ scaling as $O(w(\overline{\Delta}(\mc{C}))/\sqrt{m})$ according to Theorem \ref{theo:mainresult}, $\epsilon$
can be considered as a lower order term. The total error for estimating $x^*$ is hence proportional to the error for estimating the direction $x_u^*$. Consequently,
as far as the optimal trade-off between $m$ and $b$ is concerned, the conclusions for estimating the direction in $\S$\ref{sec:tradeoff} also apply when combining the estimates
for direction and scale. In particular, since $b = 1$ renders scale estimation impossible, $b = 2$ becomes the optimal choice for a given budget of bits.
%As  is typically unknown, ,
%replacing

\section{Experiments}\label{sec:experiments}
\vspace{-0.05in}

We here provide numerical results supporting/illustrating some of the key
points made in the previous sections. Specifically, we compare the $\ell_2$-error of the estimator $\wh{x}$ when using
one-bit respectively two-bit measurements in light of what is predicted in Table \ref{tab:tradeoff} for different classes of signals (sparse, fused sparse, group sparse and low rank matrices) and different noise models (additive noise, random bin flips and adversarial bin flips). The empirical performance of the approach to scale estimation is also investigated. A separate set of experiments is dedicated to a comparison of the estimator studied in detail herein and possible alternatives for sparse signal recovery from quantized measurements.

\vspace{-0.05in}
\subsection{One-bit vs.~two-bit measurements}

\textbf{Setup I: additive noise.} The majority of our simulations follow the model given by \eqref{eq:nonlinearCSmodel},\eqref{eq:quantizer} with $n = 500$, $\sigma \in \{0,1,2\}$ and $b \in \{1,2\}$. For $b = 2$, quantization is performed according to the Lloyd-Max problem \eqref{eq:lloydmaxproblem} for a $N(0,1)$-random variable. The number of measurements $m$ and the generation of $x^*$ varies with
$\mc{K}$.

%\vspace{-.5ex}
1) $\mc{K} = B_0(s;n)$ for $s \in \{10,20,\ldots,50\}$. The support of $x^*$ and its signs are selected uniformly at random,
while the absolute magnitude of the entries corresponding to the support are drawn from the uniform distribution on $[\beta, 2\beta]$
(for short, we write $U([\beta, 2\beta])$ in the sequel), where

\begin{equation}\label{eq:beta_m_f}
\beta = \frac{2 f}{\lambda_{1, \sigma}} \sqrt{\frac{\log n}{m}}, \quad\;\, m =  \left(\frac{3f}{\lambda_{1, \sigma}}\right)^2 s \log n,
\quad \text{and} \; \, f \in \{0.5,1,1.5,\ldots,4 \},
\end{equation}
with $f$ controlling signal strength. The resulting signal is then normalized to unit $2$-norm.

2) Fused Sparsity (cf.~$\S$\ref{sec:classes_signals}) with $s$ equally sized blocks, $s \in \{5,10,20,25,50 \}$. The entries
corresponding to each block are set to $\pm 1$ in an alternating fashion, and scaled such that $x^*$ has unit $2$-norm. The number of measurements $m$ are chosen according to \eqref{eq:beta_m_f}.

3) Group Sparsity (cf.~$\S$\ref{sec:classes_signals}) with $L = 100$ groups and $s \in \{2,4,6,8,10\}$. For $1 \leq \ell  \leq s$, we let
\begin{equation*}
x_{G_{\ell}}^* \sim U([\beta, 2\beta]) \cdot N(0, I_{n/L}), \quad \beta = \frac{2 f}{\lambda_{1, \sigma}} \sqrt{\frac{\log L}{m}}, \quad
m =  \left(\frac{3f}{\lambda_{1, \sigma}}\right)^2 s (n/L + \log(L)),
\end{equation*}
with $f$ as in \eqref{eq:beta_m_f}. We let $x_{G_{\ell}}^* = 0$ for $\ell > s$ and subsequently normalize $x^*$ to unit $2$-norm.

4) Low-rank matrices (cf.~$\S$\ref{sec:classes_signals}) with $n_1 = 50$, $n_2 = 30$, $s \in \{2,4,6,8,10\}$. We draw
a random $n_1 \times n_2$-matrix $G$ with i.i.d.~$N(0,1)$-entries and compute its SVD $G =U \Sigma V^{\T}$. We then let
\begin{equation*}
X^* = U_{s} \text{diag}(d_1,\ldots,d_s) V_s^{\T}, \quad \{d_j\}_{j=1}^s \overset{\text{i.i.d.}}{\sim} U([\beta, 2\beta]), \; \;
\beta = 2 f  \big/ \lambda_{1, \sigma}, \quad m =  (3f \big/ \lambda_{1, \sigma})^2 s (n_1 + n_2),
\end{equation*}
where $U_s$ and $V_s$ contain the first $s$ columns respectively rows of $U$ respectively $V$. Subsequently, $X^*$ is
normalized to unit Frobenius norm.

Each possible configuration for $s, f$ and $\sigma$ is replicated $20$
times. It is straightforward to compute $\wh{x}$ for 1)--4) even though $\mc{K}$ is non-convex in all cases; cf.~$\S$\ref{sec:classes_signals} and Appendix \ref{app:computation}.\\

\begin{figure}[h!]
\begin{center}
\begin{tabular}{cc}
\hspace*{-0.02\textwidth}\includegraphics[width = 0.35\textwidth]{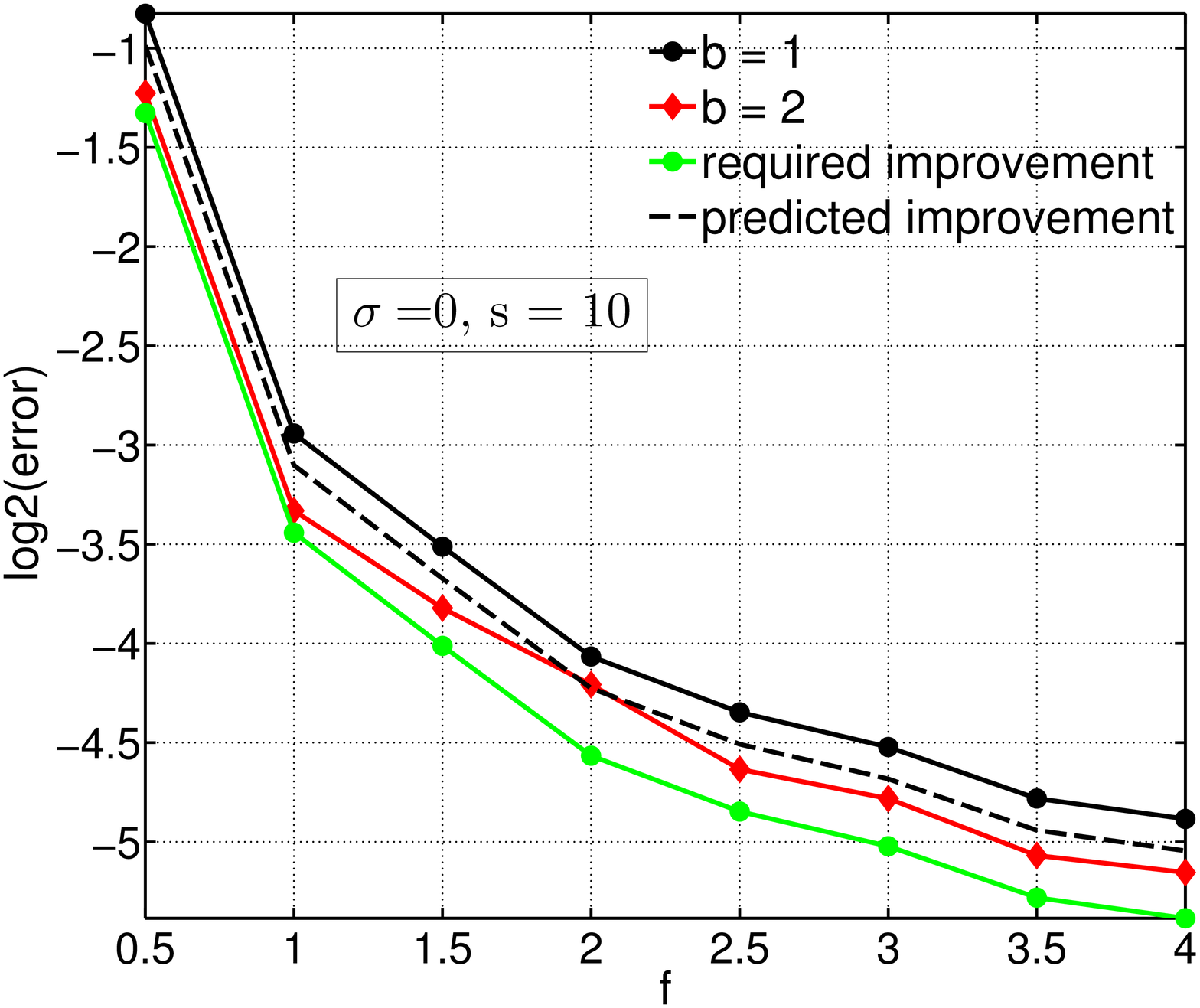}
& \includegraphics[width = 0.35\textwidth]{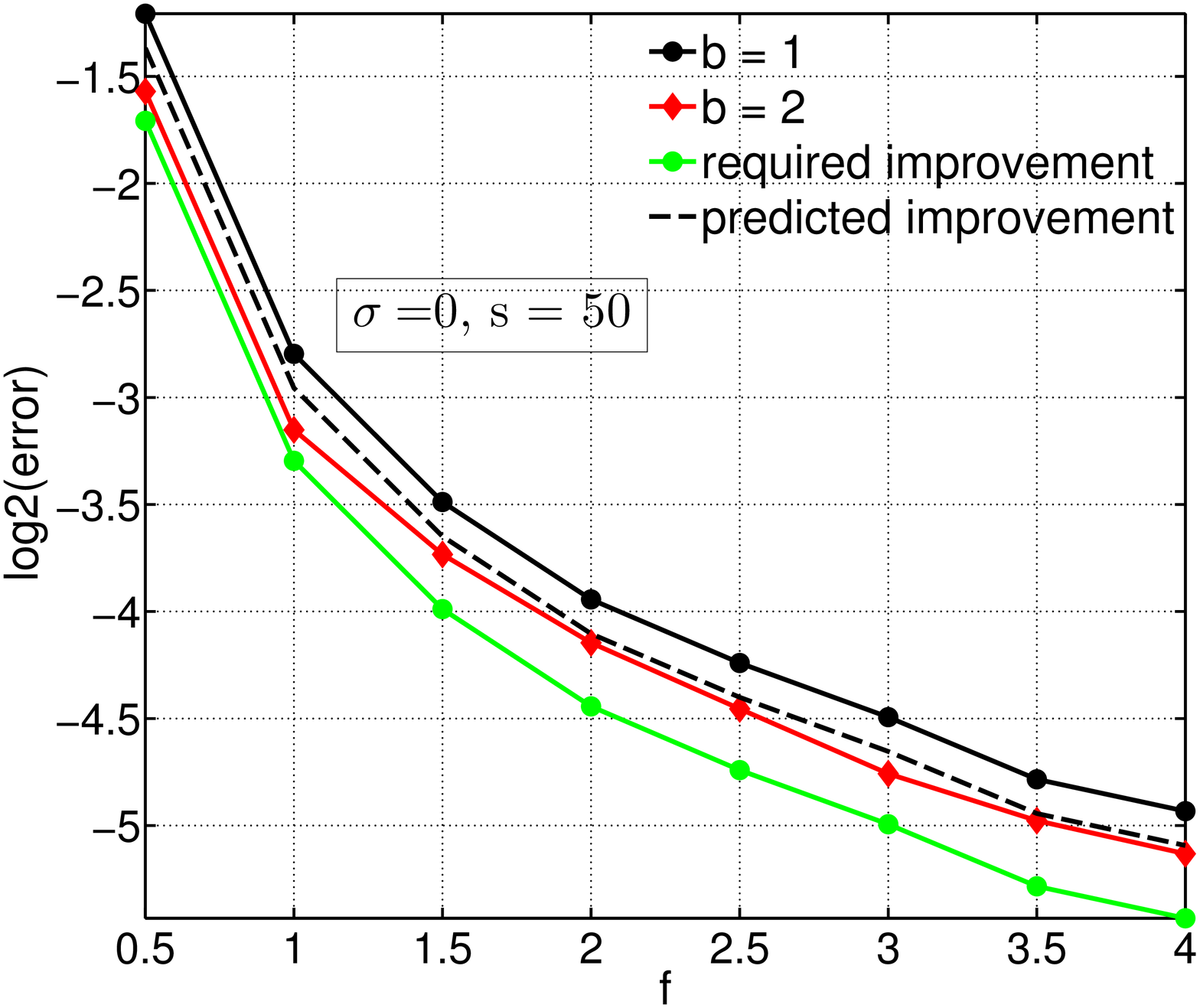} \\
\hspace*{-0.02\textwidth}\includegraphics[width =  0.35\textwidth]{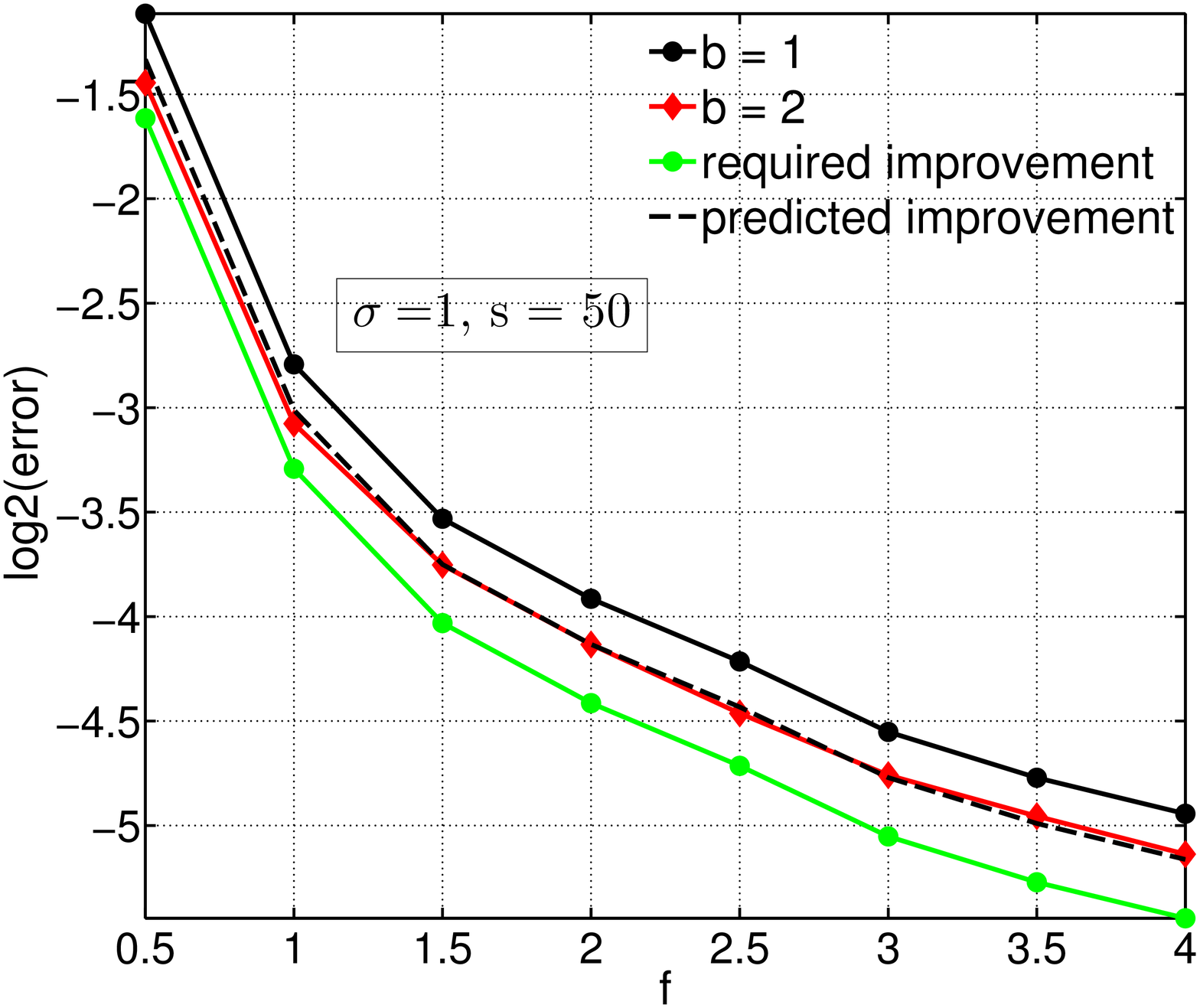} &
\includegraphics[width = 0.35\textwidth]{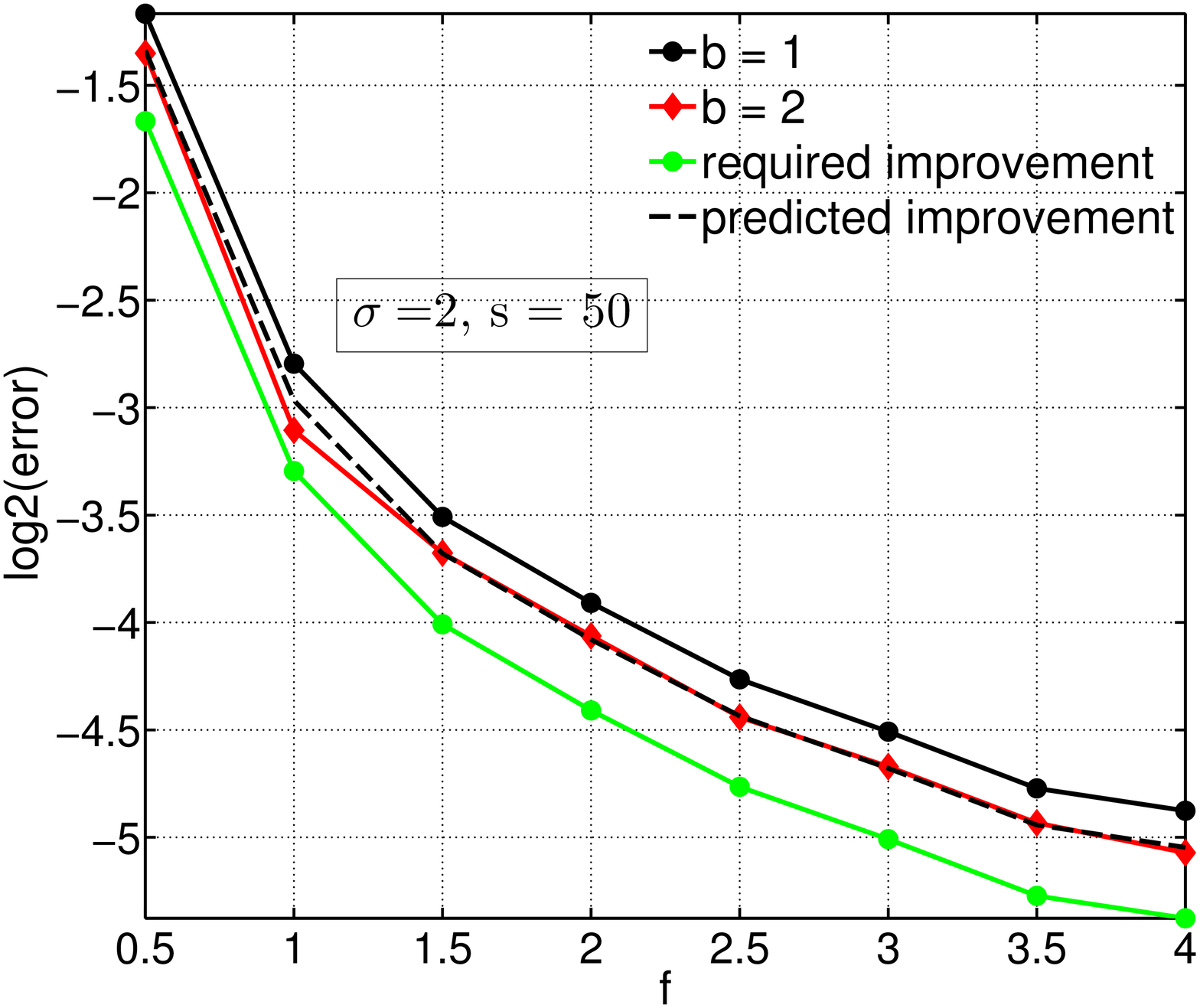}
\end{tabular}
\end{center}
\vspace*{-0.2in}
\caption{Average $\ell_2$-estimation errors $\nnorm{x^* - \wh{x}}_2$ for $\mc{K} = B_0(s;n)$ under \textbf{additive noise} (Setup I) for $b = 1$ and $b = 2$ on the $\log_2$-scale in dependence of the signal strength
  $f$. The curve ``predicted improvement'' (of $b = 2$ vs.~$b = 1$) is
  obtained by scaling the error of $b = 1$ by the factor predicted by
  the theory of Section \ref{sec:analysis} (cf.~Table \ref{tab:tradeoff}). Likewise the curve ``required improvement''
results by scaling the error of $b = 1$ by $1/\sqrt{2}$ and indicates what would be required by $b = 2$ to improve over $b = 1$ at the
level of total \#bits.}\label{fig:additivenoise_empirical}
\end{figure}

\begin{figure}[h!]
%\mbox{
\begin{tabular}{ccc}
\hspace{-0.03\textwidth}
$\begin{array}{c}
\mbox{{\centering{\footnotesize \text{fused sparsity, $\sigma = 1$, $s = 20$}}}} \\
\includegraphics[width = 0.32\textwidth]{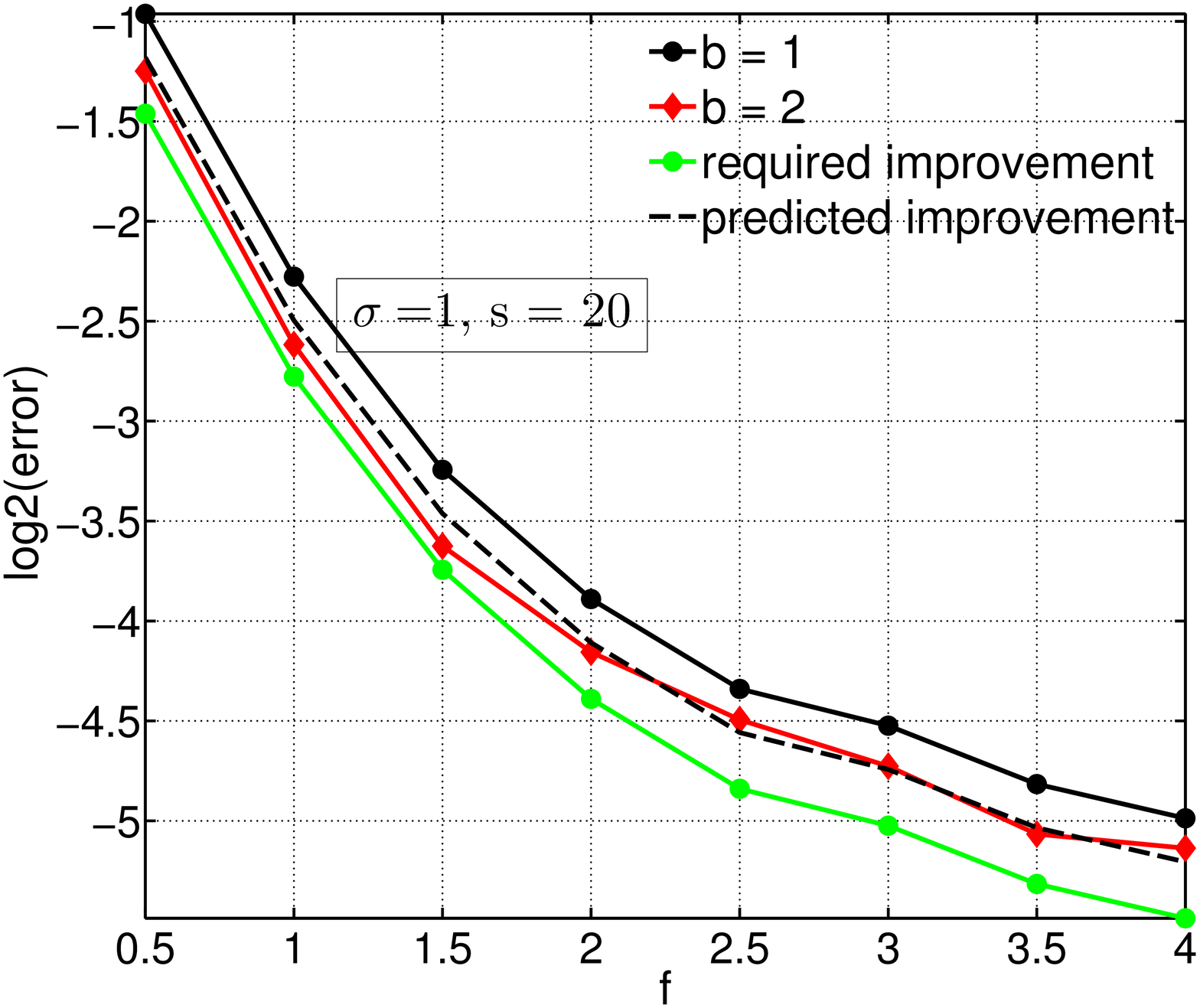}
\end{array}$ & \hspace{-0.035\textwidth}
$\begin{array}{c}
\mbox{{\footnotesize \text{group sparsity}, $\sigma = 1$, $s = 10$}} \\
\includegraphics[width = 0.32\textwidth]{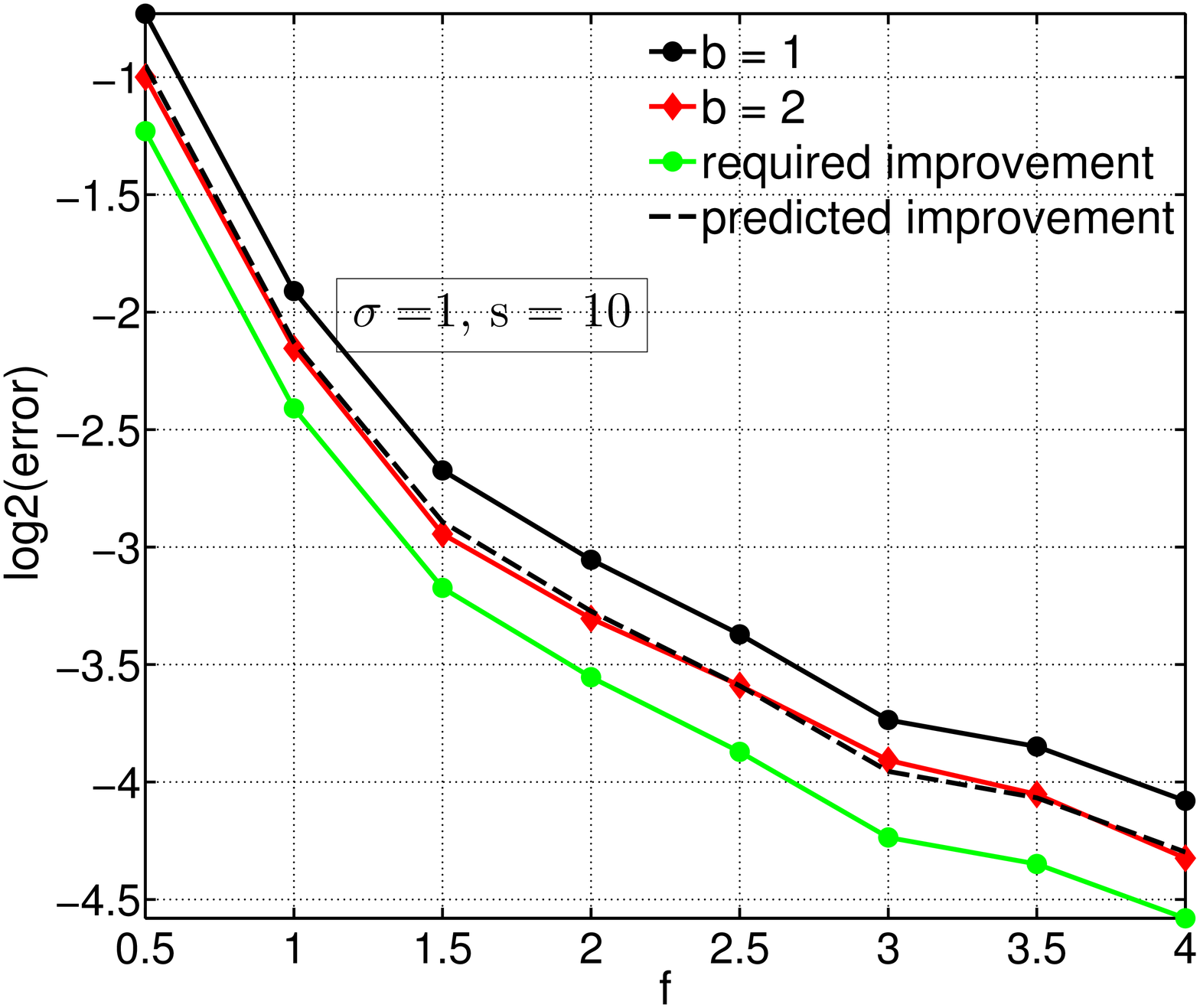}
\end{array}$
& \hspace{-0.035\textwidth}
$\begin{array}{r}
\mbox{{\footnotesize \text{low-rank matrices}, $\sigma = 1$, $s = 10$}} \\
\includegraphics[width = 0.32\textwidth]{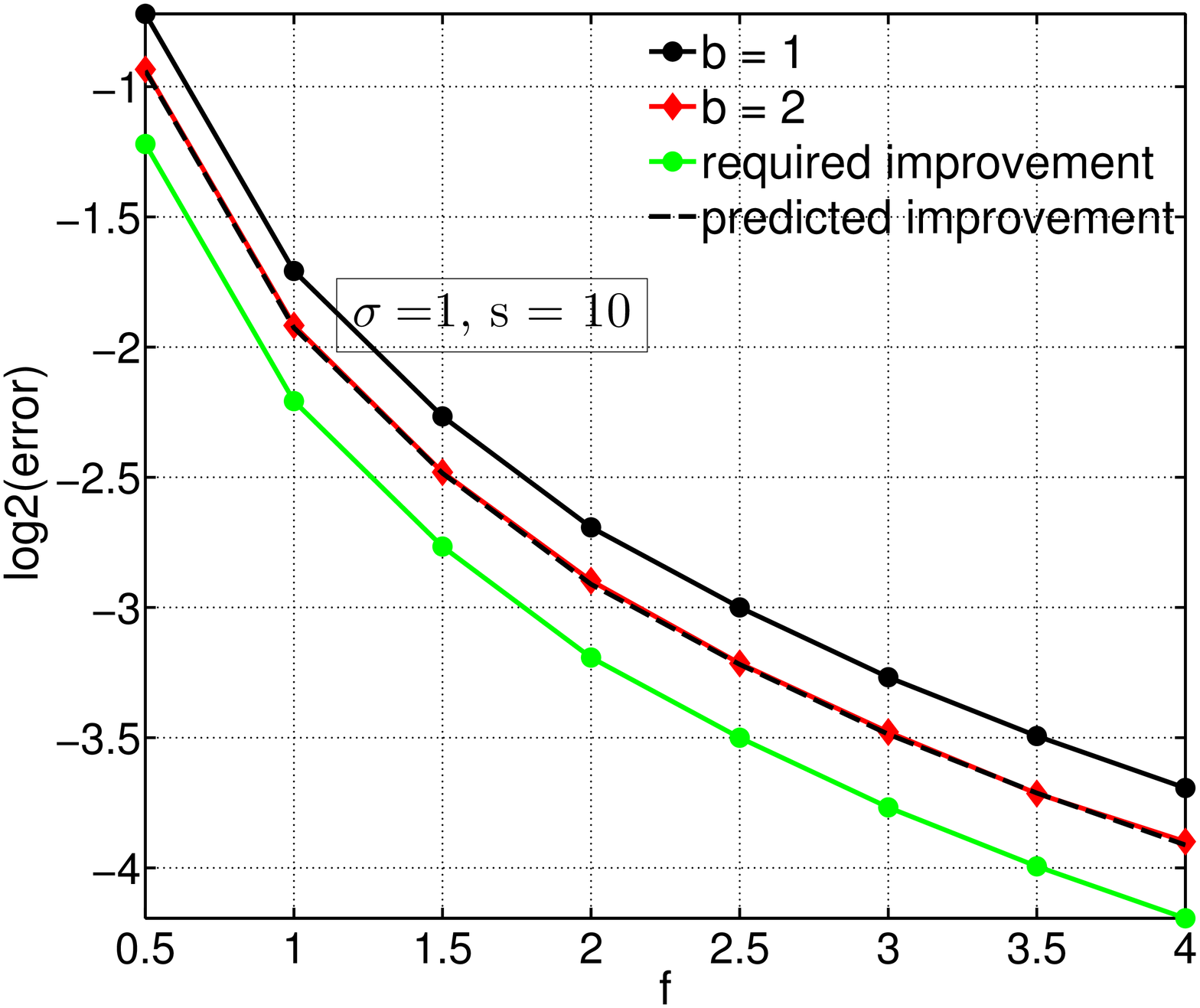}
\end{array}$
%& {\footnotesize M: group sparsity} & {\footnotesize R: low-rank matrices} \\
%   \includegraphics[width = 0.32\textwidth]{plots/1vs2group_sigmaix6_s10.eps}
%\includegraphics[width = 0.32\textwidth]{plots/1vs2lowrank_sigmaix6_s10.eps}
\end{tabular}
\vspace*{-0.1in}
\caption{Average $\ell_2$-estimation errors $\nnorm{x^* - \wh{x}}_2$ for the other three classes of signals 2)--4). The annotation of the plots follows that in
Figure \ref{fig:additivenoise_empirical}.}
\label{fig:structured_sparsity}
\end{figure}

\noindent\textbf{Setup II: beyond additive noise.} In another set of simulations, we consider random and adversarial bin flips as discussed at the end of Section \ref{sec:analysis}. We here only consider $\mc{K} = B_0(s;n)$. The setup is as in I,1) above with $\sigma = 0$ apart from the
following modifications: after quantization, the respective bin flip mechanism is applied to the observations with probability $p$, where $p \in \{0.05,0.1,\ldots,0.4\}$. Regarding $x^*$ and $m$, we follow the same scheme as for additive noise, but replace $\lambda_{1,\sigma}$ by $\lambda_{1,p}$, cf.~Appendix \ref{app:beyondadditive}.

\begin{figure}[h!]
\begin{center}
\begin{tabular}{cc}
\hspace*{-0.02\textwidth}\includegraphics[width = 0.33\textwidth]{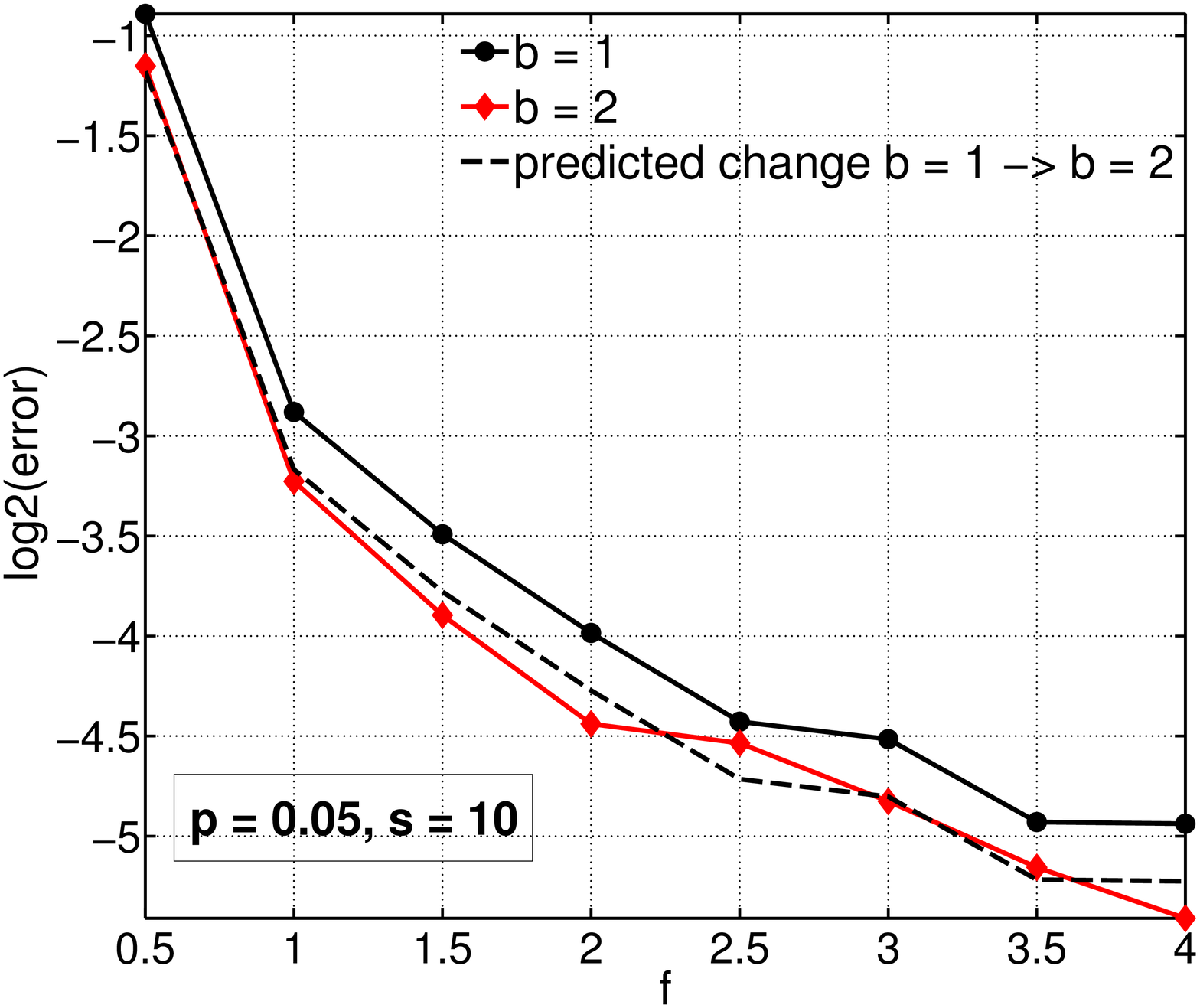}
& \includegraphics[width = 0.33\textwidth]{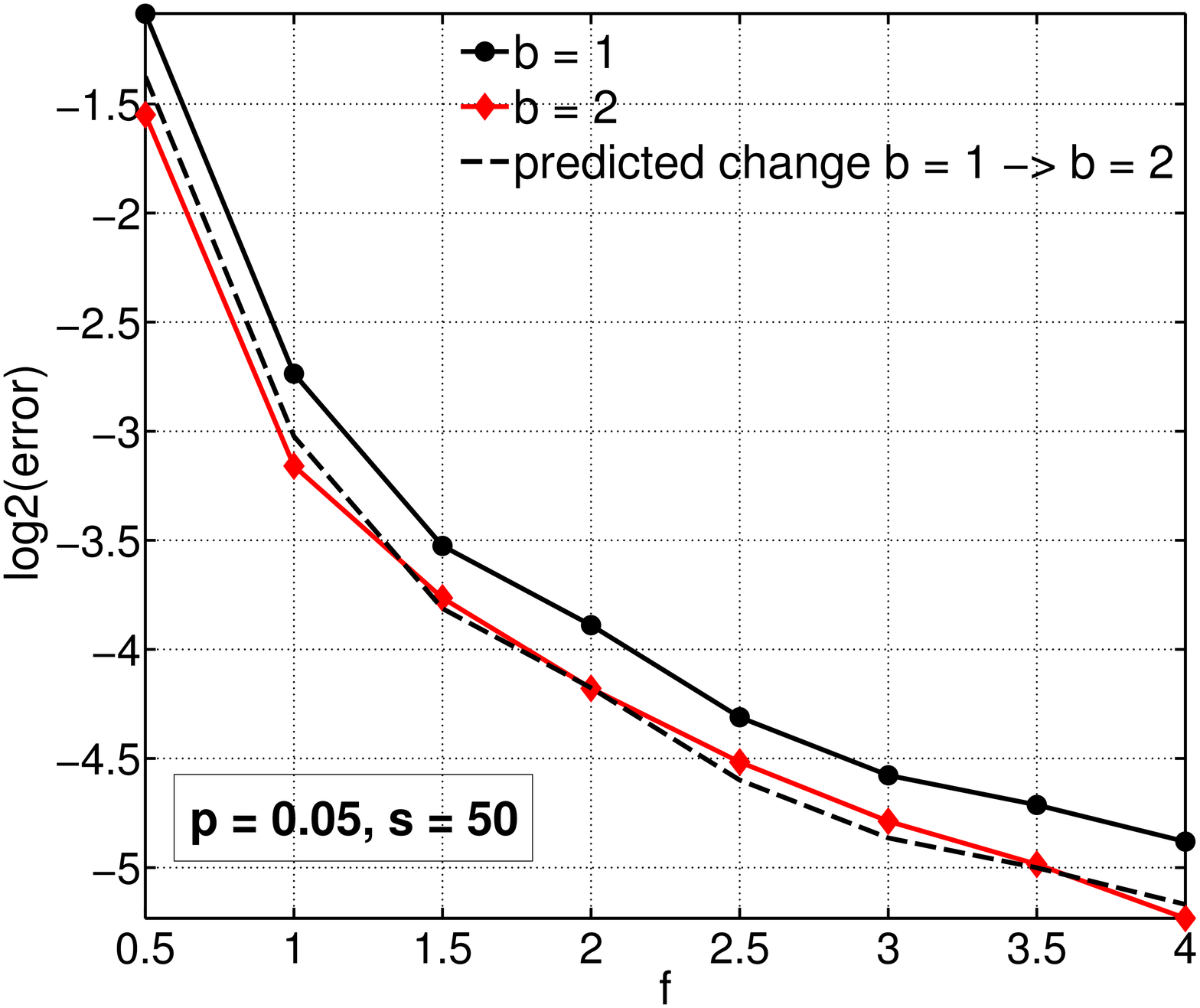} \\
\hspace*{-0.02\textwidth}\includegraphics[width =  0.33\textwidth]{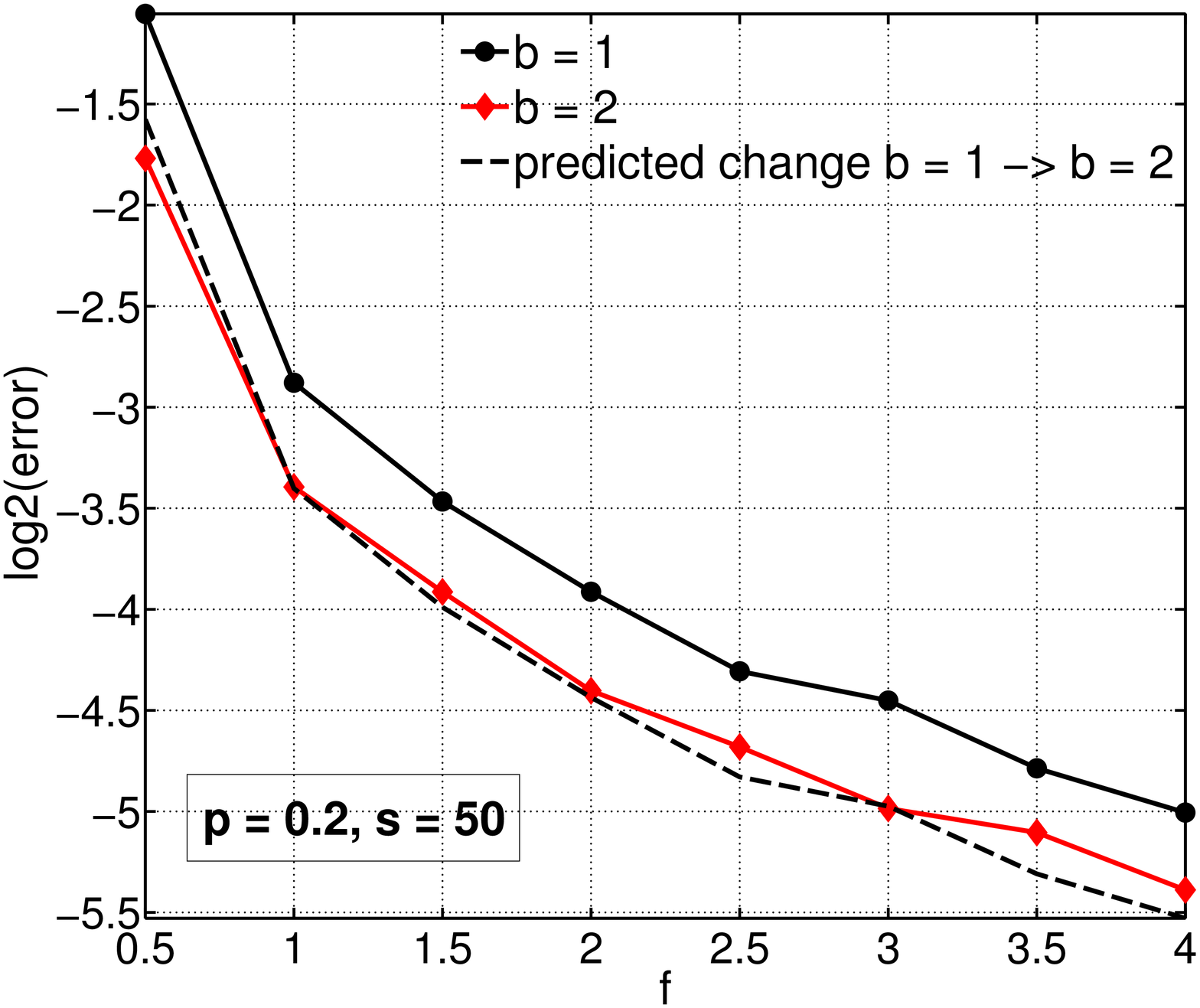} &
\includegraphics[width = 0.33\textwidth]{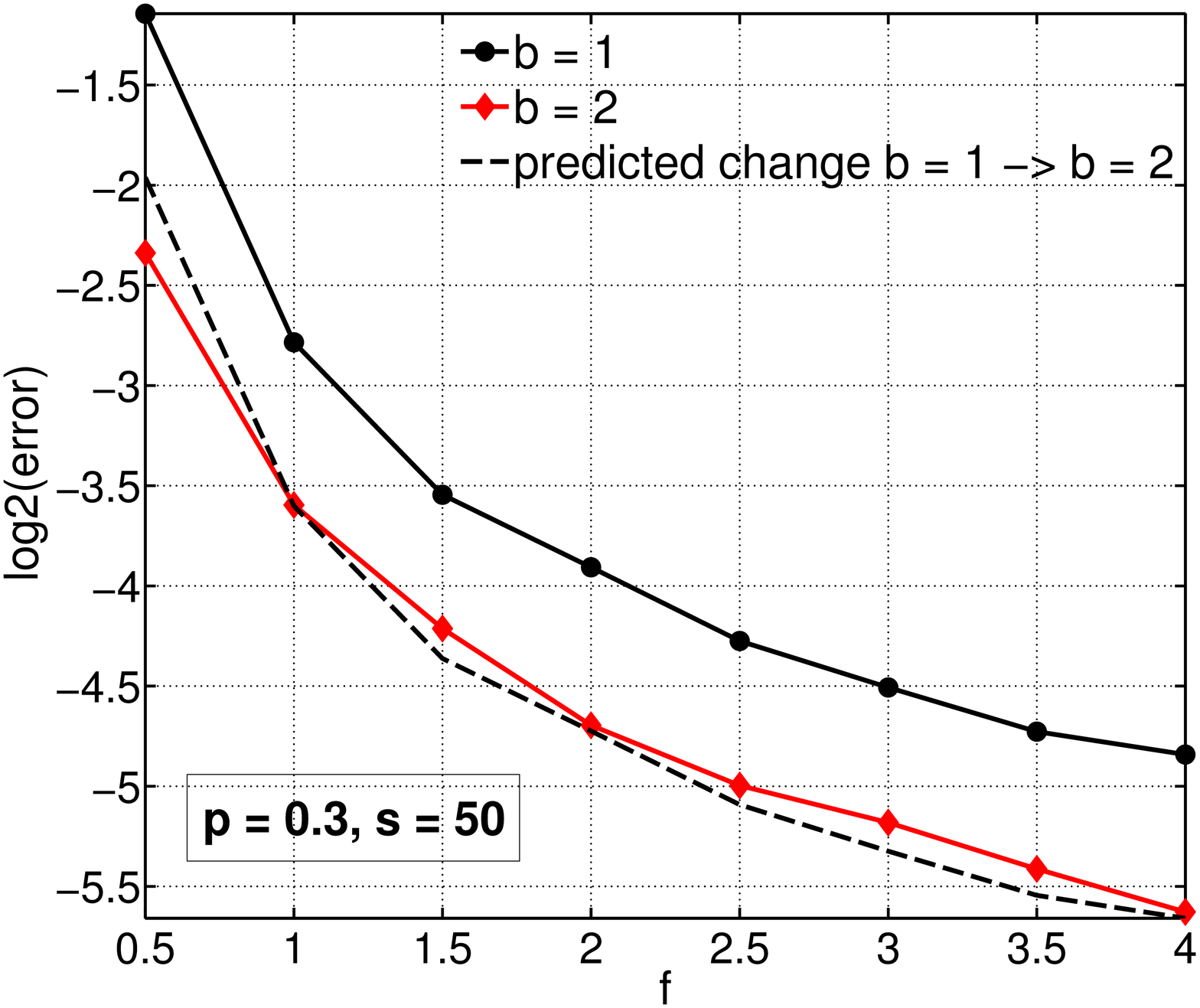}
\end{tabular}
\end{center}
\vspace{-0.0in}
\caption{Average $\ell_2$-estimation errors $\nnorm{x^* - \wh{x}}_2$ under \textbf{random bin flips} for $b =
  1$ and $b = 2$ on the $\log_2$-scale in dependence of the signal strength
  $f$. The curve ``predicted change $1 \rightarrow 2$''  is
  obtained by scaling the error of $b = 1$ by the factor predicted by
  the theory of Section \ref{sec:analysis}.}\label{fig:randombinflip_empirical}.\vspace{-0.4in}
\end{figure}

\begin{figure}[h!]
\begin{center}
\begin{tabular}{cc}
\hspace*{-0.02\textwidth}\includegraphics[width = 0.33\textwidth]{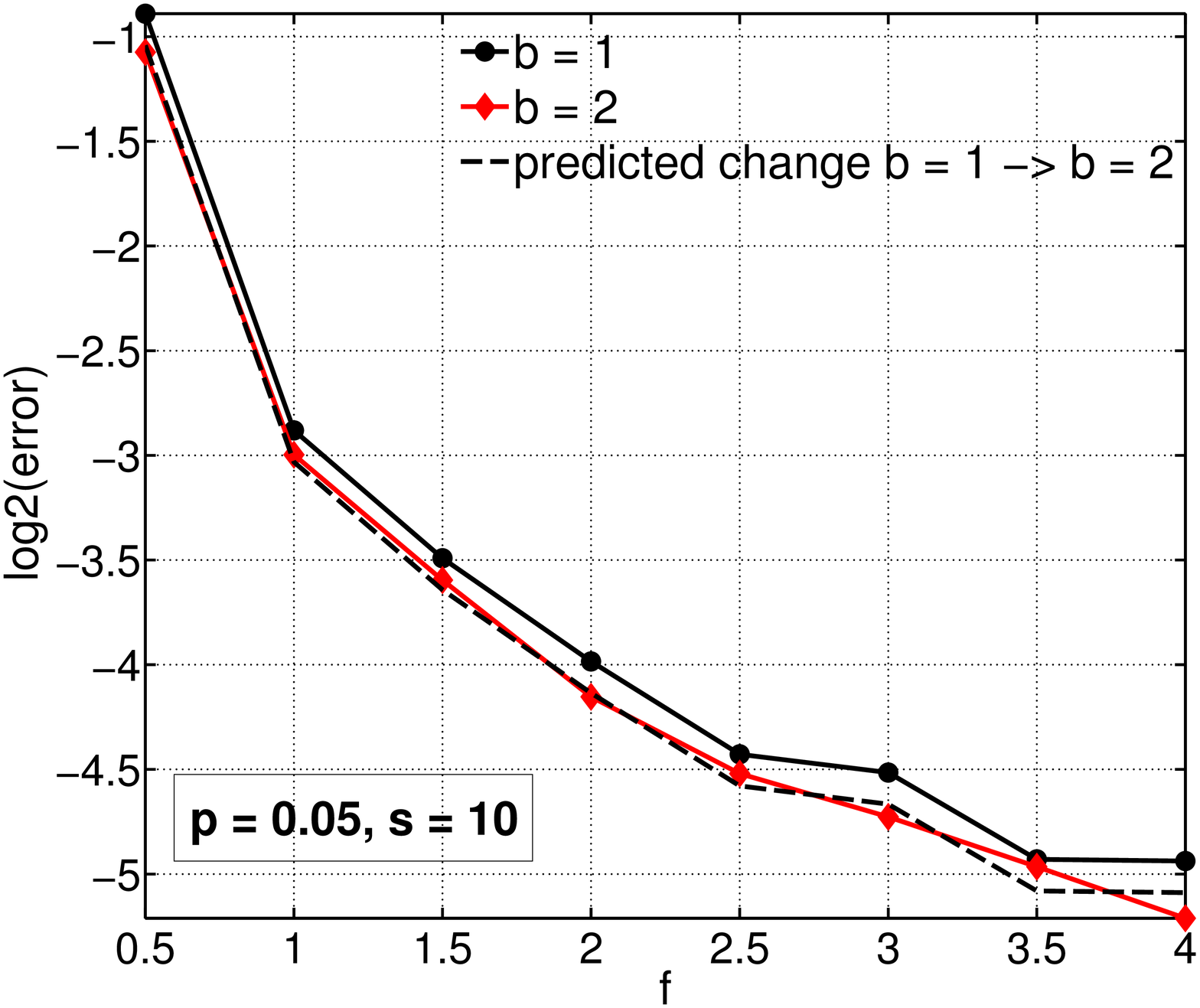}
& \includegraphics[width = 0.33\textwidth]{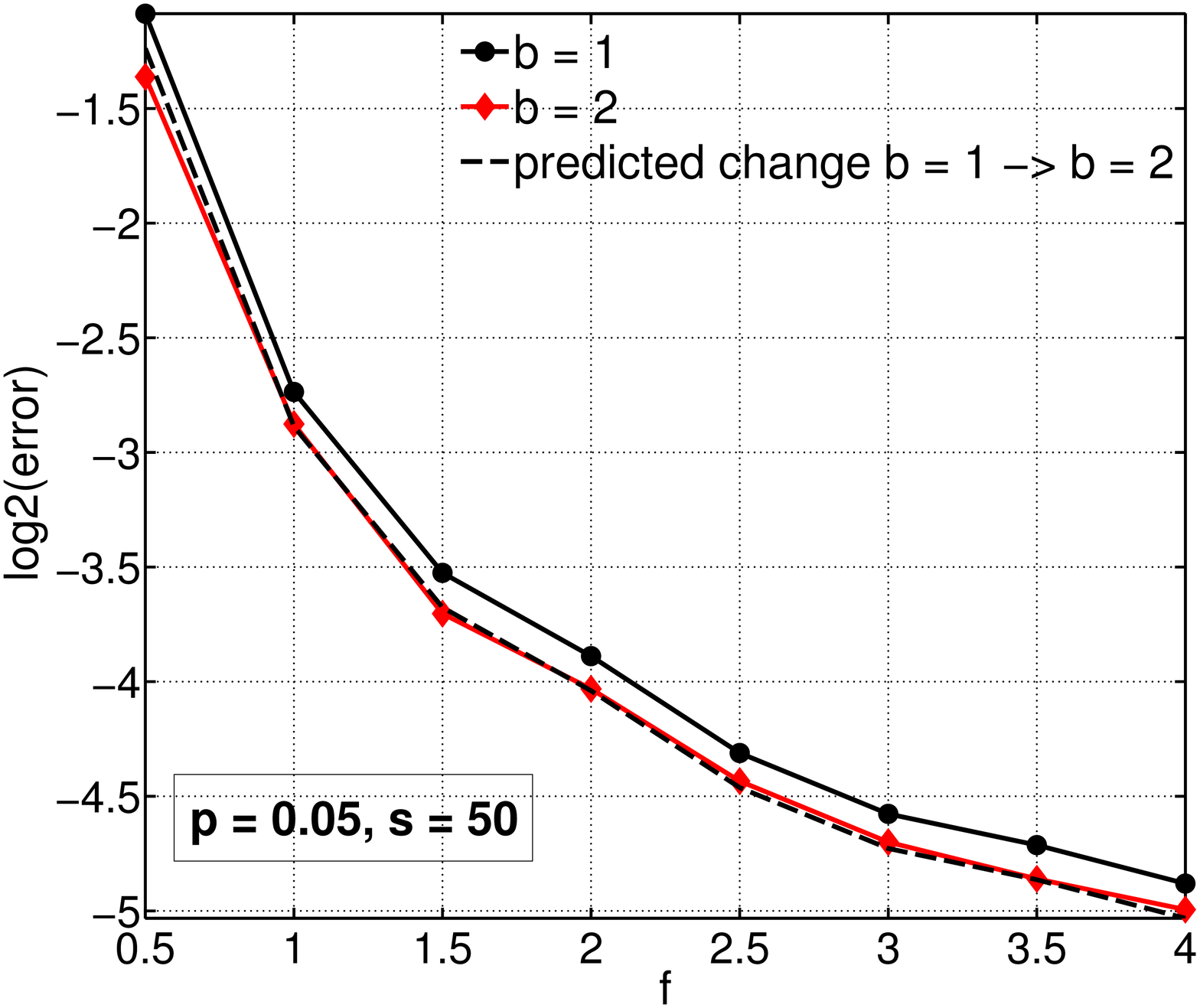} \\
\hspace*{-0.02\textwidth}\includegraphics[width =  0.33\textwidth]{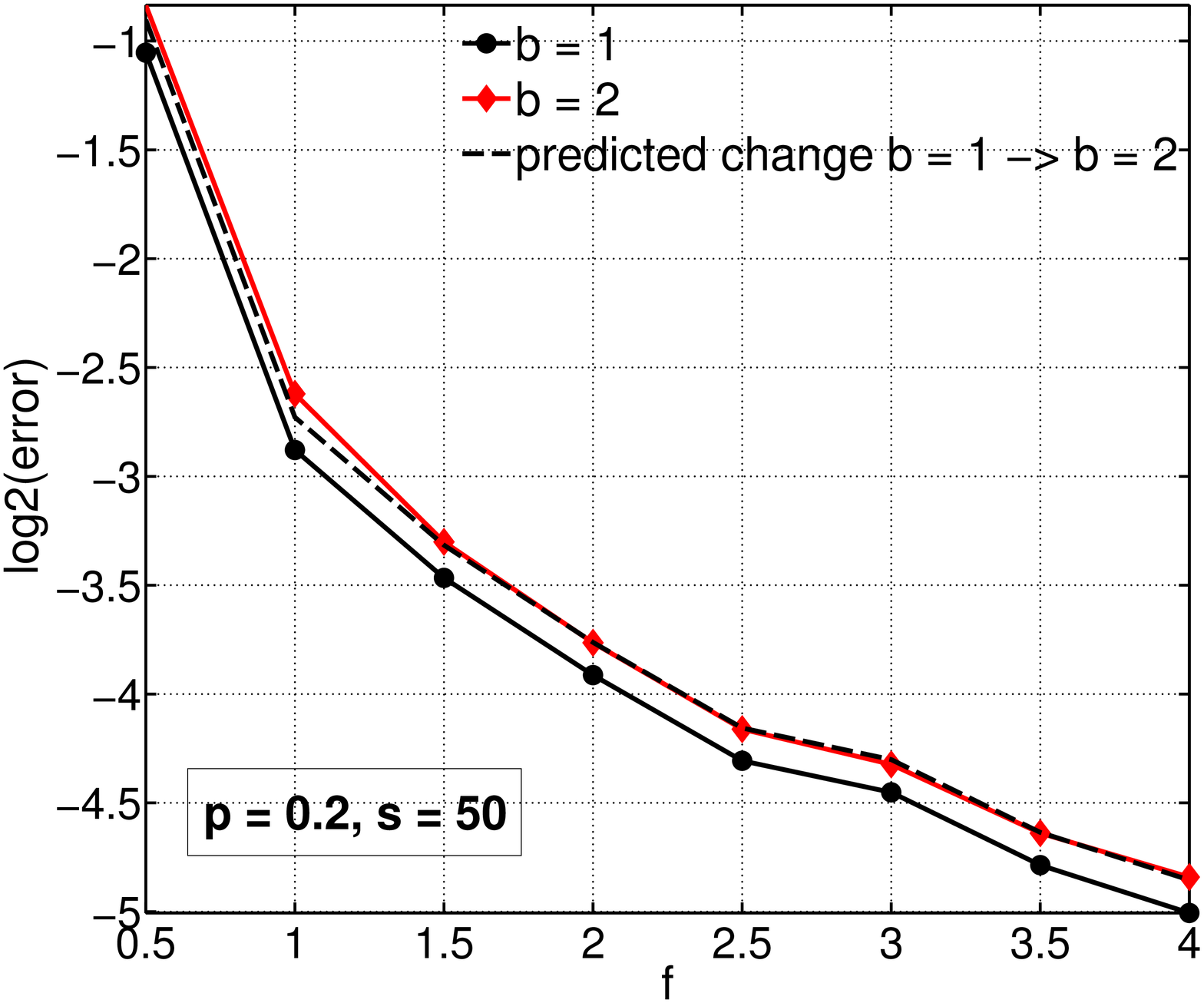} &
\includegraphics[width = 0.33\textwidth]{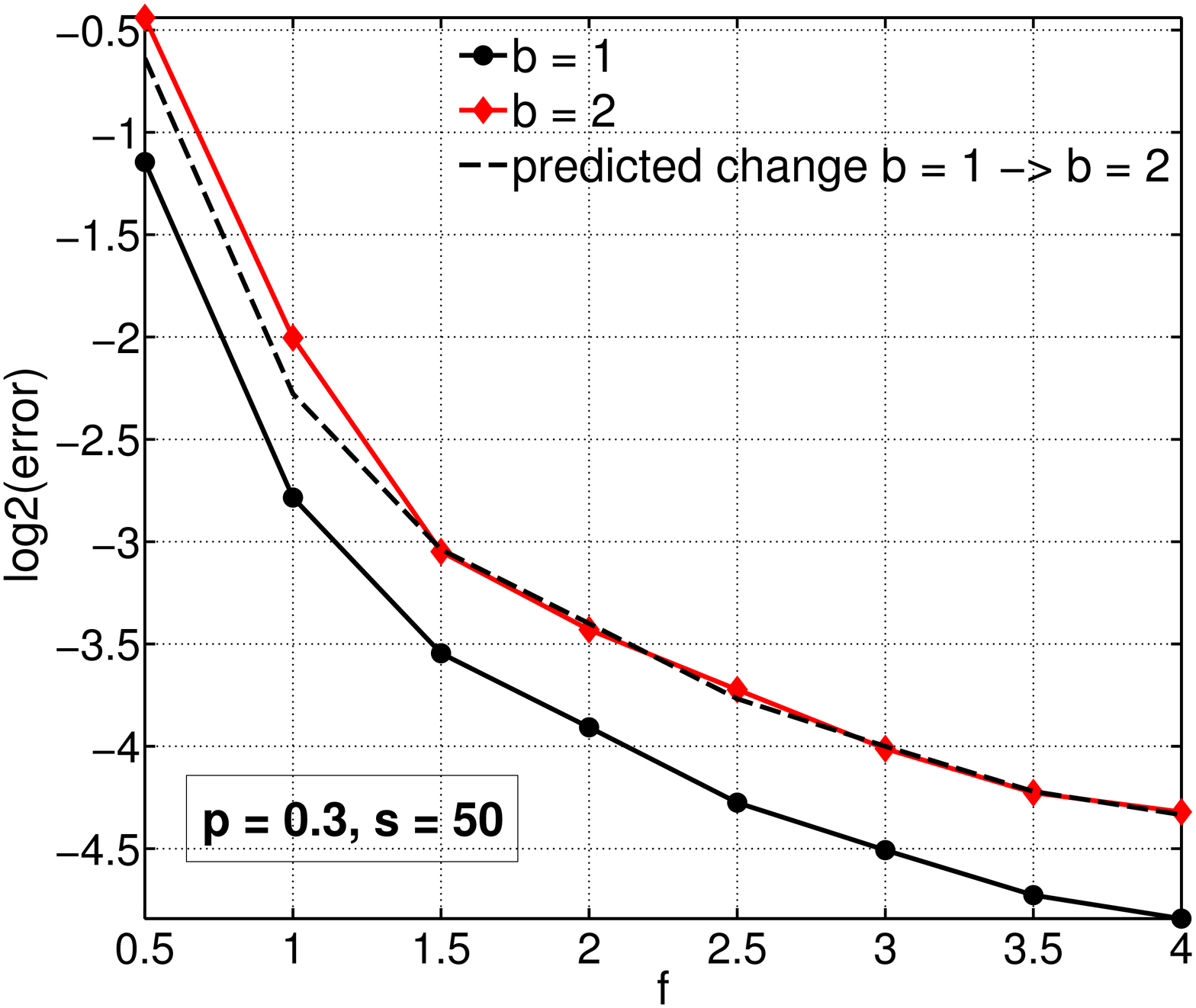}
\end{tabular}
\end{center}
\vspace{-0.3in}
\caption{Average $\ell_2$-estimation errors $\nnorm{x^* - \wh{x}}_2$ under \textbf{adversarial bin flips} for $b =
  1$ and $b = 2$ on the $\log_2$-scale in dependence of the signal strength
  $f$. The curve ``predicted change $1 \rightarrow 2$''  is
  obtained by scaling the error of $b=1$ by the factor predicted by
  the theory of Section \ref{sec:analysis}.}\label{fig:adversarialbinflip_empirical}
\end{figure}

\textbf{Conclusion}. The experiments reveal that what is predicted by the analysis concerning the relative
performance of $1$-bit and $2$-bit measurements for estimating $x^*$ closely
agrees with what is observed empirically, as can be seen in Figures \ref{fig:additivenoise_empirical}, \ref{fig:structured_sparsity}, \ref{fig:randombinflip_empirical} and \ref{fig:adversarialbinflip_empirical}. The agreement is consistent over different signal and noise models.

\newpage\clearpage

\clearpage\newpage

\subsection{Estimation of the scale and the noise level}
\begin{figure}[h!]
\begin{center}
\begin{tabular}{cc}
\includegraphics[width = 0.40\textwidth]{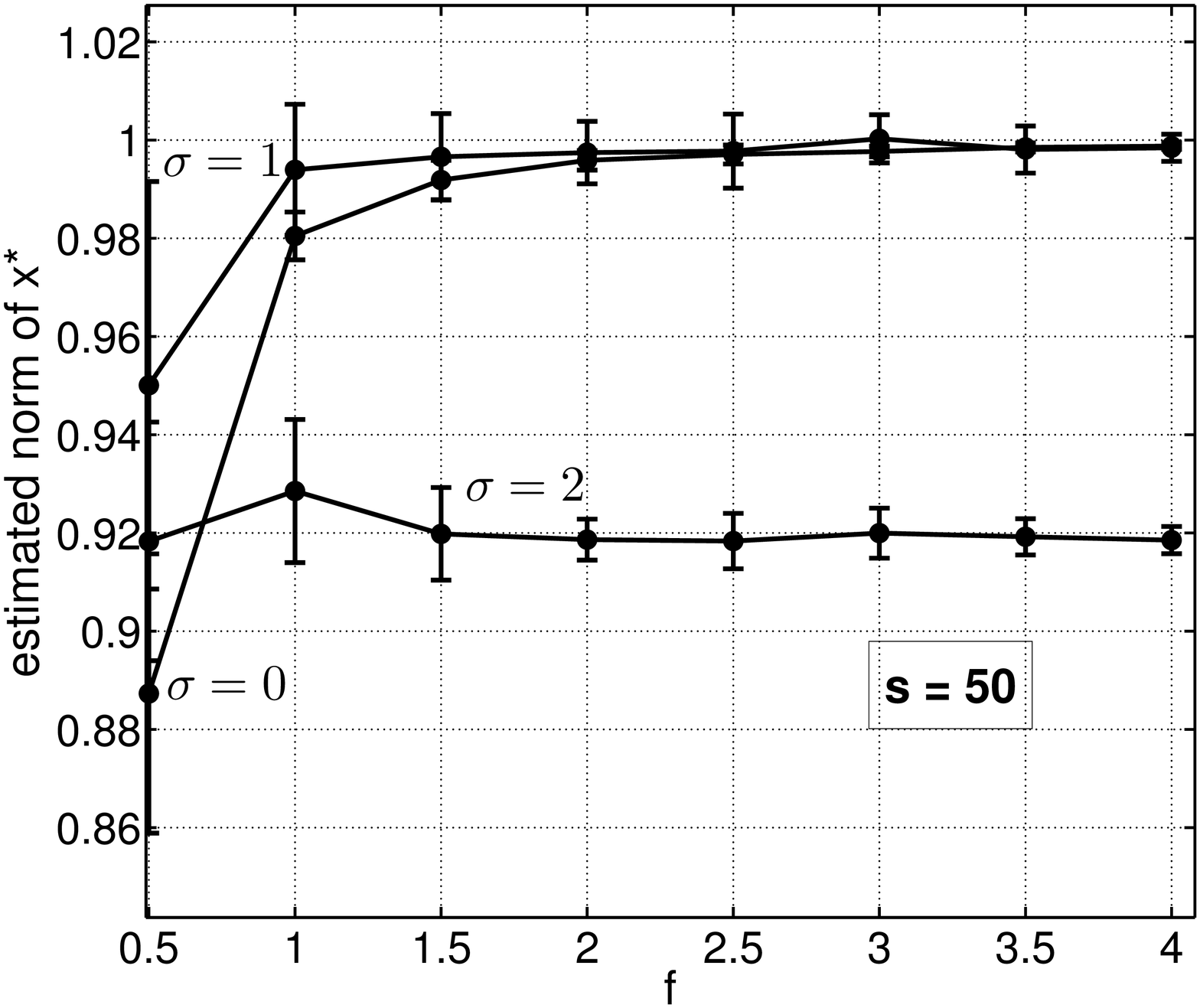}
& \includegraphics[width = 0.4\textwidth]{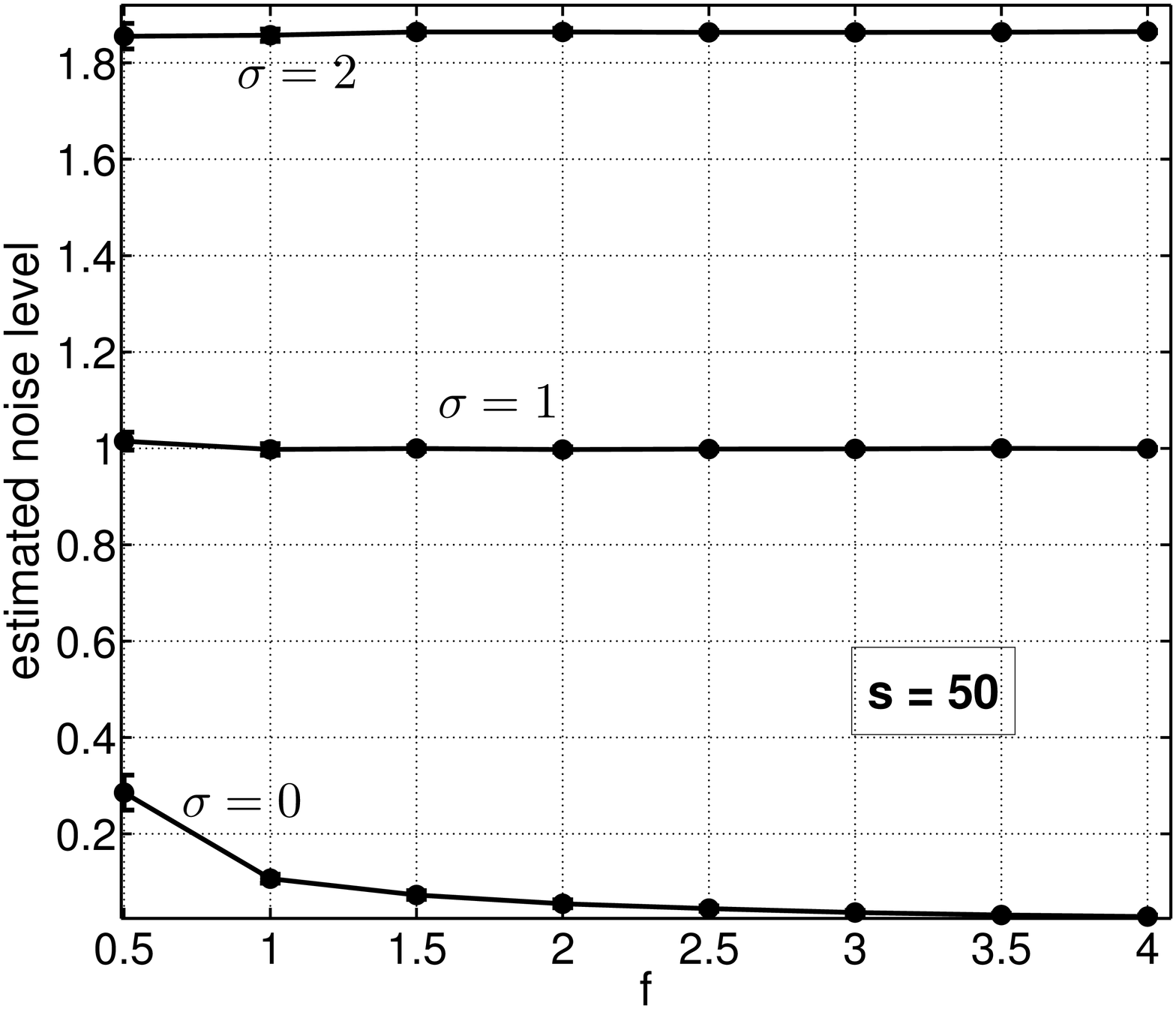}
\end{tabular}
\end{center}
\vspace*{-0.27in}
\caption{Estimation of $\psi = \nnorm{x^*}_2$ (here $1$) and $\sigma$. The
  curves depict the average of the plug-in MLE discussed in Section \ref{sec:scale} while
  the bars indicate $\pm 1$ standard deviation.}\label{fig:scale_empirical}
\end{figure}
Figure \ref{fig:scale_empirical} suggests that the plug-in MLE for $(\psi^* = \nnorm{x^*}_2, \sigma)$
as outlined in Section \ref{sec:scale} is a suitable approach, at least as long as $\psi^*/\sigma$ is not too small. For $\sigma
= 2$, the plug-in MLE for $\psi^*$ appears to have a noticeable bias as it tends to
$0.92$ instead of $1$ for increasing $f$ (and thus increasing $m$). Observe that for $\sigma = 0$, convergence to the true value $\psi^* = 1$ is slower as for
$\sigma = 1$, while $\sigma$ is over-estimated (about $0.2$) for small
$f$. The above two issues are presumably a plug-in effect, i.e.,~a consequence
of using $\wh{x}$ in place of $x_u^*$.

\subsection{Alternative recovery algorithms}

For $\mc{K} = B_0(s;n)$, we compare the empirical performance of the linear
estimator \eqref{eq:canlin} to several alternatives. Two of those are based on a more principled albeit computationally
more involved approach that tries to enforce agreement of $Q(y)$ and $Q(A \wh{x})$ w.r.t.~the Hamming distance (or a surrogate thereof), thereby using knowledge about the quantization map unlike \eqref{eq:canlin}. One may thus
expect that the performance of the latter is inferior. In summary, our experiments confirm
that this is true in low-noise settings, but not so if the noise level is
substantial. Below we briefly present the alternatives that we consider.\\

\begin{figure}[h!]
\begin{center}
\begin{tabular}{ccc}
\hspace*{-0.022\textheight}$\begin{array}{c} \boxed{b = 1} \\ \\ \\ \\ \\ \\ \\ \\ \\ \\ \\ \\
\end{array}
$ & \includegraphics[width = 0.35\textwidth]{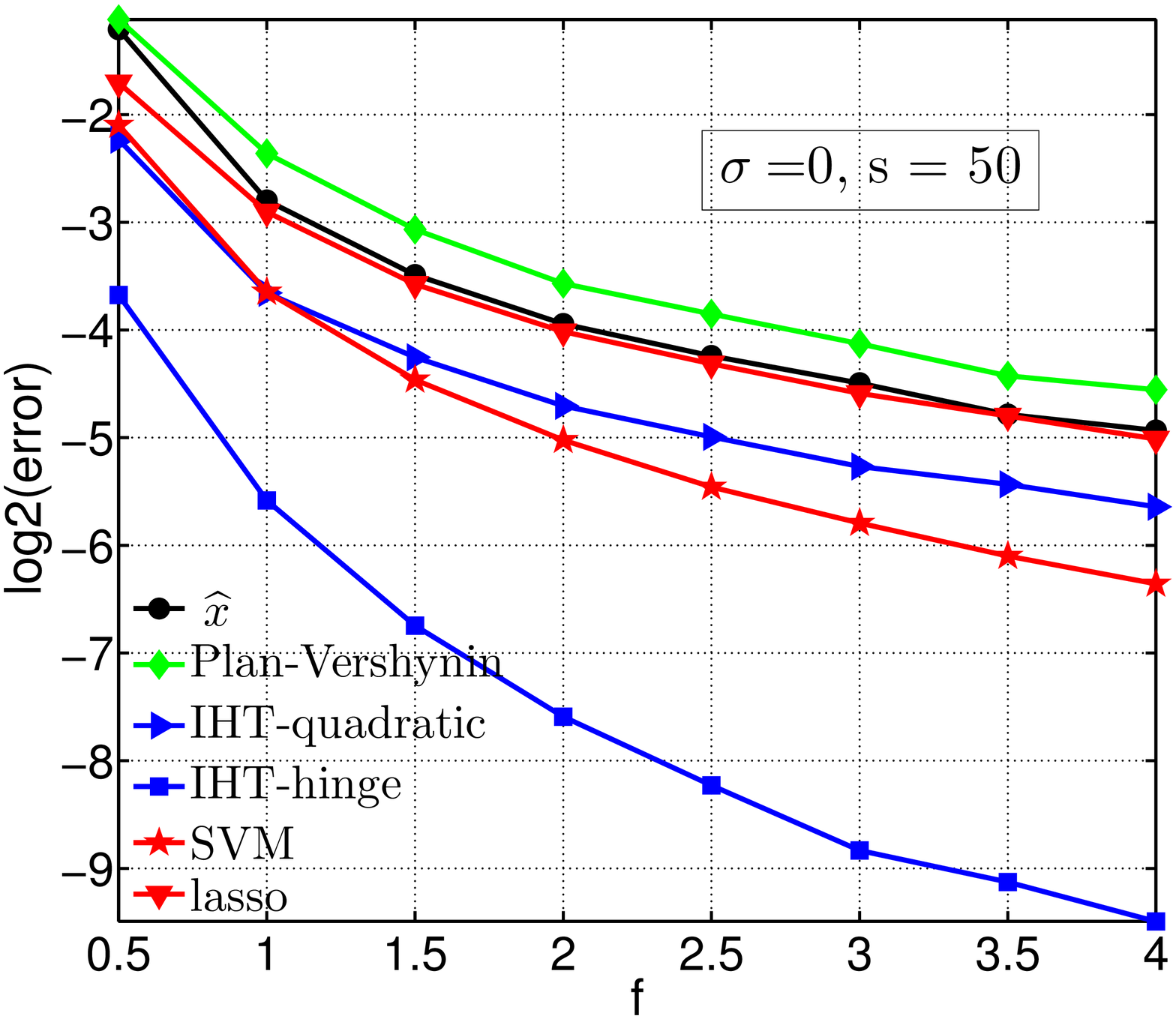}
& \includegraphics[width = 0.35\textwidth]{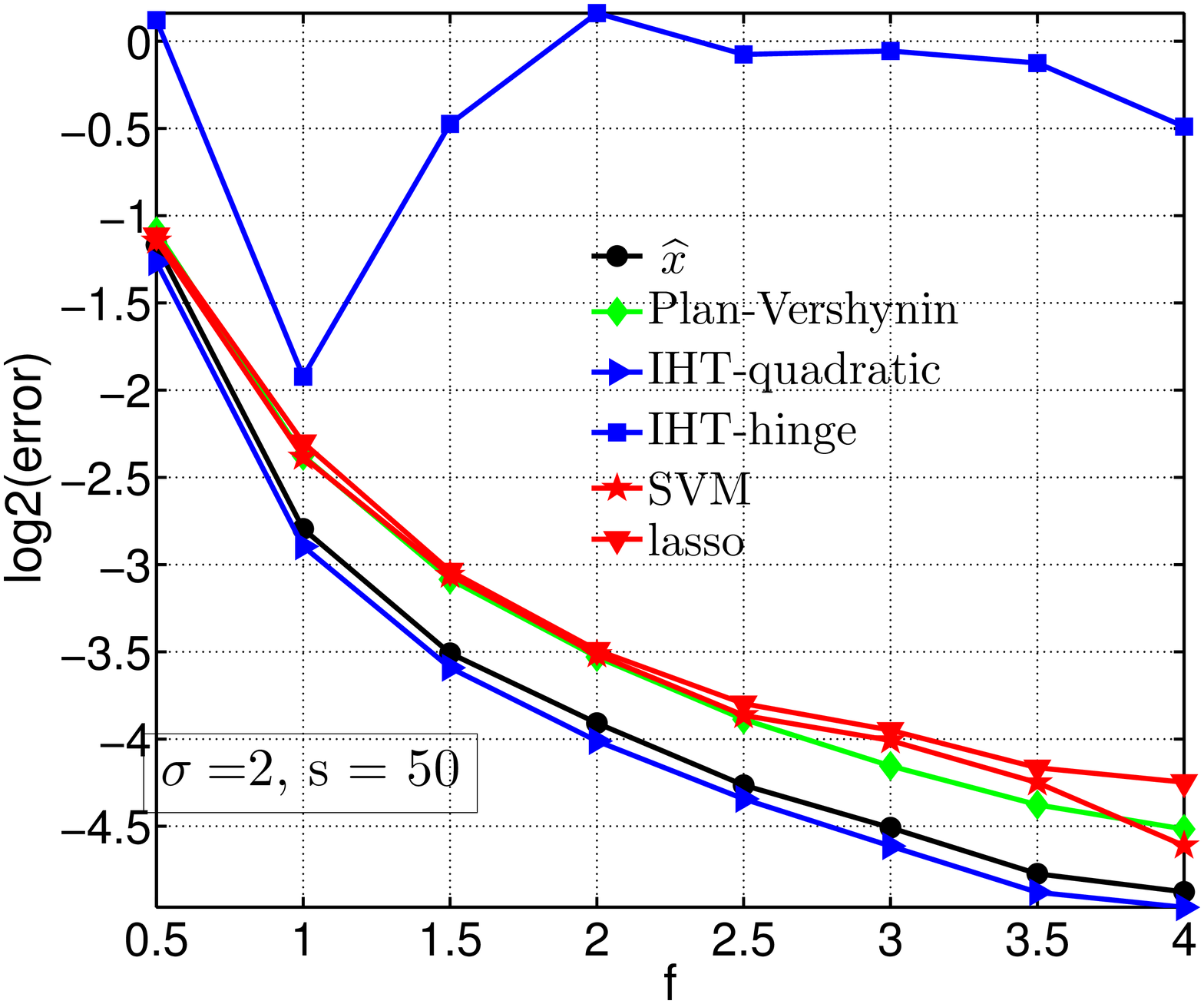} \vspace*{-0.1\textheight} \\
\hspace*{-0.022\textheight} $\begin{array}{c} \boxed{b = 2} \\ \\ \\ \\ \\ \\ \\ \\ \\ \\ \\ \\
\end{array}$ & \includegraphics[width = 0.35\textwidth]{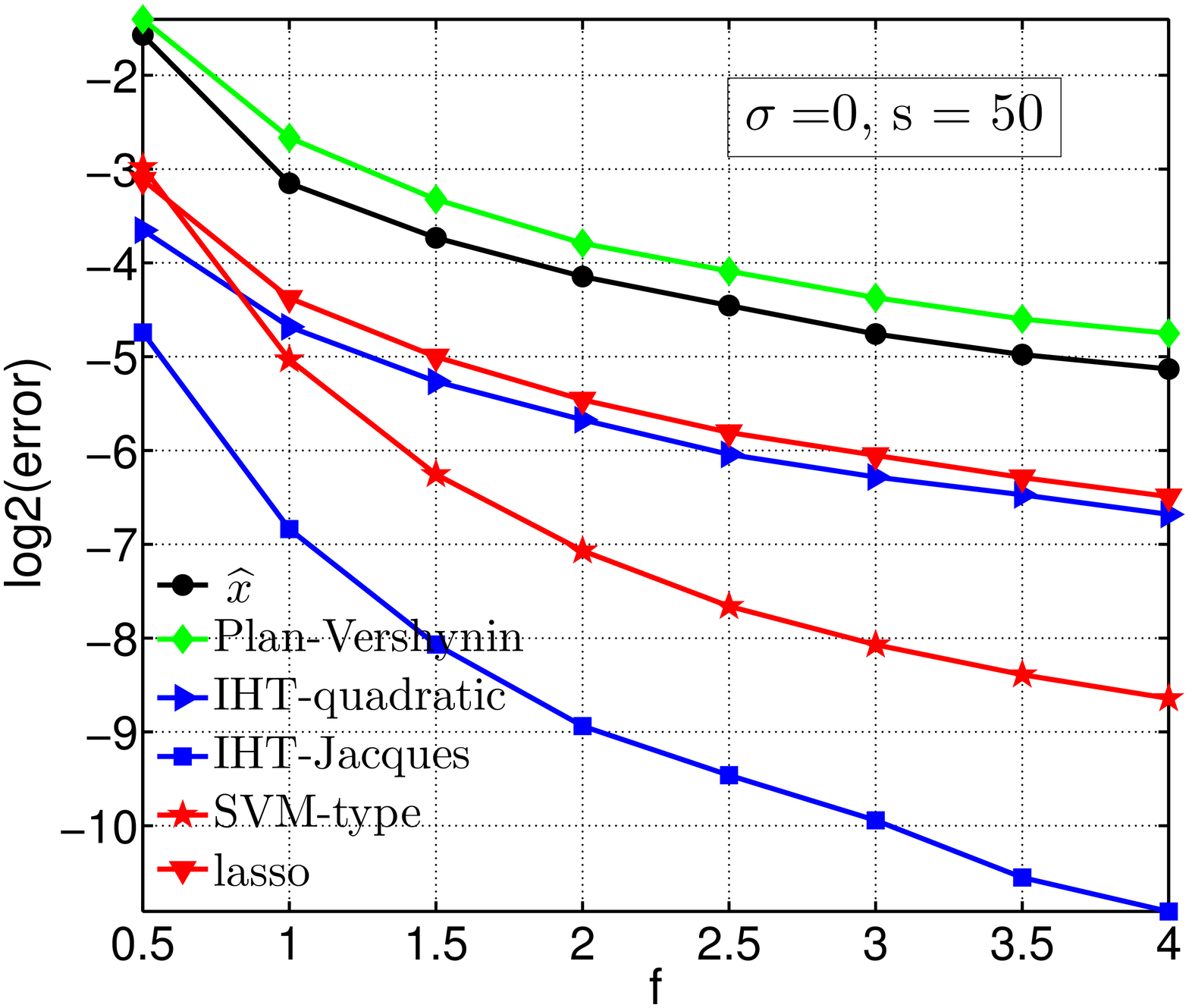}
 & \includegraphics[width = 0.35\textwidth]{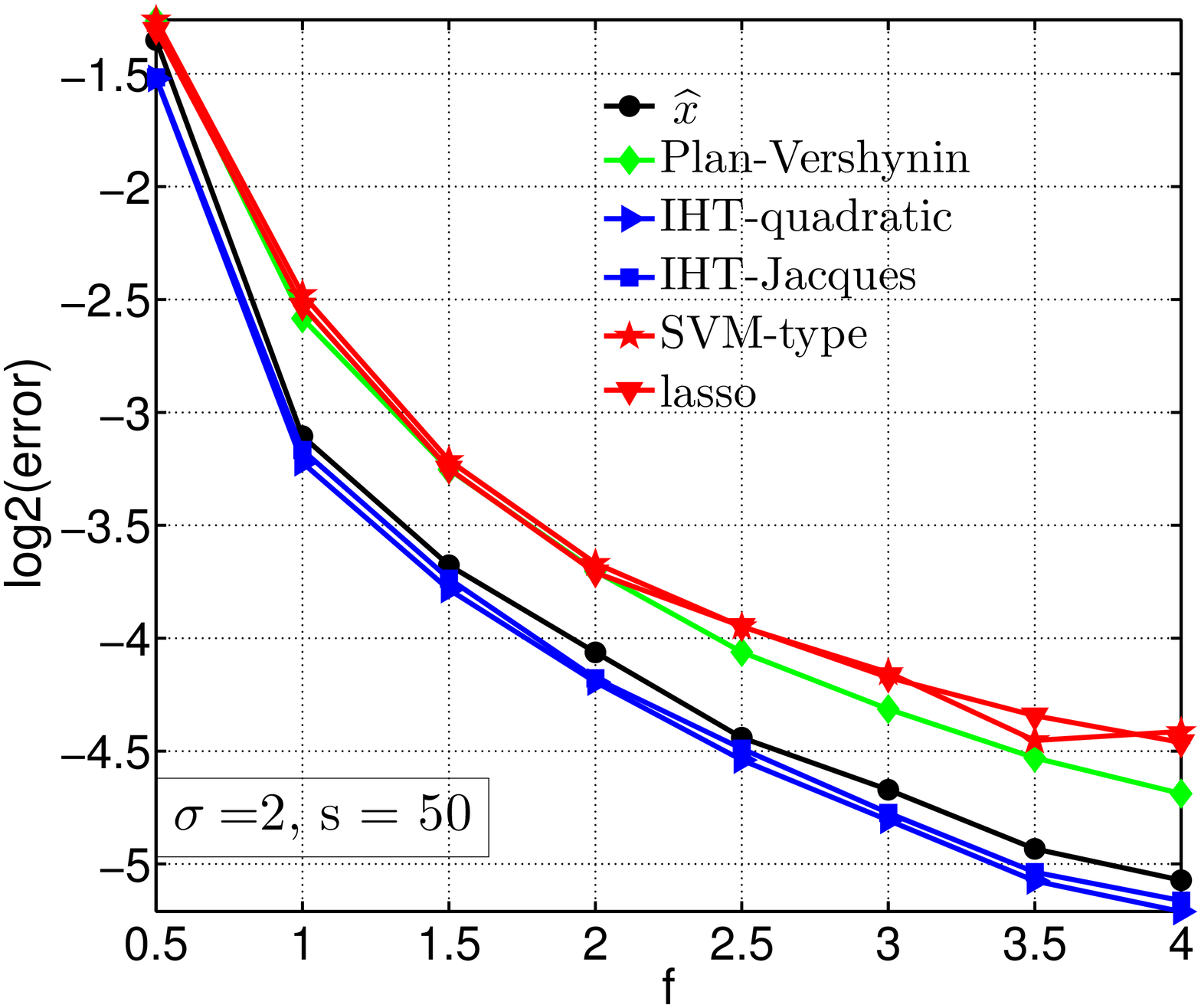}
\end{tabular}
\end{center}
\vspace*{-0.1\textheight}
\vspace*{-0.4in}
\caption{Average $\ell_2$-estimation errors of several recovery
  algorithms on the $\log_2$-scale in dependence of the signal strength
  $f\,$. We contrast $\sigma = 0$ (L) vs.~$\sigma = 2$ (R),
  $b = 1$ (T) vs.~$b = 2$ (B).}\label{fig:comparison}
\end{figure}

\textsf{Plan-Vershynin}: The approach in \cite{PlanVershynin2013a} that differs from \eqref{eq:canlin}
only in that it uses a convex relaxation of the constraint set of the form $\mc{C} = B_0(s;n) \cap B_2^n \subset \sqrt{s} B_1^n \cap B_2^n$. It thus falls under the framework of linear signal recovery outlined in $\S$\ref{sec:linearrecovery}, and its inclusion is mainly
for the sake of reference given the popularity of the work \cite{PlanVershynin2013a}. As shown in Figure \ref{fig:comparison} the performance is similar though slightly inferior to \eqref{eq:canlin}.

\textsf{IHT-quadratic}: The standard Iterative Hard Thresholding algorithm
based on quadratic loss \cite{Blumensath2009}. When using quadratic loss, one does not take into account the fact that the
observations are quantized. The estimator \eqref{eq:canlin} can be seen as one-step version of Iterative Hard Thresholding.

\textsf{Lasso}: $\ell_1$-regularized least squares \cite{Tib1996} according to the optimization problem $\min_x \nnorm{y - Ax}_2^2 + \lambda \nnorm{x}_1$.
The parameter $\lambda$ is selected based on five-fold cross-validation over the grid $\sqrt{m \log n} \cdot \{2^{-5},2^{-4},\ldots,2^2 \}$.\\

The competitors listed below do no longer fit into the framework of linear recovery but into the more sophisticated class described
in the introductory portion of this paragraph.

\textsf{IHT-hinge} ($b = 1$): The variant of Iterative Hard Thresholding for
binary observations using a hinge loss-type loss function as proposed in \cite{Jacques2013a}.

\textsf{IHT-Jacques} ($b = 2$): A variant of Iterative Hard Thresholding for
quantized observations based on a specific piecewise linear loss function
as suggested in a paper of Jacques and collaborators \cite{Jacques2013b}.

\textsf{SVM} ($b = 1$): Linear SVM with squared hinge loss and an
$\ell_1$-penalty on the weights as implemented in \texttt{LIBLINEAR}
\cite{liblinear}. The cost parameter is chosen by means of five-fold cross-validation
over the grid $\sqrt{1/(m \log m)} \cdot \{2^{-3},2^{-2},\ldots,2^3 \}$.

\textsf{SVM-type} ($b = 2$): This approach is based on the following
convex optimization problem:
\begin{equation*}
\min_{x, \{ \xi_i \}} \gamma \nnorm{x}_1 + \sum_{i = 1}^m \xi_i \; \; \; \text{subject to} \;\;
l_i - \xi_i \leq \scp{a_i}{x} \leq u_i + \xi_i, \;\;\; \xi_i \geq 0, \; \, i \in [m],
\end{equation*}
where $[l_i, u_i]$ is the bin the $i$-th observations is assigned to. The
essential idea is to enforce consistency of the observed and predicted bin
assignments up to slacks $\{ \xi_i \}$ while promoting sparsity of the
solution by means of an $\ell_1$-penalty. The parameter $\gamma$ is chosen via five-fold cross-validation
over the grid $\sqrt{m \log m} \cdot \{2^{-10},2^{-9},\ldots,2^3 \}$.

Turning to the results as depicted by Figure \ref{fig:comparison}, the
difference between a noiseless $(\sigma = 0)$ and heavily noisy setting
$(\sigma = 2)$ is perhaps most striking.

 $\sigma = 0$: \textsf{IHT-hinge} respectively \textsf{IHT-Jacques} and \textsf{SVM}-(\textsf{type})
significantly outperform $\wh{x}$. By comparing the errors for
\textsf{IHT} and \textsf{SVM}, $b = 2$ can be seen to improve over $b = 1$ at the level of the total \#
bits.

 $\sigma = 2$: the canonical linear estimator is on par with the best
performing methods. \textsf{IHT-quadratic} for $b = 2$ only achieves a moderate
reduction in error over $b = 1$, while \textsf{IHT-hinge} is supposedly affected by convergence issues that are known
to arise in a high-noise setting \cite{Jacques2015comm}.

Overall, the results suggest that a setting with substantial noise favours a
crude approach (low-bit measurements and conceptually simple recovery
algorithms).

\section{Conclusion}\label{sec:conclusion}

We have considered linear signal recovery from $b$-bit quantized measurements. The main finding is that
increasing the number of bits per measurement beyond is not efficient at the bit scale since the
reduction in estimation error at the measurement scale is not significant enough. A
compelling argument in favor of two-bit measurements is the fact that the norm of the signal
can be estimated unlike the case one-bit measurements. Compared to high-precision
measurements, two-bit measurements also exhibit strong robustness
properties. It is of interest if and under what circumstances the
conclusion may differ for other recovery algorithms. Experimental results suggest
a transition between settings with low noise on the one hand and substantial noise on the other hand.
For example, in a setting without noise we have seen improvements at the level of bits for iterative hard thresholding
algorithms when increasing $b$ from one to two. Providing theoretical support for this observation constitutes an
interesting topic for future research.

\section*{Acknowledgement}

The work is partially supported by NSF-Bigdata-1419210, NSF-III-1360971,
ONR-N00014-13-1-0764, and AFOSR-FA9550-13-1-0137. 

\newpage

\appendix

\noindent{\Large \textbf{Appendix: Proofs and Derivations}}

\section{Linearity lemma}\label{app:linearity}
The following simple result is fundamental to linear signal recovery. The lemma implies that for the non-linear model
\eqref{eq:nonlinearCSmodel}, it may still be possible to recover the signal by a linear approach depending on the nature
of the nonlinear map $Q$. Let us recall the definition of the map $\theta$ in \eqref{eq:theta} according to $\E[y_1 | a_1] = \theta(\scp{a_1}{x_u^*})$, $\lambda = \E[g \, \theta(g)]$ with $g \sim N(0,1)$, and $\eta = A^{\T} y / m$.
\begin{lemmaApp}\label{lem:expectation_eta}
For all $x \in \R^n$, we have $\E[\scp{x}{\eta}] = \lambda \scp{x}{x_u^*}$. In
particular, by considering $x = e_j$, $j \in [n]$, where $\{e_j\}_{j = 1}^n$
is the standard basis of $\R^n$, we have $\E[\eta] = \lambda x_u^*$.
\end{lemmaApp}
\begin{bew} The proof below is taken from \cite{PlanVershynin2013a}, Lemma 4.1. We include it here to make the paper self-contained. We have
\begin{align*}
\E[\scp{x}{\eta}] &= \E\left[\scp{x}{A^{\T} y/m} \right] \\
                  &= \E[\scp{A x}{y}/m] \\
                  &= \E[\scp{a_1}{x} y_1] \\
                  &= \E \E[y_1 \scp{a_1}{x} | a_1] \\
                  &= \E[\theta(\scp{a_1}{x_u^*}) \scp{a_1}{x}] \\
                  &= \E \left[\theta(\scp{a_1}{x_u^*}) \scp{a_1}{x^{\parallel} +
                    x^{\perp}} \right] \\
                  &= \scp{x}{x_u^*} \E[\theta(g) g], \; g \sim N(0,1) \\
                  &= \lambda \scp{x}{x_u^*},
\end{align*}
where in the third line from the bottom $x^{\parallel} = \scp{x}{x_u^*} x_u^*$ and $x^{\perp}$ denote the orthogonal
projection of $x$ on $x_u^*$ and its orthogonal complement, respectively. We
then use that $\scp{a_1}{x^{\perp}}$ and $\scp{a_1}{x_u^*}$ are Gaussian and uncorrelated and
hence also independent random variables.
\end{bew}
Note that $\lambda$ can be zero (e.g.,~if $\theta$ is even). This is not a concern if model \eqref{eq:nonlinearCSmodel} holds
and $Q$ is the scalar $b$-bit quantizer in \eqref{eq:quantizer} (cf.~Lemma \ref{lem:Omega_b}). For the alternative noise models in $\S$\ref{sec:tradeoff}, the situation becomes different as is explained in detail in Appendix \ref{app:beyondadditive}.

\section{Proof of Theorem \ref{theo:mainresult}}
The proof of our result relies on the previous Lemma and a series of results from the theory of empirical processes which
are relegated to separate sections in this appendix. The structure largely builds on techniques used in \cite{PlanVershynin2013a}.

\noindent  \emph{Step 1. Basic inequality.}

\noindent Since $\wh{x}$ in \eqref{eq:canlin} is a minimizer and $x_u^*$ is a feasible solution, we have
\begin{equation*}
-\scp{\eta}{\wh{x}} \leq -\scp{\eta}{x_u^*}.
\end{equation*}
Subtracting and adding $\E[\eta]$, we obtain after some re-arrangements that
\begin{equation}\label{eq:left_and_right}
\scp{x_u^* - \wh{x}}{\E[\eta]} \leq \scp{\wh{x} - x_u^*}{\eta - \E[\eta]}.
\end{equation}

\noindent  \emph{Step 2. Lower bounding the left hand side.}

\noindent From Lemma \ref{lem:expectation_eta}, we obtain that $\E[\eta] = \lambda x_u^*$. Therefore,
\begin{equation}\label{eq:lowerleft_1}
\scp{x_u^* - \wh{x}}{\E[\eta]} = \lambda \scp{x_u^* - \wh{x}}{x_u^*}
\end{equation}
On the other hand, we have that
\begin{equation}\label{eq:lowerleft_2}
\nnorm{x_u^* - \wh{x}}_2^2 \leq 2 (1 - \scp{\wh{x}}{x_u^*}) = 2 (\scp{x_u^*}{x_u^*} -
\scp{\wh{x}}{x_u^*}) = 2 (\scp{x_u^*}{x_u^* - \wh{x}}),
\end{equation}
using that $\nnorm{x_u^*}_2 = 1$ and $\nnorm{\wh{x}}_2 \leq 1$. Combining \eqref{eq:lowerleft_1}
and \eqref{eq:lowerleft_2}, the l.h.s.~of \eqref{eq:left_and_right} is bounded as
\begin{equation}\label{eq:lowerleft}
\scp{x_u^* - \wh{x}}{\E[\eta]} \geq \frac{\lambda}{2} \nnorm{x_u^* - \wh{x}}_2^2.
\end{equation}

\noindent  \emph{Step 3.1 Upper bounding the right hand side.}

\noindent Recalling the definition of $\overline{\Delta}(\mc{C}) = \overline{\Delta}(\mc{C};x_u^*)$ in \eqref{eq:tangentcone}, we have that
\begin{equation}\label{eq:upperright}
\scp{\wh{x} - x_u^*}{\eta - \E[\eta]} \leq \sup_{v \in \overline{\Delta}(\mc{C})}  \scp{v}{\eta - \E[\eta]}  \, \nnorm{\wh{x} - x_u^*}_2.
\end{equation}
The supremum on the right hand sided is controlled by first bounding its expectation and then establishing concentration.\\

\noindent \emph{Step 3.2 Upper bounding the expectation.}

\noindent In the sequel, $\{\xi_{i} \}_{i=1}^n$ denote i.i.d.~Rademacher variables taking values in $\{ \pm 1\}$ with equal probability. Invoking Lemma \ref{lem:symmetrization} with $\Gamma = \{|\scp{v}{\cdot}|,\; v \in \overline{\Delta}(\mc{C}) \}$, we obtain that
\begin{align}\label{eq:upperboundexpectation}
\E \left[\sup_{v \in \overline{\Delta}(\mc{C})} |\scp{\eta - \E[\eta]}{v}| \right] &=  \E\left[\sup_{v \in \overline{\Delta}(\mc{C})} \left|\frac{1}{m} \sm (y_i \scp{a_i}{v} - \E[y_i \scp{a_i}{v}]) \right|\right] \notag\\
& \overset{\text{Lemma \ref{lem:symmetrization}}}{\leq} 2 \E\left[\sup_{v \in \overline{\Delta}(\mc{C})} \left|\frac{1}{m} \sm   y_i \scp{\wt{a_i}}{v} \right|\right], \quad \wt{a_i} \coloneq a_i \xi_i, \; i=1,\ldots,m \notag\\
&=2 \E \nolimits_{y} \left[ \E\left[\sup_{v \in \overline{\Delta}(\mc{C})} \left|\frac{1}{m}  \scp{\sm   y_i \wt{a_i}}{v} \right| \; \Bigg| y \right] \right] \notag\\
&\overset{\substack{\textbf{(C)} \\ n \rightarrow \infty}}{=}2 \E \nolimits_{y} \left[ \E\left[\sup_{v \in \overline{\Delta}(\mc{C})} \left|\frac{\nnorm{y}_2}{m}  \scp{g}{v} \right|  \right] \right], \; \; \, g \sim N(0,I_n), \notag\\
&=\frac{2 \E[\nnorm{y}_2] w(\overline{\Delta}(\mc{C}))}{m} \notag\\
&\leq \frac{2 \sqrt{\E[\nnorm{y}_2^2}] w(\overline{\Delta}(\mc{C}))}{m} \notag\\
&=\frac{2 \Psi_{b, \sigma}(\mathbf{t}, \bm{\mu}) \, w(\overline{\Delta}(\mc{C}))}{\sqrt{m}}.
\end{align}
The fourth line from the top holds under condition \textbf{(C)} as $n \rightarrow \infty$ because $a_i$ and $y_i$ are asymptotically
independent, $i \in [m]$. In fact, for any $j \in [n]$, we have
\begin{equation*}
\E[\scp{a_i}{x_u^*} a_{ij}] =  \E[\scp{a_i}{x_u^*} \scp{a_i}{e_j}] = (x_u^*)_j \rightarrow 0 \; \; \, \text{as} \;  n \rightarrow \infty.
\end{equation*}
Since $\scp{a_i}{x_u^*}$ and $a_{ij}$ are jointly Gaussian with vanishing covariance, they are also asymptotically independent. As
$y_i$ results as a transformation of $\scp{a_i}{x_u^*}$ and $\eps_i$, it holds that $a_{ij}$ and $y_i$ are asymptotically independent as well, $j \in [n]$, $i \in [m]$. The same is true for $\wt{a}_{ij}$ and $y_i$, $j \in [n]$, $i \in [m]$. In this situation, conditional on $y$,
$\sm y_i a_{ij}$ follows a $N(0, \nnorm{y}_2^2)$-distribution, $j \in [n]$. The remaining lines follow from the rotational invariance of
the Gaussian distribution, the definitions of $\Psi_{b, \sigma}(\mathbf{t}, \bm{\mu})$ and $w(\overline{\Delta}(\mc{C}))$, and Jensen's inequality.\\

\noindent \emph{Step 3.3 Concentration.}

\noindent Using Lemma \ref{lem:concentration1} with $\Gamma = \{|\scp{v}{\cdot}|,\; v \in \overline{\Delta}(\mc{C}) \}$ and Lemma \ref{lem:concentration2}, we establish that the bound \eqref{eq:upperboundexpectation} above gives rise to an upper bound for the empirical process $\{\scp{v}{\eta - \E[\eta]}, \, v \in \overline{\Delta}(\mc{C}) \}$ in \eqref{eq:left_and_right} that holds with high probability. We start by upper bounding the right hand side of Lemma \ref{lem:concentration1}, following the strategy of the previous step, i.e.,~we condition on $y$ and then use asymptotic independence. For any $u > 0$
\begin{align}\label{eq:concentration_1}
\p \left(\sup_{v \in \overline{\Delta}(\mc{C})} \left|\frac{1}{m} \sm   \xi_i y_i \scp{a_i }{v} \right| > u/2  \right)
&= \E \nolimits_y \left[ \p \left(\sup_{v \in \overline{\Delta}(\mc{C})} \left|\frac{1}{m} \sm   \xi_i y_i \scp{a_i }{v} \right| > u/2  \; \Bigg| y \right)  \right], \notag\\
&\overset{\substack{\textbf{(C)} \\ n \rightarrow \infty}}{=} \E \nolimits_y \left[ \p \left(\frac{\nnorm{y}}{m}  \sup_{v \in \overline{\Delta}(\mc{C})} \left|\scp{g}{v} \right| > u/2   \right)  \right].
\end{align}
Consider the event $\{m^{-1/2} \nnorm{y}_2  > 2 \Psi_{b,\sigma}(\mathbf{t}, \bm{\mu}) \}$. Note that the $\{y_i \}_{i = 1}^m$ are i.i.d.~zero-mean sub-Gaussian
random variables with standard deviation $\Psi_{b,\sigma}(\mathbf{t},\bm{\mu})$. Then by Lemma \ref{lem:concentration3} and Jensen's inequality,
\begin{align}\label{eq:concentration_2}
\p(m^{-1/2} \nnorm{y}_2  > 2 \Psi_{b,\sigma}(\mathbf{t}, \bm{\mu})) \leq \exp(-c m),
\end{align}
for some universal constant $c > 0$. We work thus conditional on the event $E_0 = \{ m^{-1/2} \nnorm{y}_2  \leq 2 \Psi_{b,\sigma}(\mathbf{t}, \bm{\mu})\}$. Accordingly,
\begin{align}\label{eq:concentration_3}
\E \nolimits_y \left[ \p \left(\frac{\nnorm{y}}{m}  \sup_{v \in \overline{\Delta}(\mc{C})} \left|\scp{g}{v} \right| > u/2   \right)  \right]
&\leq \p \left(\frac{2\Psi_{b,\sigma}(\mathbf{t}, \bm{\mu})}{\sqrt{m}}  \sup_{v \in \overline{\Delta}(\mc{C})} \left|\scp{g}{v} \right| > u/2  \right) \p(E_0) + (1 - \p(E_0)) \notag\\
&\leq \p \left( \sup_{v \in \overline{\Delta}(\mc{C})} \left|\scp{g}{v} \right| > \frac{\sqrt{m} \cdot u}{4 \Psi_{b,\sigma}(\mathbf{t}, \bm{\mu})}   \right) + (1 - \p(E_0))
\end{align}
Choosing $u = 8 \Psi_{b,\sigma}(\mathbf{t}, \bm{\mu}) w(\overline{\Delta}(\mc{C}))/\sqrt{m}$, we invoke Lemma \ref{lem:concentration2} with $\mc{T} = \overline{\Delta}(\mc{C})$, $X_t = |\scp{g}{t}|, \, t \in \mc{T}$, $\E[\sup_{t \in \mc{T}} X_t] = w(\overline{\Delta}(\mc{C}))$ and $r=w(\overline{\Delta}(\mc{C}))$, and combining \eqref{eq:concentration_1}, \eqref{eq:concentration_2}, \eqref{eq:concentration_3}, we obtain that
\begin{equation}\label{eq:concentration_4}
\p \left(\sup_{v \in \overline{\Delta}(\mc{C})} \left|\frac{1}{m} \sm   \xi_i y_i \scp{a_i }{v} \right| > \frac{8}{2} \frac{\Psi_{b,\sigma}(\mathbf{t}, \bm{\mu}) w(\overline{\Delta}(\mc{C}))}{\sqrt{m}} \right) \leq \exp(-w(\overline{\Delta}(\mc{C}))/2) + \exp(-cm).
\end{equation}

\noindent \emph{Step 4. Putting together the pieces}

\noindent Comparing \eqref{eq:concentration_4} with the right hand side of \eqref{eq:lem:concentration1} in Lemma \ref{lem:concentration1} and combining that with \eqref{eq:upperboundexpectation} to bound the expectation appearing in the left hand side of \eqref{eq:lem:concentration1}
\begin{equation}\label{eq:concentration_5}
\p \left( \sup_{v \in \overline{\Delta}(\mc{C})} |\scp{\eta - \E[\eta]}{v}|  \geq \frac{12 \Psi_{b,\sigma}(\mathbf{t}, \bm{\mu}) w(\overline{\Delta}(\mc{C}))}{\sqrt{m}} \right) \leq 4(
\exp(-w(\overline{\Delta}(\mc{C}))/2) + \exp(-cm)).
\end{equation}
Combining \eqref{eq:concentration_5} with \eqref{eq:lowerleft} and \eqref{eq:upperright}, the proof is complete.

\section{Proof of Theorem \ref{theo:anisotropic}}
Let us denote $\zeta \coloneq \Sigma^{-1} \eta$, and $z^* \coloneq \Sigma^{1/2} x^*$. Without loss of generality,
suppose that the thresholds $\mathbf{t}$ are scaled such that it is equivalent to assume that $\nnorm{z^*}_2 = 1$,
cf.~the discussion following \eqref{eq:scaleissue}.

We start by noting that $\E[\zeta] = \lambda x^*$:
\begin{align*}
\E[\zeta] = \Sigma^{-1} \E[\eta] &= \Sigma^{-1} \E \left[\sum_{i = 1}^m y_i
  a_i/m \right] \\
&= \Sigma^{-1} \E[a_1 y_1] = \Sigma^{-1/2} \E \left[a_1^0 Q(\scp{a_1^0}{z^*} +
  \sigma \eps_i) \right] = \Sigma^{-1/2} \lambda z^* = \lambda x^*,
\end{align*}
where $a_{1}^0 \sim N(0, I_n)$. The penultimate inequality follows from Lemma \ref{lem:expectation_eta}. We now proceed as in the proof of Theorem \ref{theo:mainresult}. Since $x_u^*$ is feasible for the
optimization problem \eqref{eq:canlin_anisotropic}, we have $-\scp{\zeta}{\wh{x}_{\Sigma}} \leq
-\scp{\zeta}{x_u^*}$. After some manipulations, we obtain that
\begin{equation}\label{eq:basicineq_anisotropic}
\scp{x_u^* - \wh{x}_{\Sigma}}{\E[\zeta]}  \leq \scp{\wh{x}_{\Sigma} - x_u^*}{\zeta - \E[\zeta]}
\end{equation}
The left hand side of this inequality can be lower bounded as follows:
\begin{align}\label{eq:lowerlhs_anisotropic}
\scp{x_u^* - \wh{x}_{\Sigma}}{\E[\zeta]}  = \scp{x_u^* - \wh{x}_{\Sigma}}{\lambda x^*} &= \lambda \nnorm{x^*}_2 - \lambda \nnorm{x^*}_2 \scp{\wh{x}_{\Sigma}}{x_u^*} \notag\\
                                 &=  \lambda \nnorm{x^*}_2 (1 - \scp{\wh{x}_{\Sigma}}{x_u^*}) \notag\\
                                 &\geq  \frac{\lambda \nnorm{x^*}_2}{2} \nnorm{\wh{x}_{\Sigma} - x_u^*}_2^2 \notag\\
                                 &\geq \frac{\lambda}{2 \nnorm{\Sigma^{1/2}}} \nnorm{\wh{x}_{\Sigma} - x_u^*}_2^2.
\end{align}
For the last line, we have used that $\nnorm{x^*}_2 = \nnorm{\Sigma^{-1/2} z^*}_2 \geq \nnorm{z^*}_2 /
\nnorm{\Sigma^{1/2}} = 1/\nnorm{\Sigma^{1/2}}$, where $\nnorm{M}$ denotes the
spectral norm of a matrix $M$.

In order to upper bound the right hand side of inequality
\eqref{eq:basicineq_anisotropic}, we follow the approach in the proof of Theorem \ref{theo:mainresult} up to a single
modification.\\ A key step therein is to control $\E\left[\sup_{v \in
    \overline{\Delta}(\mc{C})} \left|\frac{1}{m} \sm   y_i \scp{a_i \xi_i}{v}
  \right|\right]$ in Step 3.2 above which becomes $\E\left[\sup_{v \in
    \overline{\Delta}(\mc{C})} \left|\frac{1}{m} \sm   y_i \scp{\Sigma^{-1} a_i \xi_i}{v}
  \right|\right]$ in the anisotropic case.\\

 Writing $a_i = \Sigma^{1/2} a_i^0$, where $a_i^0
\sim N(0, I_n)$ and following the approach in Step 3.2 above, we end up with the problem of controlling
$\E \left[\sup_{v \in \overline{\Delta}(\mc{C})} \left| \scp{\Sigma^{-1/2} g}{v}
    \right| \right]$. By the Sudakov-Fernique comparison inequality (e.g.,~\cite{AdlerTaylor2007},
  Theorem 2.2.3),
$$\E \left[\sup_{v \in \overline{\Delta}(\mc{C})} \left| \scp{\Sigma^{-1/2} g}{v}
    \right| \right] \leq \nnorm{\Sigma^{-1/2}} \E \left[\sup_{v \in \overline{\Delta}(\mc{C})} \left|\scp{g}{v}
    \right| \right] = \nnorm{\Sigma^{-1/2}}
  w(\overline{\Delta}(\mc{C})).$$
Thus, carrying an extra factor of
  $\nnorm{\Sigma^{-1/2}}$, we may continue with the remaining steps in the proof of Theorem \ref{theo:mainresult}. Using \eqref{eq:lowerlhs_anisotropic} and noting that $\kappa(\Sigma) =
  \nnorm{\Sigma^{-1/2}} \nnorm{\Sigma^{1/2}}$ yields the claim.

\section{Asymptotic sharpness of the leading constant}\label{app:sharpness}
We here derive \eqref{eq:asymptotics_univariate} as $m, n \rightarrow \infty$, $n/m \rightarrow 0$ and \textbf{(C)} holds.

%Note that for $\mc{K} = \R^n$, we have $\wh{x} = \eta/\nnorm{\eta}_2 = m^{-1} A^{\T} y / \nnorm{m^{-1} A^{\T} y}_2$.
We here suppress dependence of $\lambda$ and $\Psi$ on $\mathbf{t}, \bm{\mu}$, $b$ and $\sigma$. Fix $j \in [n]$ arbitrary and let $\sigma_i^2 =
\var[A_{ij} y_i] =  \E[\{ A_{ij} y_i  - \lambda x_{u,j}^* \}^2] = \E[A_{ij}
  y_i^2] - \lambda (x_{u,j}^*)^2, \; i \in [m]$, and let $s_m^2 = \sm \sigma_i^2$. Since the $\{A_{ij} y_i - \lambda
  x_{u,j}^*\}_{i = 1}^m$ are independent zero-mean sub-Gaussian random variables,
  the Lyapunov condition
\begin{equation*}
\frac{1}{s_m^3} \sm \E[|A_{ij} y_i  - \lambda x_{u,j}^*|^3] \rightarrow 0 \quad
\text{as} \; m \rightarrow \infty
\end{equation*}
is trivially satisfied. Consequently, by the Lyapunov central limit theorem (\cite{Billingsley1995},
p.~362)
\begin{equation}\label{eq:lyapunov_basic}
\frac{1}{s_m} \sm (A_{ij} y_i - \lambda x_{u,j}^*)  \overset{\mc{D}}{\rightarrow}
N(0,1) \quad \text{as} \; m \rightarrow \infty.
\end{equation}
Let us re-consider $\sigma_i^2 = \E[(A_{ij} y_i - \lambda x_{u,j}^*)^2]$. As $n
\rightarrow \infty$, $A_{ij}$ and $y_i$ become independent under \textbf{(C)} and thus $\sigma_i^2 \rightarrow \E[A_{ij}^2] \E[y_i^2] - \lambda
(x_{u,j}^*)^2 = \Psi^2 - \lambda (x_{u,j}^*)^2$, $i \in [m]$. Therefore,
\begin{equation*}
\frac{s_m^2}{m(\Psi^2 - \lambda (x_{u,j}^*)^2)} \rightarrow 1 \quad \text{as} \; m,n \rightarrow \infty.
\end{equation*}
Moreover, under \textbf{(C)}, $m(\Psi^2 - \lambda (x_{u,j}^*)^2) = m \Psi^2 -
o(m)$ as $m,n \rightarrow \infty$ and hence also
\begin{equation}\label{eq:smratio}
\frac{s_m^2}{m \Psi^2} \rightarrow 1 \quad \text{as} \, m,n \rightarrow \infty.
\end{equation}
Combining \eqref{eq:lyapunov_basic} and \eqref{eq:smratio}, it follows that
\begin{equation*}
\sqrt{m} \left( \frac{\sm A_{ij}  y_i}{m} - \lambda x_{u,j}^* \right)  \overset{\mc{D}}{\rightarrow}
N(0,\Psi^2) \quad \text{as} \; m,n \rightarrow \infty.
\end{equation*}
Now consider $\nnorm{A^{\T} y / m}_2^2$. As $m, n \rightarrow  \infty$ and
$n/m \rightarrow 0$, this quantity converges to $\lambda^2 \nnorm{x_u^*}_2^2 =
\lambda^2$ in probability. By the continuous mapping theorem and Slutsky's
theorem (\cite{vanderVaart1998}, $\S$2), we finally obtain that
\begin{equation*}
\sqrt{m} \left( \frac{\sm A_{ij}  y_i / m}{\nnorm{A^{\T} y / m}_2} -  x_{u,j}^* \right)  \overset{\mc{D}}{\rightarrow}
N \left(0, \frac{\Psi^2}{\lambda^2} \right) \quad \text{as} \; m,n \rightarrow \infty.
\end{equation*}
The result to be shown follows by noting that for $\mc{K} = \R^n$, we have $\wh{x} = \eta / \nnorm{\eta}_2$ and accordingly $\wh{x}_j = \frac{\sm A_{ij}
  y_i / m}{\nnorm{A^{\T} y / m}_2}$.

\section{Proof of Lemma \ref{lem:Omega_b}}
The proof of Lemma \ref{lem:Omega_b} requires three additional lemmas.

\begin{lemmaApp}\label{lem:integration_by_parts}
  Let $g \sim N(0,1)$ and $\zeta: \R \rightarrow \R$ be any differentiable function satisfying $|\zeta(x)
x \phi(x)| \rightarrow 0$ as $x \rightarrow \infty$, where $\phi$ denotes the standard Gaussian pdf. Then
$\E[\zeta(g) g] = \E[\zeta'(g)]$.
\end{lemmaApp}
\begin{bew} Observe that
 $\phi'(x) = -x \phi(x), \; x \in \R$. Using integration by parts we thus have
\begin{align*}
\E[\zeta(g) g] &= \int_{\R} x \zeta(x) \phi(x) \, dx = \left\{ \zeta(x) (-\phi'(x)) \right\}\bou{-\infty}{\infty}   + \int_{\R} \zeta'(x)
\, \phi(x) \; dx \\
&= \int_{\R} \zeta'(x) \, \phi(x) \; dx = \E[\zeta'(g)].
\end{align*}
\end{bew}

\begin{lemmaApp}\label{lem:integration_gaussians}
For all $\alpha, \beta > 0$ and all $\mu, \nu \in \R$, one has
\begin{equation*}
\int_{-\infty}^{\infty} \frac{1}{\alpha} \phi \left(\frac{x - \mu}{\alpha} \right)
\frac{1}{\beta} \phi \left(\frac{x - \nu}{\beta} \right) dx =
\frac{1}{\sqrt{\beta^2 + \alpha^2}}  \phi \left(\frac{\mu - \nu}{\sqrt{\beta^2 + \alpha^2}} \right).
\end{equation*}
\end{lemmaApp}
\begin{bew} Using elementary manipulations, one computes
\begin{align*}
\begin{split}
&\int_{-\infty}^{\infty}
\frac{1}{\alpha} \phi \left(\frac{x - \mu}{\alpha} \right)  \frac{1}{\beta}
\phi \left(\frac{x - \nu}{\beta} \right)
dx \\
&= \frac{1}{2 \pi \alpha \beta} \int_{-\infty}^{\infty} \exp \left(-\frac{(\mu
    - x)^2}{2 \alpha^2} \right) \exp \left(-\frac{(\nu - x)^2}{2 \beta^2} \right)
dx \\
&= \frac{1}{2 \pi \alpha \beta}  \exp \left(-{\frac { ( \mu - \nu)^{2}}{2(\alpha^2 + \beta^2)}} \right) \times \\
&\times \int_{-\infty}^{\infty} \exp \left(-\frac{1}{2} \left(  \left( {\frac {\mu}{\alpha^2}}+{\frac {\nu}{\beta^2}}
  \right)  \left( \frac{1}{\alpha^2} + \frac{1}{\beta^2} \right) - x
  \right) ^{2} \left( \frac{1}{\alpha^2}+ \frac{1}{\beta^2} \right)
 \right) \; dx \\
&= \frac{1}{2 \pi \alpha \beta} \sqrt{2 \pi} \frac{\alpha \beta}{\sqrt{\beta^2 + \alpha^2}}
\exp \left(-{\frac { ( \mu - \nu
      )^{2}}{2(\alpha^2 + \beta^2)}} \right) \\
&=\frac{1}{\sqrt{2 \pi}} \frac{1}{\sqrt{\beta^2 + \alpha^2}} \exp \left(-{\frac { ( \mu - \nu) ^{2}}{2(\alpha^2 + \beta^2)}} \right).
\end{split}
\end{align*}
\end{bew}

\begin{lemmaApp}\label{lem:conditionalexpectation} Let $h$ be a random variable with a $N(0,
  \sigma^2)$-distribution. Then for any $a,b \in \R \cup \{-\infty,\infty \}$,
  $a < b$, we have
\begin{equation*}
\E[h | h \in (a,b)] = \sigma \frac{\phi(a/\sigma) - \phi(b/\sigma)}{\Phi(b /
  \sigma) - \Phi(a / \sigma)},
\end{equation*}
where $\Phi$ denotes the standard Gaussian cdf.
\end{lemmaApp}
\begin{bew} We have
\begin{equation*}
\E[h | h \in (a,b)] = \frac{1}{\Phi(b /
  \sigma) - \Phi(a / \sigma)} \int_{a}^b \frac{x}{\sigma}\phi(x/\sigma) \, dx.
\end{equation*}
Using the change of variables $z = x / \sigma$ and the fact that $\phi'(z) = -z \phi(z)$, the result follows.
\end{bew}

Before finally turning to the proof of Lemma \ref{lem:Omega_b}, let us recall the definition of the quantization map \eqref{eq:quantizer}. In that definition we have used the
symmetry of the Gaussian distribution around $0$ so that a partitioning of $\R_+$ automatically translates into a partitioning of
$\R$. For parts of the proofs, however, it is more convenient to work with the
following alternative (albeit equivalent) definition.
\begin{defnApp}\label{defn:quant_alternative} Define
$\mc{Q}_1 = -\mc{R}_K$, $\mc{Q}_2 = -\mc{R}_{K-1}, \ldots, \mc{Q}_K = -\mc{R}_1$,
$\mc{Q}_{K+k} = \mc{R}_k$, $k \in [K]$, and $\wt{\mu} =
(-\mu_K,\ldots,\mu_1,\mu_1,\ldots,\mu_K)^{\T}$. Then an equivalent definition
of the quantization map is given by $z \mapsto Q(z) = \sum_{k = 1}^{2K}
\wt{\mu}_k I(z \in \mc{Q}_k)$. Likewise, we define $\wt{\mathbf{t}} = (-t_K,
-t_{K-1}, \ldots, t_0, t_1, \ldots, t_{K-1}, t_K)^{\T}$.
\end{defnApp}

\textbf{Expression for $\lambda$.} Recall that $\lambda = \lambda_{b,\sigma} = \lambda_{b, \sigma}(\mathbf{t},
\bm{\mu})$ is defined by $\lambda = \E[g \, \theta(g)]$, $g \sim N(0,1$), where the map
$\theta$ is in turn defined by the relation $\E[y_1 | a_1] =
\theta(z_1)$ (here and below $z_i = \scp{a_i}{x_u^*}$, $i \in [m]$). We
have
\begin{align*}
\E[y_1 | a_1] &= \sum_{k = 1}^{2^b} \wt{\mu}_k \p(y_1 \in \mc{Q}_k) \\
             &=  \sum_{k = 1}^{2^b} \wt{\mu}_k \p(z_1 + \sigma \eps_1 \in \mc{Q}_k) \\
             &=  \sum_{k = 1}^{2^b} \wt{\mu}_k  \p(z_1 + \sigma \eps_1 \in (\wt{t}_k, \wt{t}_{k+1})) \\
             &=  \sum_{k = 1}^{2^b} \wt{\mu}_k  \left\{ \Phi((\wt{t}_{k + 1} - z_1) / \sigma) - \Phi((\wt{t}_{k} - z_1)/\sigma) \right\}.
\end{align*}
We conclude that the map $\theta$ is defined by
\begin{equation*}
\theta(z) = \sum_{k = 1}^{2^b} \wt{\mu}_k  \left\{ \Phi((\wt{t}_{k + 1} - z) / \sigma) - \Phi((\wt{t}_{k} - z)/\sigma) \right\}.
\end{equation*}
Next we invoke Lemma \ref{lem:integration_by_parts} which yields $\lambda = \E[z \theta(z)] = \E[\theta'(z)]$. We have
\begin{equation*}
\theta'(z) = \sum_{k = 1}^{2^b} \wt{\mu}_k \left\{ \frac{1}{\sigma}\phi((z - \wt{t}_{k}
)/\sigma) - \frac{1}{\sigma}\phi((z - \wt{t}_{k + 1} ) / \sigma) \right\}.
\end{equation*}
With the help of Lemma \ref{lem:integration_gaussians}, we compute
\begin{align*}
\E[\theta'(z)] &= \sum_{k = 1}^{2^b} \wt{\mu}_k \int_{\R} \left\{
  \frac{1}{\sigma}\phi((z - \wt{t}_{k})/\sigma) - \frac{1}{\sigma}\phi((z - \wt{t}_{k + 1}) / \sigma)
\right\} \phi(z) \; dz. \\
&= \sum_{k = 1}^{2^b} \wt{\mu}_k \frac{1}{\sqrt{1 + \sigma^2}} \left\{ \phi \left(\frac{\wt{t}_k}{\sqrt{1 +
  \sigma^2}} \right) - \phi \left(\frac{\wt{t}_{k+1}}{\sqrt{1 +
  \sigma^2}} \right) \right\}.
\end{align*}
Applying Lemma \ref{lem:conditionalexpectation}, the last expression can be rewritten as follows:
\begin{align*}
&\sum_{k = 1}^{2^b} \wt{\mu}_k \frac{1}{\sqrt{1 + \sigma^2}} \left\{ \phi \left(\frac{\wt{t}_k}{\sqrt{1 +
  \sigma^2}} \right) - \phi \left(\frac{\wt{t}_{k+1}}{\sqrt{1 +
  \sigma^2}} \right) \right\} \\
&= \sum_{k = 1}^{2^b} \wt{\mu}_k \frac{\E[\wt{g}|\wt{g} \in (\wt{t}_k,
  \wt{t}_{k+1})]}{1+\sigma^2}  \left\{ \Phi \left(\wt{t}_{k + 1} /
  \sqrt{1+\sigma^2} \right) - \Phi \left(\wt{t}_{k}/\sqrt{1+\sigma^2} \right)\right \},\\[-0.7ex]
&\hspace{2.5in} \text{where } \ \wt{g} \sim N(0, 1+\sigma^2)
\\
&= \frac{1}{1 + \sigma^2} \sum_{k = 1}^{2^b} \wt{\mu}_k \E[\wt{g} | \wt{g} \in
\mc{Q}_k] \p(\wt{g} \in \mc{Q}_k) \\
&= \frac{1}{1 + \sigma^2} \sum_{k = 1}^{K} \mu_k \E[\wt{g} | \wt{g} \in
\mc{R}_k] \p(|\wt{g}| \in \mc{R}_k)  \\
&= \frac{1}{1 + \sigma^2}
\scp{\bm{\alpha}(\mathbf{t})}{\bm{E}(\mathbf{t}) \odot \bm{\mu}},
\end{align*}
where the penultimate line follows from the symmetry of the Gaussian
distribution around zero; at this point, we convert the partitioning of $\R$ into
$\{\mc{Q}_k\}_{k=1}^{2K}$ back to the partitioning of $\R_+$ into $\{ \mc{R}_k \}_{k = 1}^K$ (cf.~the remark preceding Definition \ref{defn:quant_alternative}). The last line
follows by comparison with the definitions in Lemma \ref{lem:Omega_b}.\\

\noindent\textbf{Expression for $\Psi$.} Recall that $\Psi = \Psi_{b,\sigma} = \Psi_{b, \sigma}(\mathbf{t}, \bm{\mu})$
is given by $\sqrt{\E[y_1^2]}$. Note that the random variable $y_1^2$ takes values in $\{\mu_k^2 \}_{k=1}^K$ with
$\p(y_1^2 = \mu_k^2) = \alpha_k(\mathbf{t})$, $k \in [K]$. Accordingly,
\begin{equation*}
\sqrt{\E[y_1^2]} = \sqrt{\sum_{k = 1}^K \alpha_k(\mathbf{t}) \mu_k^2} = \sqrt{\scp{\bm{\alpha}(\mathbf{t})}{\bm{\mu} \odot \bm{\mu}}}.
\end{equation*}

\section{Proof of Theorem \ref{theo:lloydmax}}\label{app:lloydmax}
Consider the optimization problem
\begin{equation*}
\min_{\mathbf{t},\bm{\mu}} \Omega_{b,\sigma}(\mathbf{t},\bm{\mu}) =
\min_{\mathbf{t},\bm{\mu}}
\frac{\Psi_{b,\sigma}(\mathbf{t},\bm{\mu})}{\lambda_{b,\sigma}(\mathbf{t},\bm{\mu})}.
\end{equation*}
By Lemma \ref{lem:Omega_b}, the above minimization problem is equivalent to
\begin{equation}\label{eq:ratiofun}
\min_{\mathbf{t},\bm{\mu}} R(\mathbf{t},\bm{\mu}), \quad R(\mathbf{t}, \bm{\mu})
=\frac{\sqrt{\scp{\bm{\alpha}(\mathbf{t})}{\bm{\mu}
      \odot \bm{\mu}}}}{\scp{\bm{\alpha}(\mathbf{t})}{\bm{E}(\mathbf{t}) \odot \bm{\mu}}},
\end{equation}
where the term $\sigma^2 + 1$ in $\lambda_{b,\sigma}$ has been dropped as it does not depend on
$\mathbf{t}$ or $\bm{\mu}$. We start by claiming that
\begin{equation}\label{eq:lowerbound_by_cauchyschwarz}
R(\mathbf{t}, \bm{\mu}) \geq
\frac{1}{\sqrt{\scp{\bm{\alpha}(\mathbf{t})}{\bm{E}(\mathbf{t}) \odot
    \bm{E}(\mathbf{t})}}}
\end{equation}
for all $\bm{\mu}$ with distinct, non-zero entries. The above
lower bound is attained by choosing $\bm{\mu}$ proportional to $\bm{E}(\mathbf{t})$
(note that the minimizing $\bm{\mu}$ is only defined up to a positive
constant as $R(\mathbf{t}, c \bm{\mu}) = R(\mathbf{t}, \bm{\mu})$ for all $c > 0$). Inequality \eqref{eq:lowerbound_by_cauchyschwarz} follows from the Cauchy-Schwarz inequality. Denote by
$\bm{A}(\mathbf{t})$ the diagonal matrix whose diagonal is given by the
entries of $\bm{\alpha}(\mathbf{t})$. We then have
\begin{align*}
\scp{\bm{\alpha}(\mathbf{t})}{\bm{E}(\mathbf{t}) \odot \bm{\mu}} &=
\scp{\bm{A}^{1/2}(\mathbf{t}) \bm{E}(\mathbf{t})}{\bm{A}^{1/2}(\mathbf{t})
  \bm{\mu}} \\
&\leq \sqrt{\scp{\bm{A}^{1/2}(\mathbf{t})
    \bm{E}(\mathbf{t})}{\bm{A}^{1/2}(\mathbf{t}) \bm{E}(\mathbf{t})}}
\sqrt{\scp{\bm{A}^{1/2}(\mathbf{t}) \bm{\mu}}{\bm{A}^{1/2}(\mathbf{t}) \bm{\mu}}}
 %\sqrt{}
\end{align*}
with equality holding if and only if
\begin{equation*}
\bm{A}^{1/2}(\mathbf{t}) \bm{E}(\mathbf{t}) = c \bm{A}^{1/2}(\mathbf{t})
\bm{\mu} \; \Leftrightarrow \, \bm{E}(\mathbf{t}) =  c \bm{\mu},
\end{equation*}
for some $c > 0$, where the above $\Leftrightarrow$ follows from the fact that
the entries of $\mathbf{t}$ are required to be distinct so that the matrix
$\bm{A}^{1/2}$ is regular. We conclude that
\begin{equation}\label{eq:min_thresholds}
\min_{\mathbf{t},\bm{\mu}} R(\mathbf{t},\bm{\mu}) = \min_{\mathbf{t}}
R(\mathbf{t},\bm{E}(\mathbf{t})) = \min_{\mathbf{t}} \frac{1}{\sqrt{\scp{
      \bm{\alpha}(\mathbf{t})}{\bm{E}(\mathbf{t}) \odot \bm{E}(\mathbf{t})}}}.
\end{equation}
We will now show that the above minimization problem in $\mathbf{t}$ is
equivalent to the $b$-bit Lloyd-Max quantization problem \eqref{eq:lloydmaxproblem} of a random variable
$h \sim N(0, 1+\sigma^2)$, which we re-state here for convenience:
\begin{align}\label{eq:lloydmaxproblem_app}
\begin{split}
\min_{\mathbf{t}, \bm{\mu}} \E[\{ h - Q(h;\mathbf{t}, \bm{\mu}) \}^2] &= \min_{\mathbf{t}, \bm{\mu}} \E[\{h - \sign(h) \textstyle \sum_{k = 1}^K \mu_k I(|h|
\in \mc{R}_k(\mathbf{t})\, ) \}^2]
\end{split}
\end{align}
For the above problem, it is not hard to see that for any fixed choice of
$\mathbf{t}$, the minimizing $\bm{\mu}^*(\mathbf{t})$ is given by $\mu_k^*(\mathbf{t}) = \E[h | h \in
\mc{R}_k(\mathbf{t})] = E_k(\mathbf{t})$, $k \in [K]$, where we recall that
$E_k(\mathbf{t})$ is the $k$-th component of $\bm{E}(\mathbf{t})$ as appearing
above. To finish the proof of the first part of the Theorem 1, it thus remains
to show that after substituting $\bm{\mu}^*(\mathbf{t})$ back into \eqref{eq:lloydmaxproblem_app}, the resulting minimization problem in $\mathbf{t}$ is
equivalent to \eqref{eq:min_thresholds}. We have
\begin{align*}
&\min_{\mathbf{t}} \E\left [ \left\{h - \sign(h)  \sum_{k = 1}^K I(|h| \in
\mc{R}_k(\mathbf{t}) \, ) \E[h | h \in \mc{R}_k(\mathbf{t}) ]
 \right \}^2 \right] \\
&=2  \min_{\mathbf{t}} \E \left[\sum_{k = 1}^K I(h \in \mc{R}_k(\mathbf{t})) (h - \E[h | h
  \in \mc{R}_k(\mathbf{t})])^2 \right] \\
&= 2 \min_{\mathbf{t}}  \E \left[\sum_{k = 1}^{K}  I(h \in \mc{R}_k(\mathbf{t})) \left\{ h^2  - 2h\, \E[h | h \in \mc{R}_k(\mathbf{t})] + \E[h | h \in \mc{R}_k(\mathbf{t})]^2   \right\} \right] \\
&= \E[h^2] + 2 \min_{\mathbf{t}}\Bigg\{-2\sum_{k=1}^{K} \E[h | h \in
    \mc{R}_k(\mathbf{t})] \E[I(h \in \mc{R}_k(\mathbf{t})) h]
+ \\
    &\qquad \qquad \qquad \qquad + \sum_{k=1}^{K} \p(h \in \mc{R}_k(\mathbf{t}))
    \E[h | h \in \mc{R}_k(\mathbf{t})]^2  \Bigg\} \\
&=  1 + \min_{\mathbf{t}} -\sum_{k=1}^{K} \E[h | h \in
  \mc{R}_k(\mathbf{t})]^2 \p(|h| \in \mc{R}_k(\mathbf{t})) \\
&= 1 + \min_{\mathbf{t}} -\scp{\bm{E}(\mathbf{t}) \odot
    \bm{E}(\mathbf{t})}{\bm{\alpha}(\mathbf{t})},
\end{align*}
which establishes the equivalence to \eqref{eq:min_thresholds} as claimed.

We now prove the second part of the theorem. Denote by $\mathbf{t}_0^*$ the
Lloyd-Max optimal thresholds for $\sigma = 0$, i.e.,~for a $N(0,1)$
variable. Clearly, $\mathbf{t}^* = \mathbf{t}_{\sigma}^* = \sqrt{1 + \sigma^2}
\mathbf{t}_0^*$ for any $\sigma > 0$. Evaluating $\Omega_{b,\sigma}(\mathbf{t}^*, \bm{\mu}^*)$,
we obtain in view of Lemma \ref{lem:Omega_b} and \eqref{eq:ratiofun},\eqref{eq:min_thresholds}
\begin{align*}
\Omega_{b,\sigma}(\mathbf{t}^*, \bm{\mu}^*) &= \frac{1 + \sigma^2}{
\sqrt{\scp{ \bm{\alpha}(\mathbf{t}^*)}{\bm{E}(\mathbf{t}^*) \odot
\bm{E}(\mathbf{t}^*)}}} \\
&= \frac{1 + \sigma^2}{
\sqrt{\scp{ \bm{\alpha}(\mathbf{t}_0^* \sqrt{1 +
      \sigma^2})}{\bm{E}(\mathbf{t}_0^* \sqrt{1 + \sigma^2}) \odot
\bm{E}(\mathbf{t}_0^* \sqrt{1 + \sigma^2})}}}
 \end{align*}
Evaluating the expression in the denominator, we obtain that
\begin{equation*}
\alpha_k(\mathbf{t}_0^* \sqrt{1 +
      \sigma^2}) = \p(|\wt{g}| \in \mc{R}_k(\mathbf{t}_0^* \sqrt{1 + \sigma^2}))
= \p(|g| \in \mc{R}_k(\mathbf{t}_0^*)), \; k \in [K],
\end{equation*}
where  $\wt{g} \sim N(0, 1+\sigma^2), \; g \sim N(0,1)$. Moreover, with the
help of Lemma \ref{lem:conditionalexpectation}
\begin{align*}
\bm{E}(\mathbf{t}_0^* \sqrt{1 + \sigma^2}) &= \left( \E[\wt{g} | \wt{g} \in
\mc{R}_k( \mathbf{t}_0^* \sqrt{1 + \sigma^2})] \right)_{k=1}^K \\
&=\sqrt{1 + \sigma^2} \Big( \E[g | g \in
\mc{R}_k( \mathbf{t}_0^*)] \Big)_{k = 1}^K.
\end{align*}
Putting together the pieces, we obtain that
\begin{align*}
\Omega_{b,\sigma}(\mathbf{t}^*, \bm{\mu}^*) &=
\frac{1 + \sigma^2}{\sqrt{\scp{\bm{\alpha}_0(\mathbf{t}_0^*)}{{(1+ \sigma^2)
    \bm{E}_0(\mathbf{t}_0^*) \odot \bm{E}_0(\mathbf{t}_0^*)}}}}
= \frac{\sqrt{1 + \sigma^2}}{\sqrt{\lambda_{b,0}(\mathbf{t}_0^*, \bm{\mu}_0^*)}}
\end{align*}
where the $\bm{\alpha}_0(\mathbf{t})$, $\bm{E}_0(\mathbf{t})$ and
$\lambda_{b,0}(\mathbf{t}, \bm{\mu})$ refer to the definitions of $\bm{\alpha}(\mathbf{t}),
\bm{E}(\bm{t}), \lambda_{b,\sigma}(\mathbf{t}, \bm{\mu})$ for $\sigma = 0$. This
completes the proof.
%\end{bew}

\section{Proof of Proposition \ref{prop:scale}}
%\begin{bew}
In the sequel, we derive tail bounds of the form
\begin{align*}
\p(\wh{\psi} \geq (1 + \eps) \psi^*) &\leq \exp(-c m \eps^2),\\
\p(\wh{\psi} \leq (1 - \eps) \psi^*) &\leq \exp(-2 c m \eps^2).
\end{align*}
for $\eps \in (0,1)$ and $c = 2  \{ \phi'(t/\psi^*) \}^2$. This implies that the
probability of the event $\left \{ \left|\frac{\wh{\psi}}{\psi^*} - 1 \right| > \eps \right \}$
is upper bounded by $2 \exp(-c m \eps^2)$.

\textbf{1) Upper tail}
\begin{align*}
\p(\wh{\psi} \geq (1 + \eps) \psi^*) &= \p \left(\frac{t_1}{\Phi^{-1}\left(\frac{1}{2}(1 + \frac{m_1}{m}) \right)} \geq  (1 + \eps) \psi^*  \right) \\
                                   &= \p \left(\frac{m_1}{m} \leq 2 \Phi \left(\frac{t_1}{(1 + \eps) \psi^*} \right) - 1 \right) \\
&= \p \left(\frac{m_1}{m} - \E \left[\frac{m_1}{m} \right] \leq 2 \left\{ \Phi \left(\frac{t_1}{(1 + \eps) \psi^*} \right) - \Phi \left(\frac{t_1}{\psi^*}   \right) \right\} \right)
\end{align*}
We have
\begin{align*}
 \Phi \left(\frac{t_1}{(1 + \eps) \psi^*} \right) - \Phi \left(\frac{t_1}{\psi^*}   \right) &= -\int_{t_1/ (\psi^* (1 + \eps))}^{t_1/\psi^*} \phi(u) \; du \\
&\leq -\phi(t_1 / \psi^*) \frac{t_1}{\psi^*} \frac{\eps}{\eps + 1} \\
&\leq -\phi(t_1 / \psi^*) \frac{t_1}{\psi^*} \frac{\eps}{2} = \phi'(t_1/\psi^*) \frac{\eps}{2}.
\end{align*}
for $\eps \in (0,1)$. Thus
\begin{align*}
\p(\wh{\psi} \geq (1 + \eps) \psi^*) \leq \p \left(\frac{m_1}{m} - \E \left[\frac{m_1}{m} \right] \leq \eps \phi'(t_1/\psi^*)\right)
\end{align*}

\textbf{2) Lower tail}\\[1ex]
Similarly, we obtain that
\begin{align*}
\p(\wh{\psi} \leq (1 - \eps) \psi^*) &\leq \p \left(\frac{m_1}{m} - \E \left[\frac{m_1}{m} \right] \geq 2 \left\{\Phi \left(\frac{t_1}{(1 - \eps) \psi^*}  \right) - \Phi \left(\frac{t_1}{\psi^*} \right) \right\} \right)
\end{align*}
We have
\begin{align*}
\Phi \left(\frac{t_1}{(1 - \eps) \psi^*}  \right) - \Phi \left(\frac{t_1}{\psi^*} \right) &= \int_{t_1/\psi^*}^{t_1/( \psi^* (1 - \eps))} \phi(u) \; du
&\geq \phi(t_1/\psi^*) \frac{t_1}{\psi^*} \frac{\eps}{1 - \eps}
&\geq -\phi'(t_1/\psi^*) \eps.
\end{align*}
Thus,
\begin{equation*}
\p(\wh{\psi} \leq (1 - \eps) \psi^*) \leq  \p \left(\frac{m_1}{m} - \E \left[\frac{m_1}{m} \right] \leq 2 \eps (-\phi'(t_1/\psi^*)) \right).
\end{equation*}\\[1ex]
Note that $m_1$ is a Binomial random variable. Applying Hoeffding's inequality
to \textbf{1)} and \textbf{2)}, we obtain that
\begin{align*}
\p(\wh{\psi} \geq (1 + \eps) \psi^*) &\leq \exp\left( -2 m \eps^2 \{ \phi'(t_1/\psi^*) \}^2 \right) \\
\p(\wh{\psi} \leq (1 - \eps) \psi^*) &\leq \exp\left( -4 m \eps^2 \{ \phi'(t_1/\psi^*) \}^2 \right).
\end{align*}
which proves the claim made above. %by choosing $c = 2  \{ \phi'(t/\psi) \}^2$ and $c' = 2 c$.
%\end{bew}

\section{Equivalence of $\wh{x}$ and $\wh{x}_{\lambda}$}\label{app:equivalence}
We here derive that $\wh{x}_{\lambda} / \nnorm{\wh{x}_{\lambda}}_2 = \wh{x}$ if $\nnorm{\wh{x}_{\lambda}}_2 > 0$ and
$\wh{x}_{\lambda} = \wh{x} = 0$ otherwise, with $\wh{x}$ and $\wh{x}_{\lambda}$ defined in
\eqref{eq:canlin} and \eqref{eq:projectedmarginal}, respectively, provided that $\mc{K}$ is a cone (i.e.,~$\alpha \mc{K} = \mc{K}$ for all $\alpha > 0$).

Let us consider the minimization problem that defines $\wh{x}_{\lambda}$. Since
$\mc{K}$ is a cone, we have for any $z \in \R^n$
\begin{equation}\label{eq:projectK}
\min_{x \in \lambda \mc{K}} \nnorm{x - z}_2^2 =
\min_{x \in \mc{K}} \nnorm{x - z}_2^2 =
\min_{\substack{u \in \mc{K}_1 \\ \alpha \geq 0}} \nnorm{\alpha u - z}_2^2,
\quad \text{where} \; \mc{K}_1 \coloneq \mc{K} \cap \mathbb{S}^{n-1}.
\end{equation}
The latter optimization problem is equivalent to
\begin{equation}\label{eq:min_alpha_u}
\min_{\substack{u \in \mc{K}_1 \\ \alpha \geq 0}} \frac{1}{2} \alpha^2 -
\alpha \scp{u}{z}.
\end{equation}
Note that for any fixed $u$, the minimizing $\alpha$ is given by
$\alpha^*(u) = \max\{\scp{u}{z}, 0\}$.

\textbf{Case 1}:\\
If $\max_{u \in \mc{K}_1} \alpha^*(u) = 0$, then $0$ is the unique minimizer
of \eqref{eq:projectK}. This immediately implies that $0$ is also a minimizer
of $\min_{x \in \mc{C}} -\scp{x}{z}$, where we recall that $\mc{C} = \mc{K} \cap B_2^n$.

\textbf{Case 2}:\\
Otherwise, we then must have $\min_{z \in \mc{C}} -\scp{x}{z} = \min_{x
  \in \mc{K} \cap \mathbb{S}^{n-1}} -\scp{x}{z} = \min_{x \in \mc{K}_1}
-\scp{x}{z}$. On the other hand, substituting the expression for
$\alpha^*(u)$ back into \eqref{eq:min_alpha_u}, we obtain the optimization problem $\min_{u \in \mc{K}_1} -\frac{1}{2} \scp{u}{z}^2$, which is equivalent to the previous one. Hence $\wh{x}_{\lambda}$ and $\wh{x}$ only differ by a scalar multiple.

\section{Computation of $\wh{x}$ for $\mc{K} = B_0(s;n)$}\label{app:computation}

\begin{lemmaApp}\label{lem:marginal_nonlinear} Consider the optimization problem in
\eqref{eq:canlin} with $\mc{K} = B_0(s;n)$, i.e.
\begin{equation*}
\wh{x} \in \argmin_{x \in B_0(s;n) \cap B_2^n} -\scp{\eta}{x}
\end{equation*}
Suppose that $\eta$ has distinct entries so that $|\eta_{(1)}| > \ldots > |\eta_{(n)}|$, where $|\eta_{(k)}|$ denotes the
k-th largest value of $\{|\eta_j|, \; j \in [n] \}$, $k \in [n]$. It then holds that $\wh{x} = \wt{x} / \nnorm{\wt{x}}_2$, where $\wt{x}_j = \eta_j I(|\eta_j| \geq |\eta_{(s)}|), \; j \in [n]$.
\end{lemmaApp}
We note that if the $\{a_i \}_{i=1}^n$ are Gaussian, $\eta$ has distinct entries with probability one.
\begin{bew} Let $\emptyset \neq S \subseteq \{1,\ldots,n\}$. Then
for any unit vector $x$ supported on $S$, $\scp{\eta}{x} \leq
\nnorm{\eta_S}_2$ which is attained by setting $x_S = \eta_S /
\nnorm{\eta_S}_2$. Consequently,
\begin{equation*}
\min_{x \in B_0(s;n) \cap B_2^n} -\scp{\eta}{x}
= \min_{S: |S| \leq s} -\nnorm{\eta_S}_2.
\end{equation*}
The optimization problem on the right hand side can be solved by
finding the index set of the $s$ largest component (in absolute magnitude) in
$\eta$. This yields the claim.
\end{bew}

\section{Gaussian widths for the examples in $\S$\ref{sec:classes_signals}}\label{app:classes_signals}

1) \emph{Sparsity.}

To make this paper self-contained and since examples 2) and 3) depend on this result as well, we repeat the argument from the proof of Lemma 2.3 in \cite{PlanVershynin2013a} at this point, and provide
explicit constants. For $1 \leq r \leq n$, note that $B_0(r;n) \cap \mathbb{S}^{n-1}$ is the union over the unit spheres of all $\binom{n}{r}$ subspaces of dimension $r$ in $\R^n$. Accordingly, we obtain that
\begin{align*}
\E \left[\sup_{v \in B_0(r;n) \cap \mathbb{S}^{n-1}} |\scp{g}{v}| \right] &= \E \left[\sup_{\substack{S \subseteq [n], |S|=r \\
                                                                                    v: \nnorm{v}_2 = 1, v_{S^c} = 0} } |\scp{g}{v}| \right]
\leq \E \left[\max_{S \subseteq [n], |S|=r} \nnorm{g_S}_2 \right].
\end{align*}
The latter expectation can be controlled by integrating and using a tail bound combined with the union bound:
\begin{align*}
&\E \left[\sup_{S \subseteq [n], |S|=r} \nnorm{g_S}_2 \right] = \int_{0}^{\infty} \p\left(\max_{S \subseteq [n], |S|=r} \nnorm{g_S}_2 > u \right) \; du \\
                                                                &= \int_0^{a}  \p \left(\max_{S \subseteq [n], |S|=r} \nnorm{g_S}_2 > u \right) \, du    + \int_{a}^{\infty} \p \left(\max_{S \subseteq [n], |S|=r} \nnorm{g_S}_2 > u \right) \, du,
\end{align*}
for any $a > 0$. Choosing $a = \sqrt{r} + \sqrt{2\log{\binom{n}{r}}}$, we obtain that
\begin{align*}
\E \left[\sup_{S \subseteq [n], |S|=r} \nnorm{g_S}_2 \right] &\leq \sqrt{r} + \sqrt{2\log{\binom{n}{r}}} + \int_{0}^{\infty} \p \left(\max_{S \subseteq [n], |S|=r} \nnorm{g_S}_2 > a + u \right) \, du  \\
&\leq   \sqrt{r} + \sqrt{2\log{\binom{n}{r}}}  + \int_{0}^{\infty} \binom{n}{r} \p \left(\nnorm{(g_1,\ldots,g_r)^{\T}}_2 > a + u \right) \, du   \\
\end{align*}
by applying the union bound. From $\E[\nnorm{(g_1,\ldots,g_r)^{\T}}_2] \leq \sqrt{r}$ and the well-known tail bound $\p(\nnorm{(g_1,\ldots,g_r)^{\T}}_2 - \E[\nnorm{(g_1,\ldots,g_r)^{\T}}_2] > t) \leq \exp(-t^2 / 2)$ for any $t \geq 0$ (cf.~Proposition 5.34 in \cite{Vershynin2010}), we obtain that
\begin{equation*}
\int_{0}^{\infty} \binom{n}{r} \p \left(\nnorm{(g_1,\ldots,g_r)^{\T}}_2 > a + u \right) \, du  \leq \int_0^{\infty} \exp(-u^2 / 2) \; du = \sqrt{\pi/2}.
\end{equation*}
Using that $\binom{n}{r} \leq \exp(r \log(e n / r))$, putting together the pieces and setting $r = 2s$, the derivation is complete.

2) \emph{Fused Sparsity.}

Let us start by fixing some notation. For $k \in [n]$, denote by $\mc{P}(k;n)$ the set of all collection of subsets of $[n]$ defining a partitioning into $k$ blocks, i.e.,~
\begin{align*}
\mc{P}(k;n) = \Big\{\, \{ B_{\ell} \}_{\ell = 1}^k&:\;\exists \{ a_{\ell}
  \}_{\ell = 1}^k \subset \mathbb{N},\; 1 = a_1 \leq \ldots \leq a_k\leq n \,\;\\
  &\text{s.t.}  \; B_{\ell} = \{j \in [n]:\, a_{\ell} \leq j  \leq a_{\ell+1} - 1 \}, \; \ell=1,\ldots,k-1,\quad B_k = \{a_k,\ldots,n\}.   \Big\}
\end{align*}
Moreover, for $B \subseteq [n]$ let $\mathbf{1}_{B} \in \R^n$ denote the
indicator vector of the index set $B$, i.e.,~$(\mathbf{1}_{B})_j = 1$ if $j \in
B$ and $(\mathbf{1}_{B})_j = 0$ otherwise.

Note first that $\Delta(\mc{C}) \subset \text{PC}(2s;n)$. We hence have
\begin{align*}
\overline{\Delta}(\mc{C}) \subset \text{PC}(2s;n) \cap \mathbb{S}^{n-1} &= \left\{x
  \in \R^n:\; \sum_{\ell = 1}^{2s} \bm{1}_{B_{\ell}} c_{\ell}, \, c_{\ell} \in \R, \, \{ B_{\ell}
  \}_{\ell = 1}^{2s} \in \mc{P}(2s;n), \; \nnorm{x}_2 =  1   \right\} \\
&=  \left\{x \in \R^n: x = \sum_{\ell = 1}^{2s}
  \frac{\bm{1}_{B_{\ell}}}{\sqrt{|B_{\ell}|}} \theta_{\ell}, \; \{ B_{\ell}
  \}_{\ell = 1}^{2s} \in \mc{P}(2s;n),  \; \nnorm{\theta}_2 = 1 \right\}.
\end{align*}
Accordingly, we obtain that
\begin{align*}
w(\overline{\Delta}(\mc{C})) = \E \left[\sup_{z \in \overline{\Delta}(\mc{C})}
  \left| \scp{g}{z} \right| \right] = \E \left[ \sup_{\substack{\theta \in
      \mathbb{S}^{2s-1} \\ \{ B_{\ell} \}_{\ell=1}^{2s} \in \mc{P}(2s;n)}}
  \left| \sum_{\ell = 1}^{2s}  \scp{g}{\bm{1}_{B_{\ell}} \big /\sqrt{|B_{\ell}|}}
    \theta_{\ell} \right|     \right]
\end{align*}
For $\{ B_{\ell} \}_{\ell = 1}^{2s} \in \mc{P}(2s; n)$ arbitrary, consider the
random vector $g_{\{ B_{\ell} \}_{\ell = 1}^{2s}} =
\left(\scp{g}{\bm{1}_{B_\ell}/\sqrt{|B_{\ell}|}} \right)_{\ell =
  1}^{2s}$. Observe that $g_{\{ B_{\ell} \}_{\ell = 1}^{2s}} \sim N(0,
I_{2s})$. Therefore
\begin{align*}
\E \left[ \sup_{\substack{\theta \in \mathbb{S}^{2s-1} \\ \{ B_{\ell}
      \}_{\ell=1}^{2s} \in \mc{P}(2s;n)}} \left| \sum_{\ell = 1}^{2s}
  \scp{g}{\bm{1}_{B_{\ell}} \big /\sqrt{|B_{\ell}|}} \theta_{\ell} \right|    \right] &= \E \left[\sup_{\substack{\theta \in \mathbb{S}^{2s-1} \\ \{ B_{\ell}
      \}_{\ell=1}^{2s} \in \mc{P}(2s;n)}} \left| \scp{g_{\{ B_{\ell} \}_{\ell =
        1}^{2s}}}{\theta} \right| \right] \\ &= \E \left[\sup_{\{ B_{\ell}
      \}_{\ell=1}^{2s} \in \mc{P}(2s;n)} \norm{g_{\{ B_{\ell} \}_{\ell =
        1}^{2s}}}_2 \right]
\end{align*}
Noting that $|\mc{P}(2s;n)| \leq \binom{n-1}{2s-1}$, the last expectation can
be bounded as in the derivation of 1) above.

3) \emph{Group sparsity.}

We have $\Delta(\mc{C}) \subset B_{0,\mc{G}}(2s)$ and hence $w(\overline{\Delta}(\mc{C})) \leq w(B_{0,\mc{G}}(2s) \cap \mathbb{S}^{n-1})$. Analogously to 1), we have
\begin{align*}
w(B_{0,\mc{G}}(2s) \cap \mathbb{S}^{n-1}) = \E \left[\sup_{v \in B_{0,\mc{G}}(2s) \cap \mathbb{S}^{n-1}} |\scp{g}{v}| \right] \leq \E \left[\sup_{S \in [L], |S| \leq 2s} \nnorm{g_{I(S)}}_2 \right],\;\,
I(S) \coloneq \cup_{\ell \in S} G_{\ell}.
\end{align*}
Following along the lines of i), we obtain
$$w(B_{0,\mc{G}}(2s)) \leq \sqrt{2s \max_{\ell \in [L]} |G_l|} + \sqrt{4s \log(e L / 2s)} + \sqrt{\pi/2},$$
which concludes the derivation.

4) \emph{Low-rank matrices.}

We have
\begin{equation*}
\overline{\Delta}(\mc{C}) \subset \{X \in \R^{n_1 \times n_2}:\,\text{rank}(X) \leq 2r,\; \nnorm{X}_F \leq 1\} \subset \{X \in \R^{n_1 \times n_2}:\,\nnorm{X}_* \leq \sqrt{2r},\; \nnorm{X}_F \leq 1\},
\end{equation*}
where $\nnorm{\cdot}_*$ denotes the Schatten-one norm (the sum of the singular values).
Let $G$ be an $n_1$-by-$n_2$ random matrix whose entries are i.i.d.~$N(0,1)$. From the containment above, we obtain that
\begin{equation*}
w(\overline{\Delta}(\mc{C})) \leq \E \left[\sup_{X \in \R^{n_1 \times n_2}:\, \nnorm{X}_F \leq 1, \nnorm{X}_* \leq \sqrt{2r}} |\scp{G}{X}| \right] \leq \sqrt{2r} \E[\nnorm{G}],
\end{equation*}
by the duality of the Schatten-one norm and the spectral norm. Gordon's Theorem (Theorem 5.3.2 in \cite{Vershynin2010}) states that $\E[\nnorm{G}] \leq \sqrt{n_1} + \sqrt{n_2}$, which completes the derivation.

5) \emph{$\ell_1$-ball constraint.}

We have
\begin{equation*}
\overline{\Delta}(\mc{C}) = \left\{\frac{x - x_u^*}{\nnorm{x - x_u^*}}_2: \; \, x \in B_2^n \setminus \{x_u^*\}, \; \,
\nnorm{x}_1 \leq \nnorm{x_u^*}_1 \right\}.
\end{equation*}
Denote by $S$, $|S| = s$, the support of $x_u^*$. For any $x$ with $\nnorm{x}_1
\leq \nnorm{x_u^*}_1$, we have $\nnorm{x_{S^c}}_1 \leq \nnorm{x_S -
  (x_u^*)_S}_1$ and thus
\begin{equation*}
\nnorm{x - x_u^*}_1 \leq \nnorm{x_{S^c}}_1 + \nnorm{x_{S} - (x_u^*)_S}_1 \leq 2
\nnorm{x_{S} - (x_u^*)_S}_1 \leq 2 \sqrt{s} \nnorm{x - x_u^*}_2.
\end{equation*}
It follows that
\begin{equation*}
\overline{\Delta}(\mc{C}) \subset \{v:\nnorm{v}_2 = 1, \;\nnorm{v}_1 \leq 2
\sqrt{s} \;\,\}  \subset \sqrt{2} \{v:\nnorm{v}_2 \leq 1, \;\nnorm{v}_1
\leq \sqrt{2s} \} \subset 2 \sqrt{2} \conv \{B_0(2s;n) \cap \mathbb{S}^{n-1}\},
\end{equation*}
where the last containment is proved in \cite{PlanVershynin2013b}, Lemma 3.1
therein. Since $w(\conv \{B_0(2s;n) \cap \mathbb{S}^{n-1}\}) = w(B_0(2s;n)
\cap \mathbb{S}^{n-1})$, the derivation is complete.

\section{Derivations  for the paragraph ``Beyond additive noise''  in $\S$\ref{sec:tradeoff}}\label{app:beyondadditive}
We fix $\sigma = 0$ and the corresponding Lloyd-Max optimal choices
$\mathbf{t} = \mathbf{t}_0^*$, $\bm{\mu} = \bm{\mu}_0^*$ so that
$\mu_k = \E[g | g \in \mc{R}_k], \; g \sim N(0, 1)$ with $\mc{R}_k =
\mc{R}_k(\mathbf{t}_0^*)$, $k \in [K]$.

\emph{Mechanism (I).} In order to evaluate $\lambda = \lambda_{b,p}$, we first need to derive an expression for the
corresponding map $\theta$. Recalling Definition \ref{defn:quant_alternative}, we have
\begin{equation*}
\E[y_1|a_1] = (1-p) \sum_{k=1}^{2^b} \wt{\mu}_k I(\scp{a_1}{x^*} \in \mc{Q}_k) + p \frac{1}{2^{b}-1} \sum_{k=1}^{2^b} \wt{\mu}_k I(\scp{a_1}{x^*} \notin \mc{Q}_k)
\end{equation*}
and thus
\begin{equation*}
\theta(z) = (1-p) \sum_{k=1}^{2^b} \wt{\mu}_k I(z \in \mc{Q}_k) + p \frac{1}{2^{b}-1} \sum_{k=1}^{2^b} \wt{\mu}_k I(z \notin \mc{Q}_k).
\end{equation*}
This is immediate from the definition of Mechanism (I) which keeps the original bin with probability $1 - p$, and selects
one of the remaining $2^{b}-1$ bins uniformly at random with probability $p$.

It follows that for $g \sim N(0,1)$
\begin{align*}
\lambda_{b,p} = \E[g \, \theta(g)] &= \sum_{k=1}^{2^b} \wt{\mu}_k \left\{(1 - p) \E \left[g I(g \in \mc{Q}_k)  \right] +
                                                       p \frac{1}{2^{b}-1} \E[g I(g \notin \mc{Q}_k)] \right\} \\
                               &= \sum_{k=1}^{2^b} \wt{\mu}_k \left\{(1 - p) \E \left[g I(g \in \mc{Q}_k)  \right] +
                                                       p \frac{1}{2^{b}-1} \E[g (1 - I(g \in \mc{Q}_k))] \right\} \\
                               &= \sum_{k=1}^{2^b} \wt{\mu}_k \left\{ (1 - p) - \frac{p}{2^{b} - 1} \right\} \E \left[g I(g \in \mc{Q}_k)  \right] \\
&=\sum_{k = 1}^K  \p(|g| \in \mc{R}_k)  \E[g | g \in \mc{R}_k]^2 \left\{ (1 - p) - \frac{p}{2^{b} - 1} \right\} \\
&= \scp{\bm{\alpha}_0(\mathbf{t}_0^*)}{\bm{E}_0(\mathbf{t}_0^*) \odot \bm{E}_0(\mathbf{t}_0^*)} \left\{ (1 - p) - \frac{p}{2^{b} - 1} \right\}
\\
&=\lambda_{b,0} \left\{ (1 - p) - \frac{p}{2^{b} - 1} \right\},
\end{align*}
where $\bm{\alpha}_0(\mathbf{t}_0^*)$ and $\bm{E}_0(\mathbf{t}_0^*)$ are defined at the end of Appendix \ref{app:lloydmax}. From the last expression we deduce the breakdown point $\bar{p}_b = 1 - 1/2^b$.

$\Psi_{b,p}$ can be evaluated by using the expression in Lemma \ref{lem:Omega_b}. The only thing that changes under Mechanism (I) are the probabilities $\bm{\alpha}(\mathbf{t})$ which become
\begin{equation*}
\alpha_k(\mathbf{t}) = \p(|g| \in \mc{R}_k(\mathbf{t})) (1-p) + \frac{p}{2^{b-1}}\sum_{l \neq k} \p(|g| \in \mc{R}_l(\mathbf{t})), \quad k \in [K].
\end{equation*}

\emph{Mechanism (II).} Following the same route as for Mechanism (I), one derives
\begin{equation*}
\theta(z) = (1-p) \sum_{k=1}^{2^b} \wt{\mu}_k I(z \in \mc{Q}_k) + p \left\{-\mu_K I(z \geq 0) + \mu_K I(z < 0) \right\}
\end{equation*}
and accordingly for $g \sim N(0,1)$
\begin{align*}
\lambda_{b,p} &= \E[g \, \theta(g)] = (1 - p) \sum_{k = 1}^K  \p(|g| \in \mc{R}_k)  \E[g | g \in \mc{R}_k]^2  - p \mu_K \E[g|g > 0] \\
             &=(1-p) \lambda_{b,0} - p \mu_K \sqrt{2/\pi}
\end{align*}
so that the breakdown points results as $\bar{p}_b = \lambda_{b,0}/(\lambda_{b,0} + \mu_K \sqrt{2/\pi})$. As for
Mechanism (I), $\Psi_{b,p}$ is obtained by evaluating the changes in $\bm{\alpha}(\mathbf{t})$. We have
\begin{align*}
&\alpha_k(\mathbf{t}) =  (1-p) \p(|g| \in \mc{R}_k(\mathbf{t})) , \quad k \in [K-1],\\
&\alpha_K(\mathbf{t}) =p \sum_{k=1}^{K-1} \p(|g| \in \mc{R}_k(\mathbf{t}))  +  \p(|g| \in \mc{R}_K(\mathbf{t})).
\end{align*}

\section{Auxiliary results from the theory of empirical processes}\label{sec:auxresults}
The first Lemma follows from a result known as symmetrization in the theory of empirical processes, cf.~Lemma 2.3.1 in \cite{VdVWbook}.
\begin{lemmaApp}\label{lem:symmetrization} Let $\Gamma$ be a set of measurable functions and $\{ Z_i \}_{i = 1}^m$ random variables and $\{ \xi_i \}_{i = 1}^m$ be i.i.d.~Rademacher random variables, i.e.,~$\p(\xi_1 = 1) = \p(\xi_1 = -1) = 1/2$. Then:
\begin{equation*}
\E \left[\sup_{\gamma \in \Gamma} \left| \frac{1}{m} \sm (\gamma(Z_i) - \E[\gamma(Z_i)]) \right| \right]
\leq 2 \E \left[\sup_{\gamma \in \Gamma} \left| \frac{1}{m} \sm \xi_i \gamma(Z_i) \right| \right]
\end{equation*}
\end{lemmaApp}
\noindent The next lemma is essentially derived in \cite{LedouxTalagrand}, $\S$6.1. We here also provide a detailed proof for the sake
of completeness.
\begin{lemmaApp}\label{lem:concentration1} In the setting of the previous Lemma, we have
\begin{align}\label{eq:lem:concentration1}
&\p \left(\sup_{\gamma \in \Gamma} \left| \frac{1}{m} \sm (\gamma(Z_i) - \E[\gamma(Z_i)]) \right| >  2 \E \left[\sup_{\gamma \in \Gamma} \left| \frac{1}{m} \sm (\gamma(Z_i) - \E[\gamma(Z_i)]) \right| \right] + u \right) \\
&\leq 4 \p \left(\sup_{\gamma \in \Gamma} \left|\frac{1}{m} \sm \xi_i \gamma(Z_i) \right| > u/2 \right) \notag.
\end{align}
\end{lemmaApp}
\begin{bew} Let $Z_i'$ be an independent copy of $Z_i$, $i \in [m]$. Observe that for any $a, t > 0$, we have that
\begin{align*}
&\p \left(\sup_{\gamma \in \Gamma} \left| \frac{1}{m} \sm (\gamma(Z_i) - \E[\gamma(Z_i)]) \right| \leq a \right)
\p \left(\sup_{\gamma \in \Gamma} \left| \frac{1}{m} \sm (\gamma(Z_i') - \E[\gamma(Z_i')]) \right| >  a + t \right)  \\ \leq
&\p \left(\sup_{\gamma \in \Gamma} \left| \frac{1}{m} \sm (\gamma(Z_i) - \E[\gamma(Z_i)])  - \frac{1}{m} \sm (\gamma(Z_i') - \E[\gamma(Z_i')]) \right| > t \right)
\end{align*}
Choose $a = 2 \E \left[\sup_{\gamma \in \Gamma} \left| \frac{1}{m} \sm (\gamma(Z_i) - \E[\gamma(Z_i)]) \right| \right]$. Then by Markov's inequality, the first term in the left hand side of the above inequality is
lower bounded by $1/2$ and thus
\begin{align*}
&\p \left(\sup_{\gamma \in \Gamma} \left| \frac{1}{m} \sm (\gamma(Z_i') - \E[\gamma(Z_i')]) \right| >  2 \E \left[\sup_{\gamma \in \Gamma} \left| \frac{1}{m} \sm (\gamma(Z_i) - \E[\gamma(Z_i)]) \right| \right] + t \right) \\
&\leq 2   \p \left(\sup_{\gamma \in \Gamma} \left| \frac{1}{m} \sm (\gamma(Z_i) - \E[\gamma(Z_i)])  - \frac{1}{m} \sm (\gamma(Z_i') - \E[\gamma(Z_i')]) \right| > t \right) \\
&\leq 2  \p \left(\sup_{\gamma \in \Gamma} \left| \frac{1}{m} \sm \xi_i (\gamma(Z_i) - \gamma(Z_i')) \right| > t \right) \\
&\leq 2  \p \left(\sup_{\gamma \in \Gamma} \left| \frac{1}{m} \sm \xi_i
    \gamma(Z_i) \right| + \sup_{\gamma \in \Gamma} \left|\frac{1}{m} \sm \xi_i \gamma(Z_i') \right| > t \right) \\
&\leq 4  \p \left(\sup_{\gamma \in \Gamma} \left| \frac{1}{m} \sm \xi_i \gamma(Z_i) \right| > t/2 \right)
\end{align*}
\end{bew}
\noindent The next Lemma is taken from \cite{Boucheron2013}, Theorem 5.8.
\begin{lemmaApp}\label{lem:concentration2} Let $\mc{T}$ be a totally bounded subset of some metric space and let $(Z_t)_{t \in \mc{T}}$ be an almost surely continuous Gaussian process
indexed by $\mc{T}$ such that $\sigma_{Z}^2 \coloneq \sup_t \E[Z_t^2] < \infty$. Then $\forall r > 0$
\begin{equation*}
\p \left(\sup_{t} Z_t  - \E \left[\sup_{t} Z_t \right] \geq r \right) \leq \exp(-r^2 / (2\sigma_Z^2)).
\end{equation*}
\end{lemmaApp}
\noindent The following result can be found in \cite{Rudelson2013}.
\begin{lemmaApp}\label{lem:concentration3} Let $\mathbf{Z} = (Z_i)_{i = 1}^m$ be a sub-Gaussian random vector whose components $(Z_i)_{i = 1}^m$ are i.i.d.~zero-mean and variance $\sigma_Z^2$. Then:
\begin{equation*}
\p(\nnorm{\mathbf{Z}} > \E[\nnorm{\mathbf{Z}}] + t) \leq \exp(-c t^2 / \sigma_Z^2),
\end{equation*}
where $c > 0$ is a universal constant depending only on the sub-Gaussian norm of $Z_1/\sigma_Z$.
\end{lemmaApp}

%Define $\wt{\rho}_j = x_j^*/\sqrt{1 + \sigma^2}$

{
\bibliographystyle{plain}%model1a-num-names
\bibliography{bbitmarginal_longversion}
}

\end{document}